\documentclass[usenatbib]{mn2e}
\usepackage{epsfig}
\usepackage{rotating}
\bibpunct{(}{)}{,}{a}{}{,} 

\newcommand{\HI}{H\,{\sc i}}
\newcommand{\HII}{H\,{\sc ii}}

\newcommand{\kms}{~km\,s$^{-1}$}
\newcommand{\kkms}{km\,s$^{-1}$}

\newcommand{\Msun}{~M$_{\odot}$}
\newcommand{\LLsun}{L$_{\odot}$}
\newcommand{\Lsun}{~L$_{\odot}$}

\title[The Local Volume HI Survey: Galaxy Kinematics]
      {The Local Volume HI Survey: Galaxy Kinematics\thanks{The observations 
       were obtained with the Australia Telescope which is funded by the
       Commonwealth of Australia for operations as a National Facility
       managed by CSIRO.}}

\author[E.~Kirby et al.]
       {Emma M. Kirby$^{1,2}$, 
        B\"arbel Koribalski$^2$, 
        Helmut Jerjen$^1$,
        \'Angel L\'opez S\'anchez$^{2,3}$ \\
        $^1$Research School of Astronomy and Astrophysics, 
            Australian National University, Cotter Rd, Weston, 
            ACT 2611, Australia \\
        $^2$Australia Telescope National Facility, CSIRO Astronomy \& 
            Space Science, P.O. Box 76, Epping, NSW 1710, Australia \\
        $^3$Australian Astronomical Observatory, PO BOX 296, Epping, NSW 1710, Australia\\
}
 
\date{Received date; accepted date}

\begin{document}

\maketitle

 \begin{abstract}

We present a detailed analysis of the neutral hydrogen kinematics of 12 nearby dwarf irregular galaxies
observed as part of the Local Volume HI Survey (LVHIS) conducted at the Australia Telescope Compact  Array. 
For each galaxy we measure the disk parameters (inclination, position angle) and the \HI\ rotation curve. 
Six galaxies in our sample (AM0605-341, Argo Dwarf, ESO059-G001, ESO137-G018, ESO174-G?001, 
ESO308-G022) have their atomic hydrogen distribution studied for the first time. AM0605-341was found to have an extension of redshifted \HI\ which we propose is due to a tidal interaction with NGC2188.  There is evidence that ESO215-G?009 has extraplanar \HI\ gas. We also 
compare the global galaxy properties, in particular the integrated \HI\ flux density and velocity widths of the 
observed \HI\ spectra with the results from the low angular resolution \HI\ Parkes All Sky 
Survey (HIPASS). We discuss under what circumstances the 21cm emission line profile can accurately predict
the galaxyÕs rotation velocity, an observational parameter crucial to study the classical and baryonic 
Tully-Fisher relations. 
\end{abstract}

 
\section{Introduction}
The kinematics of the gas and stars in galaxies are often very complex,
reflecting both internal and external processes such as turbulence,
infall/outflow, rotation, tidal interactions and ram pressure stripping.
The rotational component, which dominates in most spiral galaxies,
provides a direct measure of the gravitational potential. In contrast,
small dwarf galaxies often show similar rotational and non-rotational 
amplitudes (e.g., \citealt{begum03}) that sometimes are consequence of interactions \citep{lopezsanchez10}. 
The rotation curve, i.e. the rotational velocity as a function of galaxy
radius, gives an accurate estimate of the total galaxy mass out to the
largest measured radial extent.
When deriving a galaxy's rotation curve from the data, projection effects 
as well as the shape and orientation of the disk/orbits must be taken 
into account. In particular, the majority of spiral galaxies display mild
or strong bars in the inner region and warps in the outer regions and these effects can make the derivation of an accurate rotation curve problematic.

Rotation curves have long been used as a tool to study the bulk motion of the 
\HI\ disk. Following several early studies of 21-cm (\HI) rotation curves,
e.g. by \cite{rogstad72} and \cite{roberts73}, \cite{bosma81}
carried out the first comprehensive analysis of relatively high resolution 
\HI\ rotation curves of 35 galaxies with a range of morphological types.
\cite{begeman89} described a new method, the tilted-ring analysis, to derive 
rotation curves and demonstrated it for the extended \HI\ disk of the spiral 
galaxy NGC~3198. This method is still the most commonly used for modelling
mildly inclined galaxy disks.

The Local Volume (LV; $D<10$ Mpc) is the ideal place to study the kinematics 
of galaxies. More than  550 galaxies are currently known to reside in this 
volume \citep{karachentsev08}, approximately 85\% of which are dwarf galaxies 
\citep{karachentsev04}. Due to their proximity, independent distances are 
available for most LV galaxies, and they can be studied with high spatial 
resolution. 

A large range of \HI\ rotation curve studies already exist and more are under 
way. While we are working towards a large set of excellent galaxy rotation 
curves, each individual galaxy requires suitable data and a lot of time to accurately
model their kinematics and mass distribution. Dwarf galaxies are particularly
difficult to model, because of their small size, low rotation velocity,
non-rotational motions such as turbulence, infall/outflow, etc. as well as
large neighbours. \cite{cote00} presented rotation curves for eight dwarf 
irregular galaxies located in the nearby Sculptor and Centaurus\,A groups, 
several of which are also discussed in this paper (see also \citealt{vaneymeren09}). \cite{swaters09} just published the \HI\ rotation curves of 62 late-type 
dwarf galaxies based on data from the WHISP project (see \citealt{swaters02}) 
while \cite{verheijen01a} derived rotation curves for 43 spiral galaxies in
the nearby Ursa Major cluster. Several high-resolution, high-sensitivity \HI\ surveys of nearby galaxies are 
currently underway which together will provide a comprehensive picture of the 
local galaxy dynamics and mass distribution. These are (1) `The \HI\ 
Nearby Galaxy Survey' (THINGS; \citealt{walter08}) which provides a detailed 
analysis of the rotation curves of 19 nearby galaxies \citep{deblok08,oh08}, (2) Little THINGS \citep{hunter07} which is obtaining deep HI-line maps of dwarf galaxies (3) the `Faint Irregular GMRT Galaxies Survey' (FIGGS; \citealt{begum07}) 
which provides detailed \HI\ velocity fields \citep{Begum06} and rotational
velocities (Begum et al. 2008a) for a large number of dwarf galaxies (4) VLA-ANGST (ACS Nearby Galaxy Survey Treasury; \citealt{ott10}) which is obtaining high resolution ($\sim$6Ó) \HI\ maps of 36 nearby galaxies and (5) WRST-LVHIS which a new northern hemisphere extension to the LVHIS survey. 
Most of the THINGS targets are well-resolved, large spiral galaxies, while
the FIGGS targets are faint dwarf irregular galaxies observed with very high 
velocity and angular resolution. The `Local Volume HI Survey' (LVHIS; 
Koribalski et al. 2010) will initially provide medium resolution ($\sim45''$)
\HI\ data for a complete sample of $\sim$70 southern galaxies (see Section~2) 
aimed at studying their large-scale gas distribution and kinematics. High 
resolution ($10''$) data will be published after further data processing.

This paper  is organised as follows: in Section~\ref{s:HIsample} we describe 
the 12 nearby galaxies that were selected for the rotation curve analysis,
followed by a brief outline of the radio observations and data reduction in 
Section~\ref{s:HIobsdata}. In Section~\ref{s:spectra} we discuss the galaxy
properties as measured and derived from interferometric and single dish \HI\ 
data. Rotation curves, $v(r)$, and disk orientation parameters are presented 
for each galaxy in Section~\ref{s:rotcurs}. In Section~\ref{s:rotcurspect} we
compare the rotation curve and spectral properties of our sample galaxies. 
The kinematics of individual galaxies are presented in 
Section~\ref{s:individualkinematics} and the Tully-Fisher relation discussed in Section~\ref{s:tfr}. Finally the results are summarised in 
Section~\ref{s:HIsummary}.

\section{The Galaxy Sample}\label{s:HIsample}
The `Local Volume \HI\ Survey' (LVHIS\footnote{http://www.atnf.csiro.au/research/LVHIS/}; \citealt{koribalski08, koribalski09}) is a large project which aims to provide \HI\ distributions, \HI\ velocity  fields and 20-cm radio continuum maps for all galaxies within 10 Mpc. For observations with the Australia Telescope Compact Array (ATCA) we initially chose only those LV galaxies that are detected in the \HI\ Parkes All-Sky Survey (HIPASS; \citealt{barnes01,koribalski04}) and reside south of approximately  --30 degrees declination. For a brief overview of the LVHIS observations and 
data reduction see Section~\ref{s:HIobsdata}.

While we would like to derive rotation curves and mass models for all $\sim$70 LVHIS galaxies observed with the ATCA, the typical angular resolution of the  \HI\ maps used here allows us to study only those galaxies with \HI\ diameters 
larger than $\sim$5\arcmin. Furthermore, a number of large, nearby galaxies have already been extensively
studied and good rotation curves are available in the literature  (see for example Circinus: \citealt{jones99}, ESO215-G?009: \citealt{warren04}, ESO245-G005, ESO381-G020, ESO325-G011, ESO444-G084  and UGC442:  \citealt{cote00}, M83: \citealt{tilanus93}, NGC247: \citealt{carignan90a}, NGC253: \citealt{puche91a}, NGC300: \citealt{puche90, westmeier10}, NGC625: \citealt{cannon04}, NGC1313: \citealt{ryder95}, NGC1512: \citealt{koribalski09a}, NGC1705: \citealt{meurer98}, NGC2188: \citealt{domgeorgen96}, NGC2915: \citealt{bureau99}, NGC4945: \citealt{ott01}, NGC5102: \citealt{vanwoerden93}, NGC5128: \citealt{schiminovich94}, NGC5253: \citealt{kobulnicky95}, NGC6822: \citealt{weldrake03}, NGC7793: \citealt{carignan90}). A tilted-ring analysis is most appropriate for galaxies with inclination angles
of $i$ = 30 -- 80 degrees. Edge-on galaxies ($i > 80$ degrees require a 
different method (the envelope tracing method of \citealt{sofue96,sofue97}) to derive robust 
rotation curves as the observed line-of-sight crosses a large range of 
projected velocities. 
For nearly face-on galaxies, non-rotational and rotational motions in the 
line-of-sight can be of similar amplitude, leading to large uncertainties in 
the derived rotation curve \citep{lewis75,meyer08}. 
Following \cite{deblok08} we chose a lower limit of $i$  = 30 degrees.
Using the above criteria, we select 12 LVHIS galaxies for our 
kinematical study.

\begin{figure*} 
\begin{tabular}{ccccc}
  \mbox{\epsfig{file=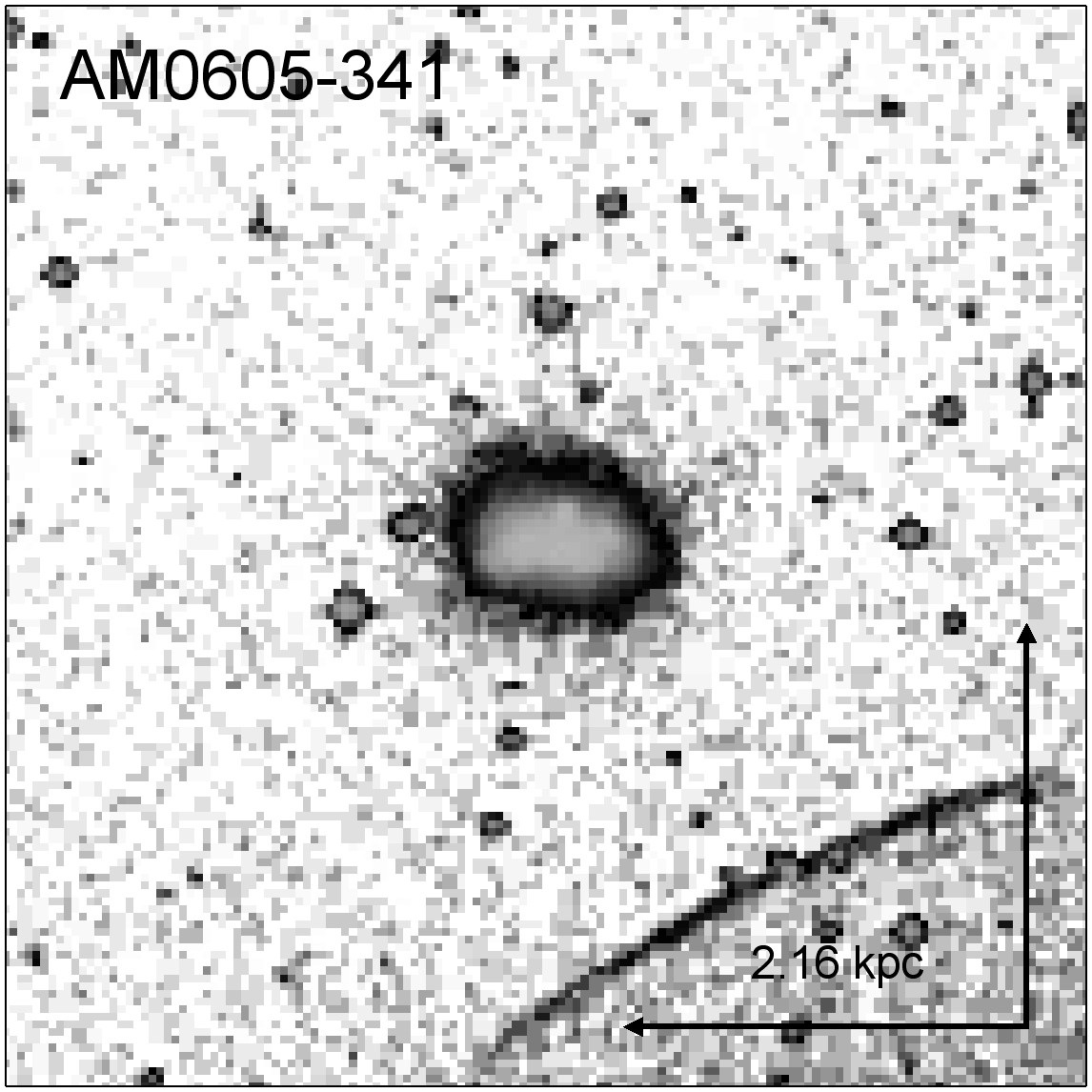,width=4cm}} &
  \mbox{\epsfig{file=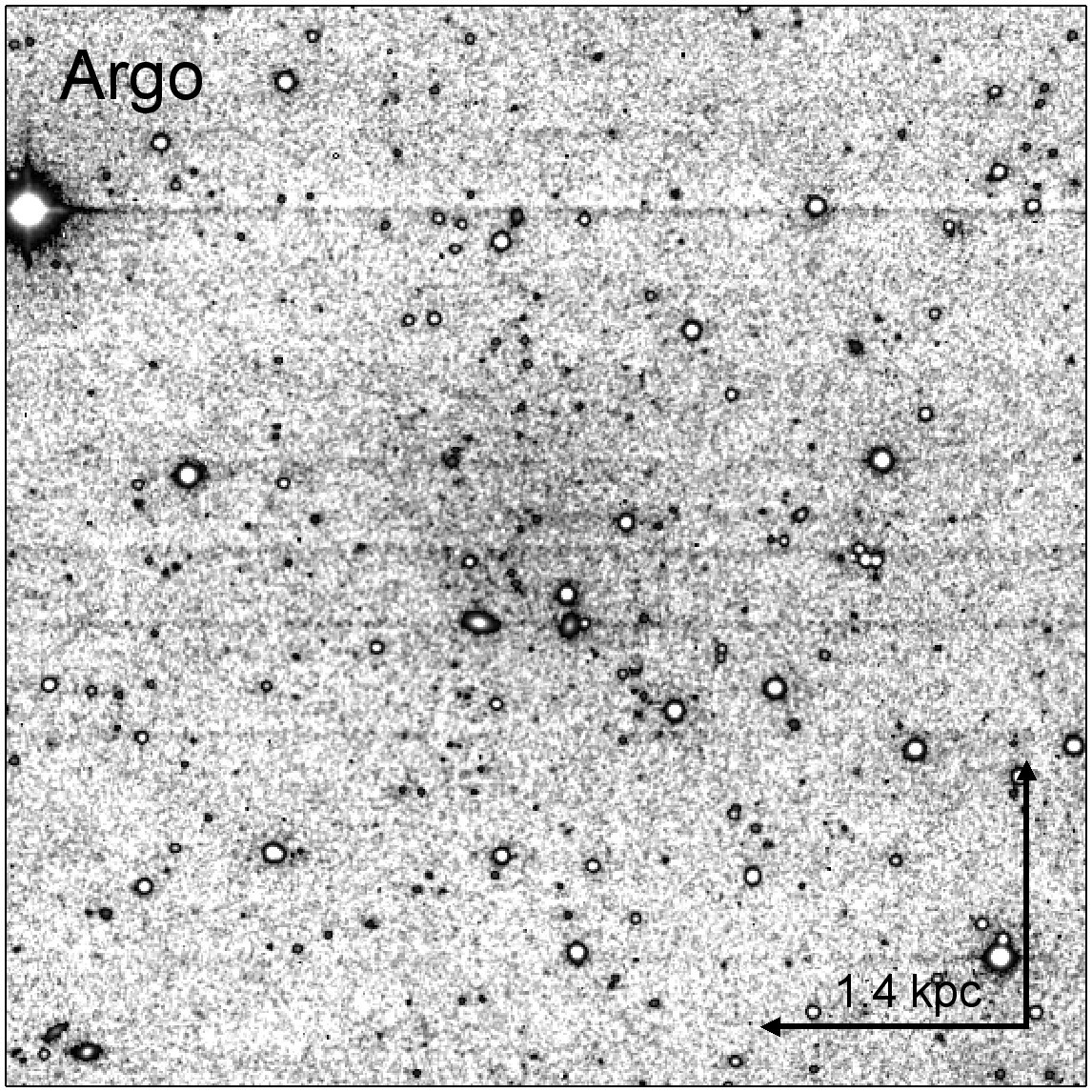,width=4cm}} &
  \mbox{\epsfig{file=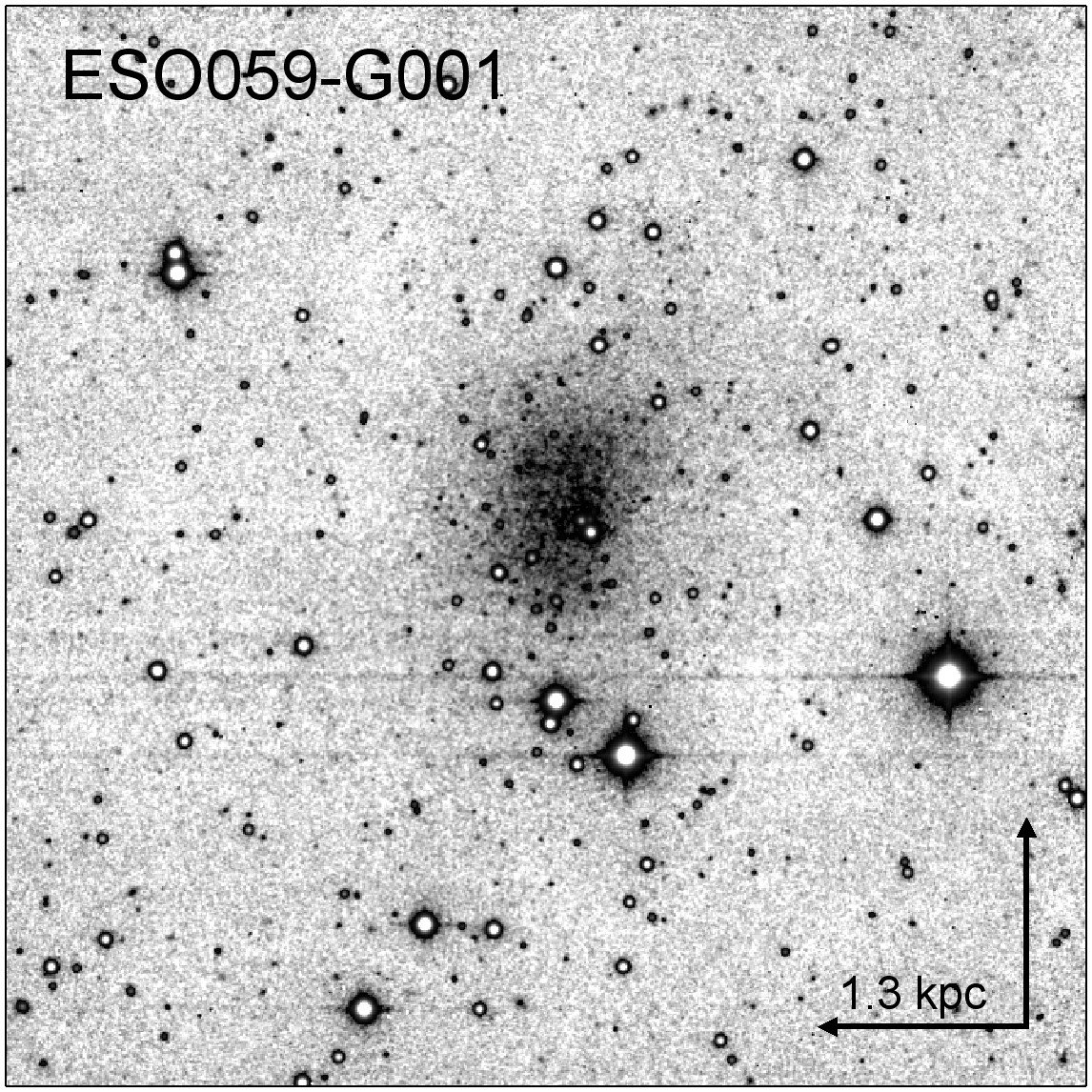,width=4cm}}  &
  \mbox{\epsfig{file=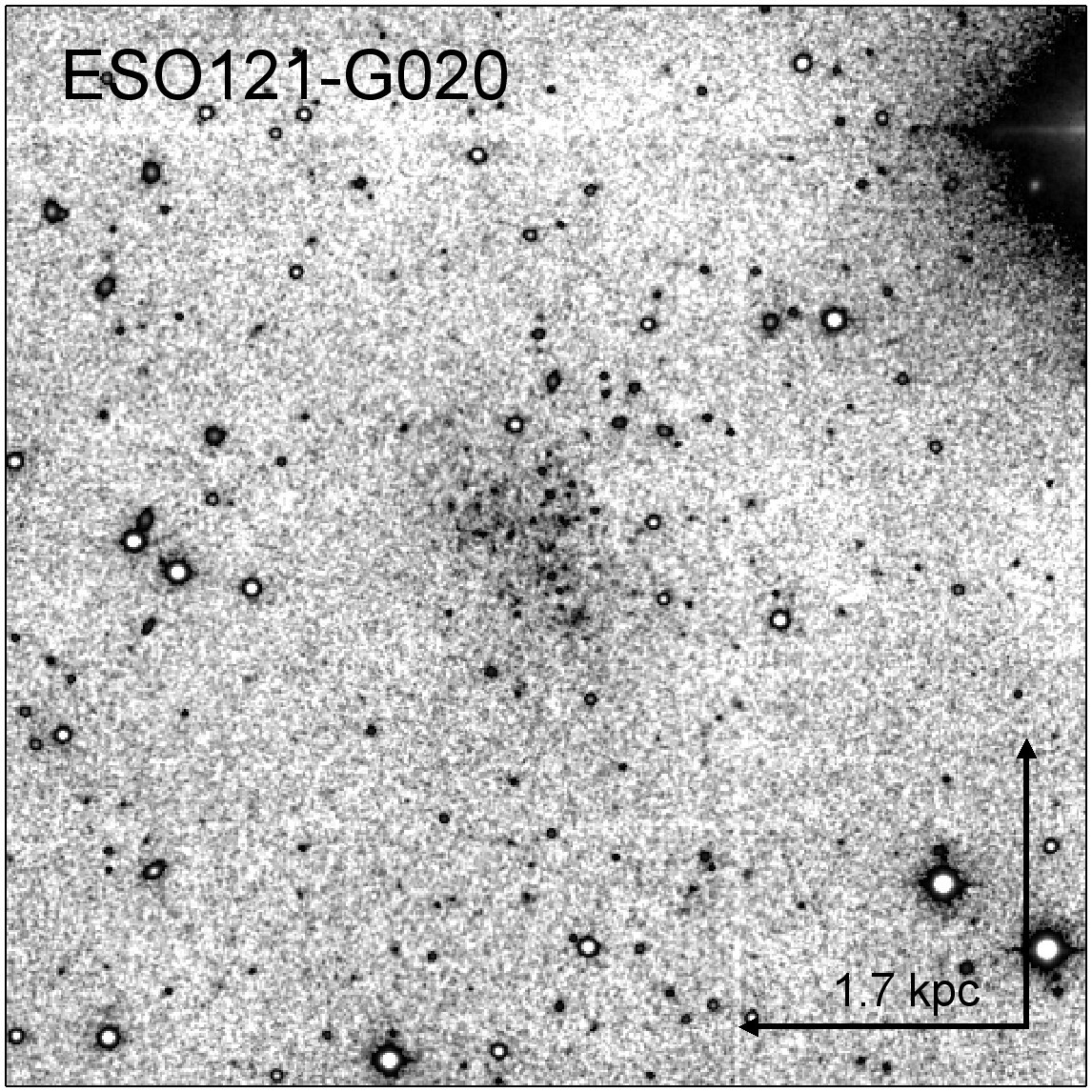,width=4cm}} \\
  \mbox{\epsfig{file=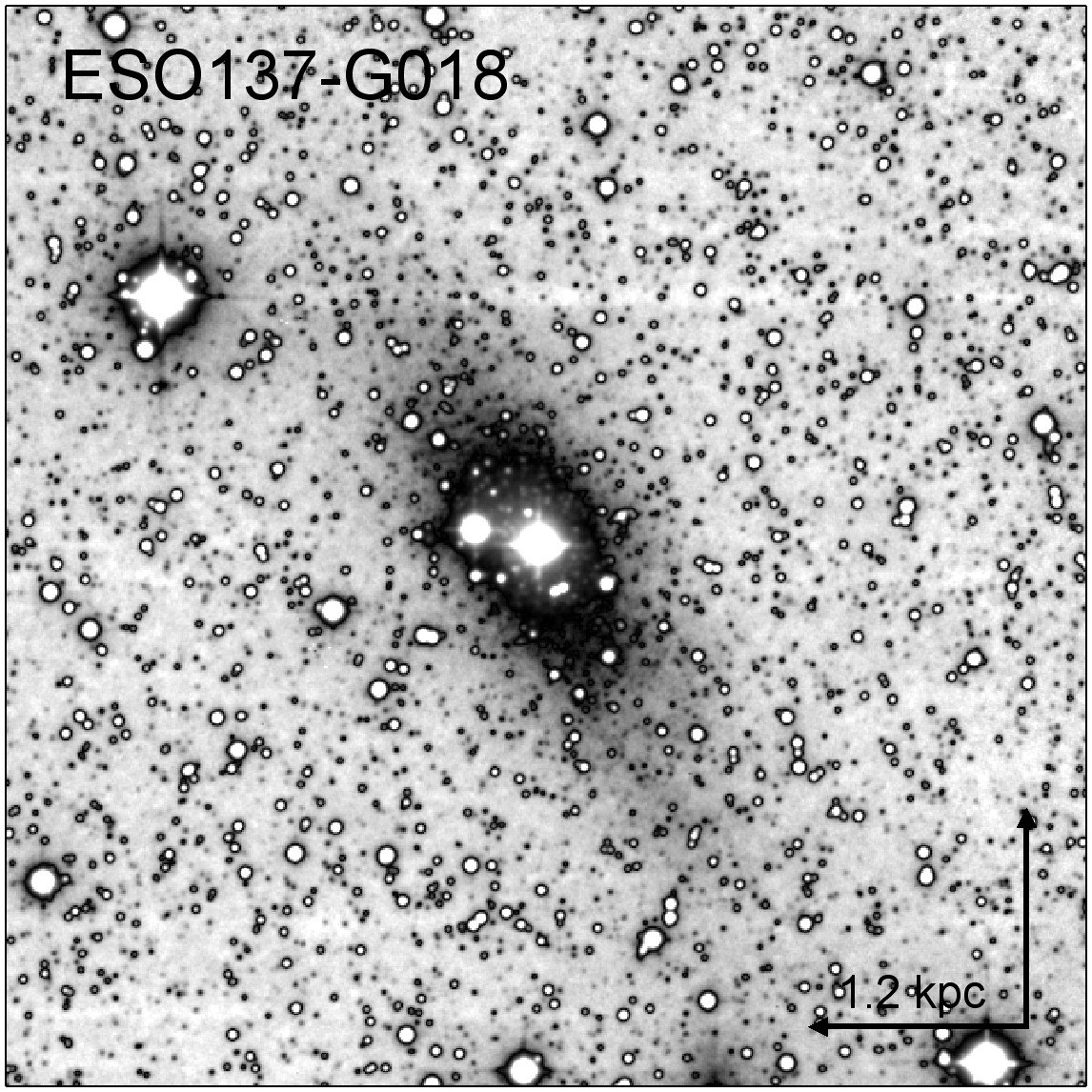,width=4cm}} &
  \mbox{\epsfig{file=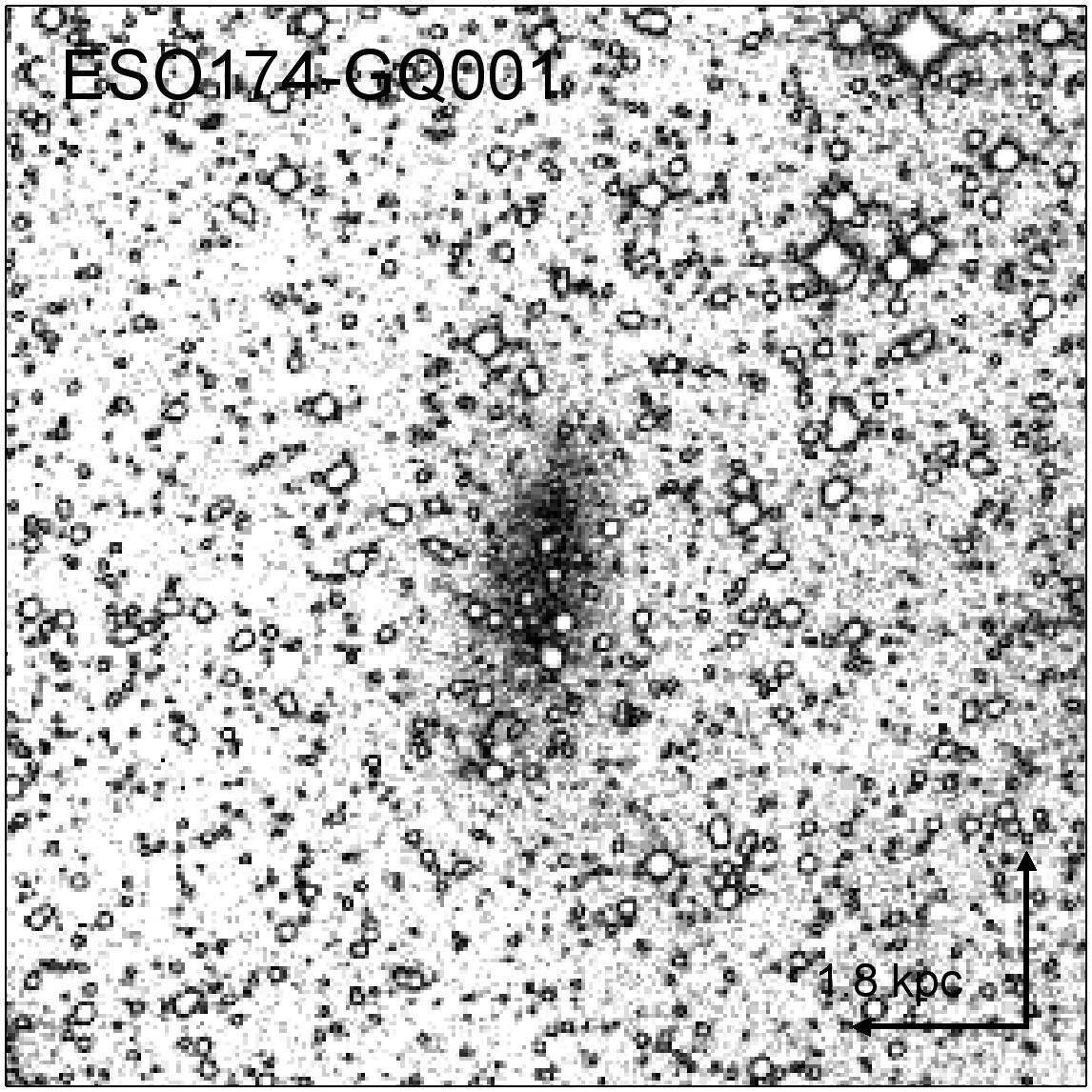,width=4cm}} &
  \mbox{\epsfig{file=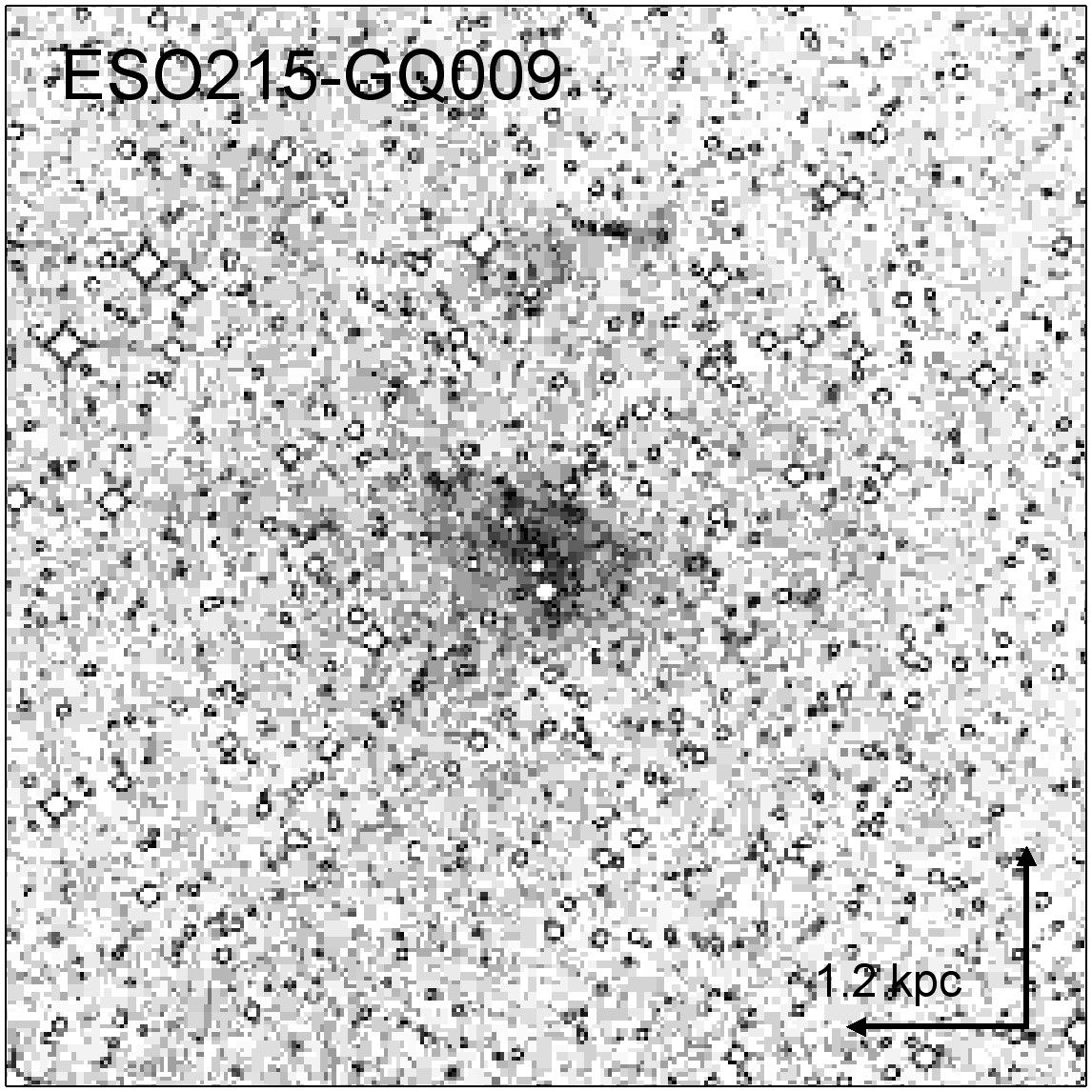,width=4cm}} &
  \mbox{\epsfig{file=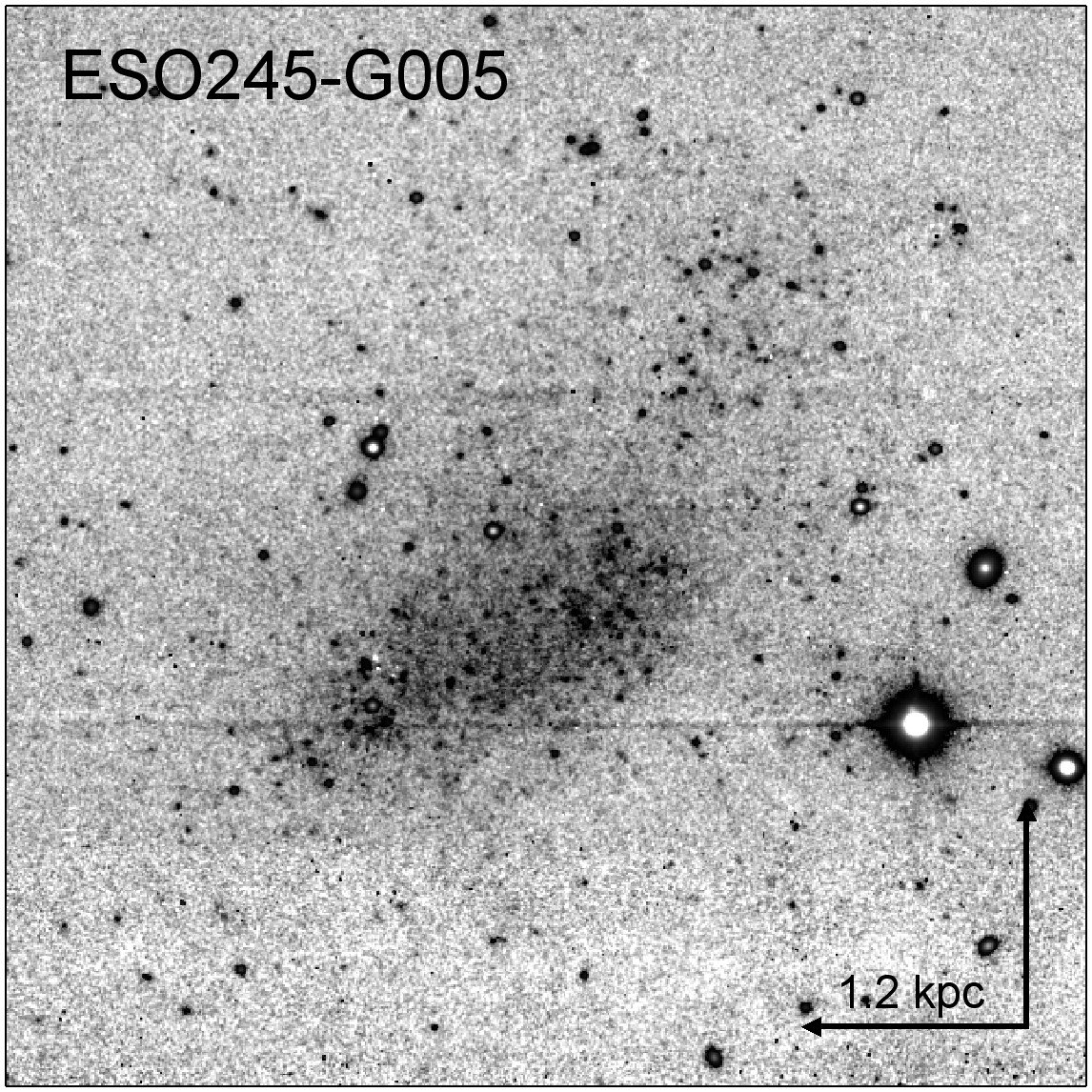,width=4cm}} \\
  \mbox{\epsfig{file=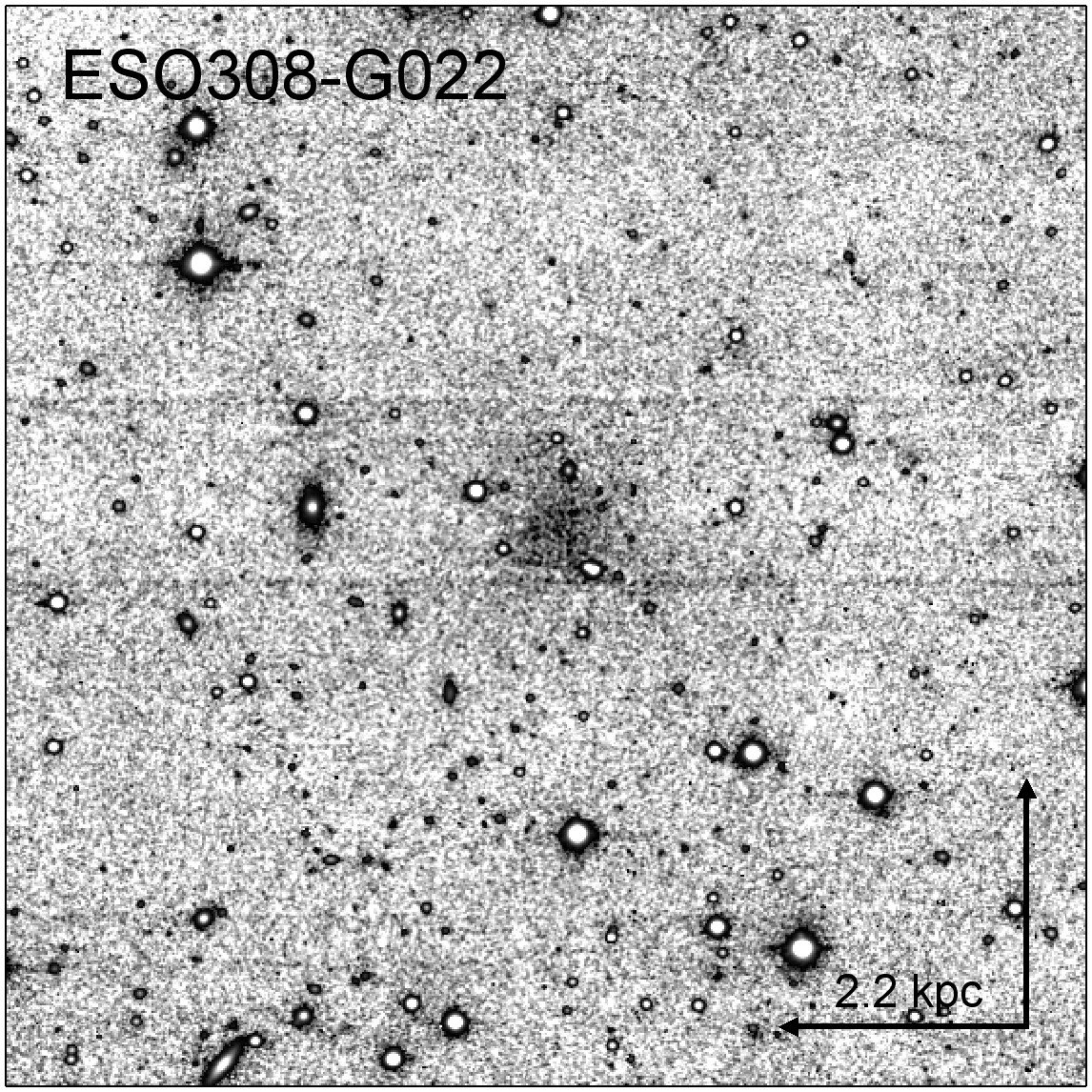,width=4cm}} &
  \mbox{\epsfig{file=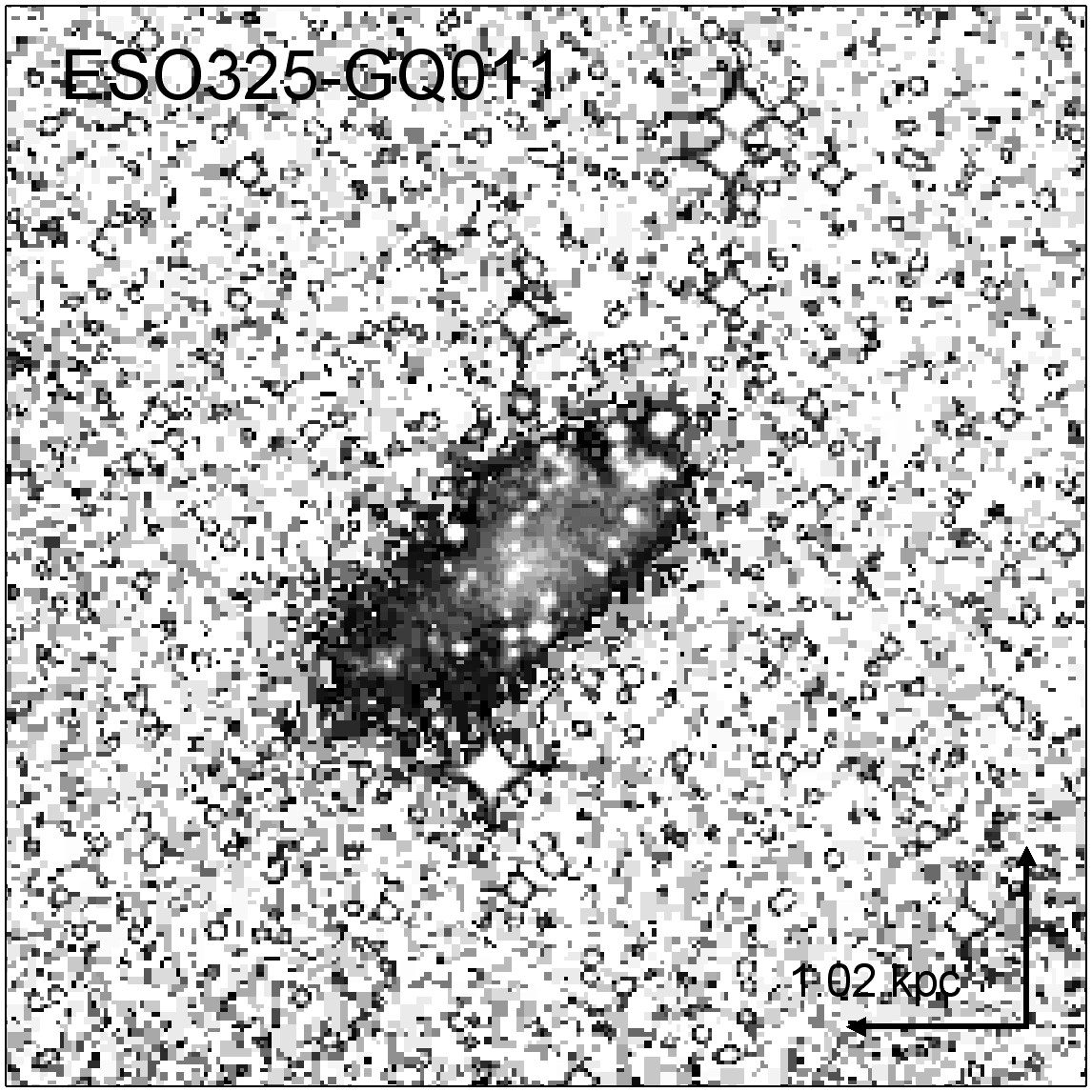,width=4cm}} &
  \mbox{\epsfig{file=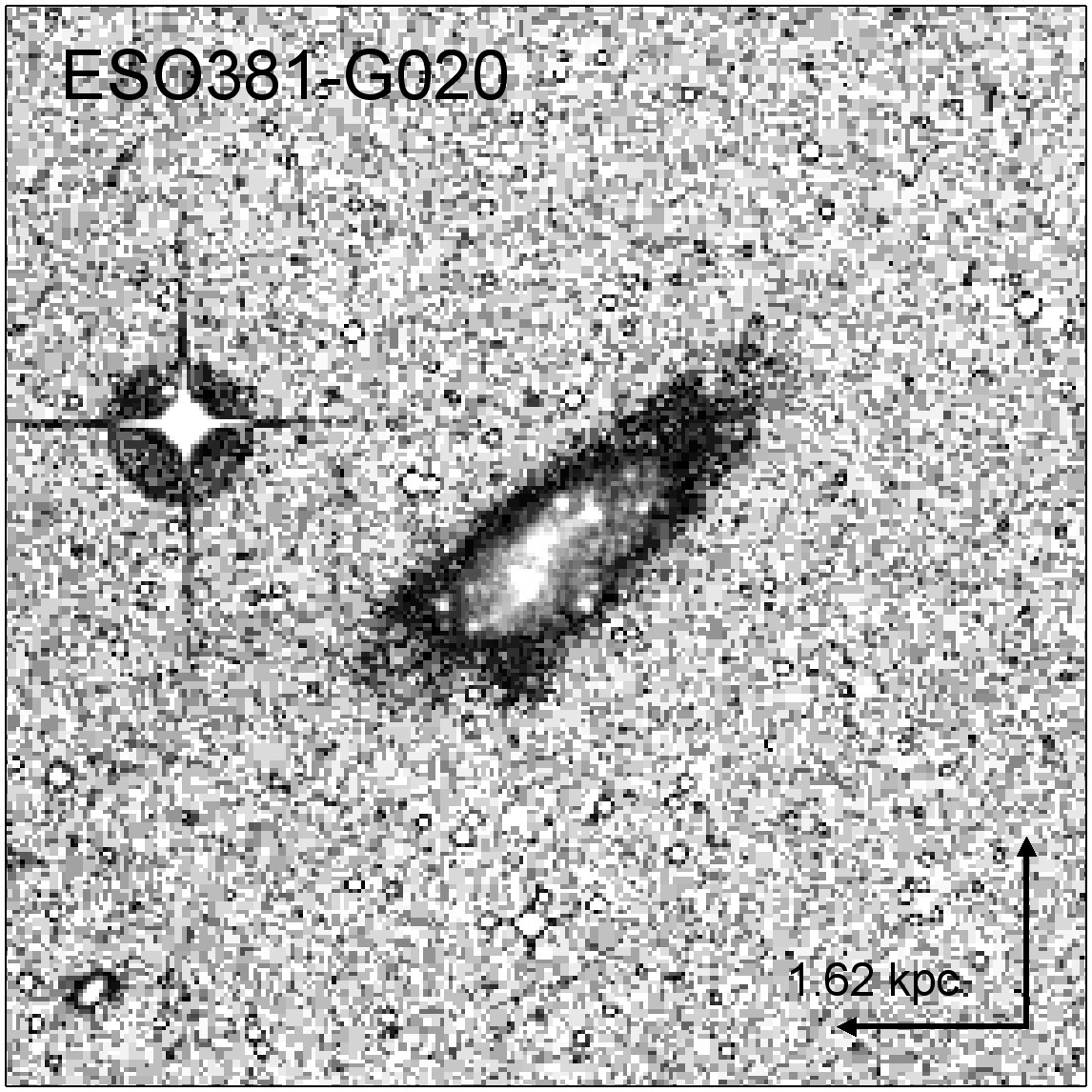,width=4cm}} &
  \mbox{\epsfig{file=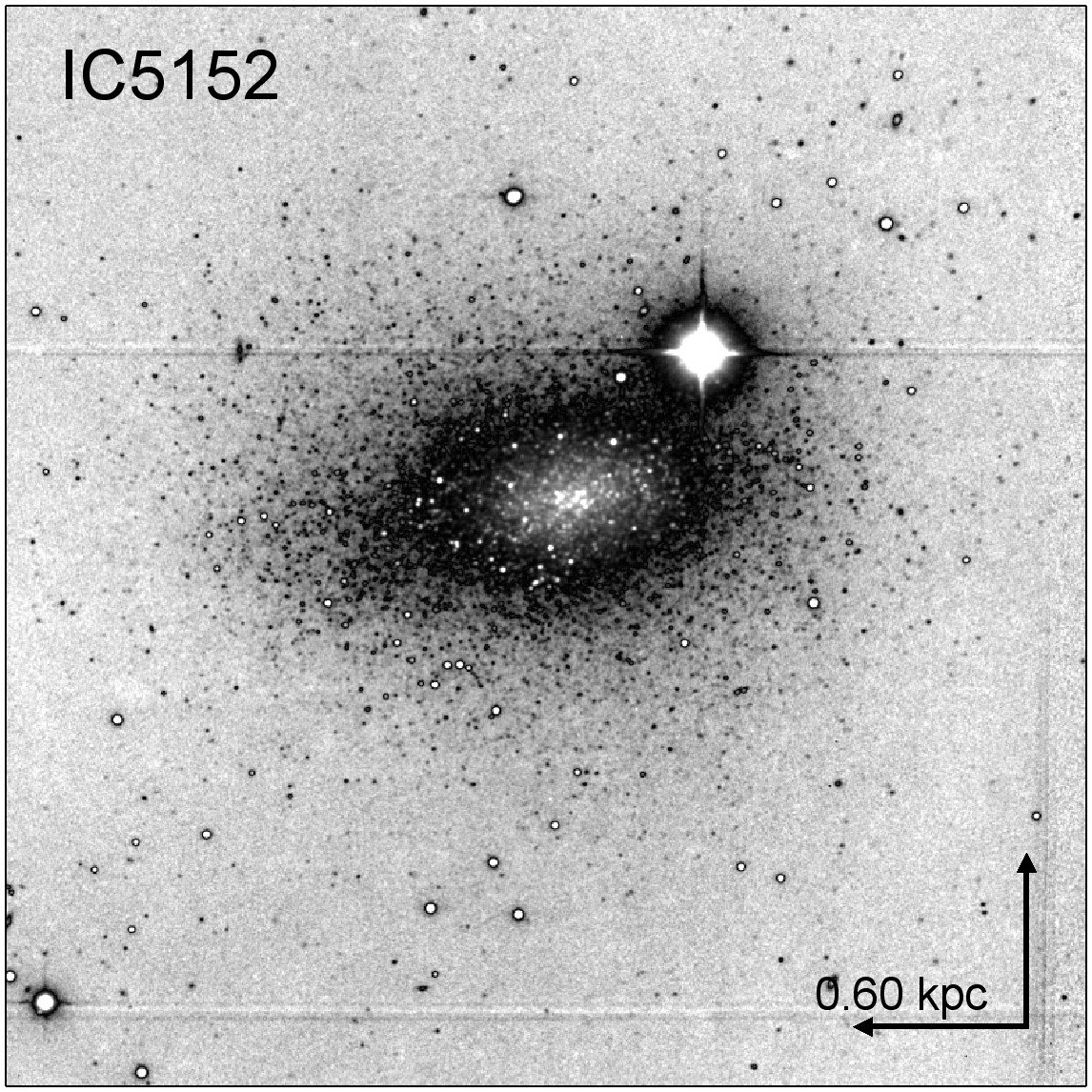,width=4cm}} \\
 \end{tabular}
\caption{Near-IR $H$-band images \citep{kirby08} for seven sample galaxies. 
  For another five galaxies (AM0605--341, ESO174-G?001, ESO215-G?009, 
  ESO325-G?011 and ESO381-G020) no $H$-band images are currently available.
  Instead we show their $B$-band (468 nm) images from the Digitized Sky Survey.
  The displayed scaling bar equals 1 arcmin; the corresponding linear scale is 
  also indicated. North is up and East is to the left. The $H$- and $B$-band 
  intensity is represented by a grayscale, which goes from white (low-) to 
  black (medium-) and then back to white (high-intensity).}
\label{fig:opt}
\end{figure*}

In Figure~\ref{fig:opt} we show deep ($24<\mu_{lim}<26$\,mag arcsec${}^{-2}$) near-IR, $H$-band images, if
available from \cite{kirby08}, or $B$-band (optical) images of the selected
galaxies. We prefer near-IR over optical images to reveal the full extent and
shape of the stellar disk because (a) dust attenuation is minimal and (b) the 
observed light emission is not dominated by that of short-lived giant O and B 
stars. Deep $H$-band  ($1.65 \mu m$) images are, unfortunately, not yet 
available for AM0605-341, ESO174-G?001, ESO215-G?009, ESO325-G?011 and 
ESO381-G020, so $B$-band (468 nm) Digitised Sky Survey (DSS) images are 
displayed instead. Based on their stellar distribution we find that all 12 
sample galaxies are dwarf irregular galaxies, some are of Magellanic type
 and some have central bars. 
 
The basic optical properties of the sample galaxies are given in  
Table~\ref{tab:basicprops}. The columns are organised as follows: 
\begin{itemize}
 \item[] {\emph{Column}} (1) --  galaxy name.
 \item[] {\emph{Columns}} (2) and (3) --  equatorial coordinates of the
     centre of the optical emission for the epoch of J2000, obtained from 
     the ESO-LV catalog, \citep{lauberts89} and the \cite{arp87} catalog. 
 \item[] {\emph{Columns}} (4) and (5) --  distance to the galaxy and an 
     indication of the method used: tip of the red giant branch (TRGB) or 
     Hubble flow distance (H), $D = v_{LG}/ H_{0}$, where $v_{LG}$ is the Local Group 
     velocity calculated using the HIPASS derived heliocentric velocity (see 
     Table~\ref{tab:hipassprops}) and coordinate transformation of 
     \cite{karachentsev96}. Here, $H_0 = 73$\kms\,Mpc$^{-1}$ is adopted  
     (Wilkinson Microwave Anisotropy Probe, WMAP; \citealt{spergel07}). 
     An uncertainty of 10\% is adopted \citep{karachentsev06}.
 \item[] {\emph{Columns}} (6) and (7) -- the position angle and inclination 
     of the outer isophotes in degrees. For the galaxies with $H$-band images available, the surface brightness at which these values were measured is typically 26 mag arcsec${}^{-2}$. When DSS images were used, the surface brightness is higher. The position angle is listed taken in the anti-clockwise direction, between the north direction on the sky and the major axis. It is important to note that these parameters are defined by the optical morphology and may not be equivalent to the position angle and inclination defined kinematically.
 \item[] {\emph{Column}} (8)  -- the total $B$-band magnitude and its source. 
     An uncertainty of 0.2 mag is adopted.
 \item[] {\emph{Column}} (9)  -- the total $H$-band magnitude obtained from 
    \cite{kirby08}.
 \item[] {\emph{Column}} (10) -- Reddening estimate, $E(B-V)$, which has an uncertainty of 16\% 
    \citep{schlegel98}. 
  \item[] {\emph{Column}} (11) --  the absolute $B$-band magnitude calculated as $M_B=m_B - 5\log D - 25 - A_B$ where $A_B=4.32\cdot E(B-V)$ \citep{schlegel98}.
\end{itemize}

The basic \HI\ properties of our sample were established by HIPASS \citep{koribalski04, meyer04}. In Table~\ref{tab:hipassprops} we list:
\begin{itemize}
 \item[] {\emph{Column}} (1) --  galaxy name.
 \item[] {\emph{Column}} (2) --  HIPASS name.
  \item[] {\emph{Column}} (3) -- Integrated (spatially and spectrally) \HI\ flux density, $F_{HI}$, and its uncertainty in  Jansky kilometres per second.
 \item[] {\emph{Column}} (4) -- \HI\ heliocentric velocity, $v_{hel}$, measured at the midpoint of the 50\% level of peak flux ~\citep{koribalski04}.
 \item[] {\emph{Column}} (5) and (6) -- The \HI\  velocity line widths, $w_{50}$ and $w_{20}$, measured at the 50\% and 20\% level respectively. 
 \item[] {\emph{Column}} (7) -- Reference used.
 \item[] {\emph{Column}} (8) -- The \HI\ mass in solar units calculated as $M_{HI}=2.36\times10^5D^2F_{HI}~M_{\odot}$ \citep{roberts75,roberts94}, with $D$ in Mpc and  $F_{HI}$ in Jy km/s.
 \item[] {\emph{Column}} (9) -- The \HI\ mass-to-light ratio,  $M_{HI}/L_{B,0}$, in solar units. Here $L_{B,0}=D^2 10^{10-0.4(m_B-A_B-M_{B,\odot})} ~L_{\odot,B}$ where $M_{B,\odot}$ is the absolute solar B magnitude taken as 5.48 \citep{bessell98} and $A_B=4.32\cdot E(B-V)$ \citep{schlegel98}. The mass-to-light ratio is independent of the galaxy's distance.
 \end{itemize}

\begin{table*} 
\caption{Optical Properties}
\label{tab:basicprops} 
\begin{tabular}{lcccccccccc}
\hline
\multicolumn{1}{c}{Galaxy}  &  R.A. & Decl.&  Distance & Method& P.A. &  incl. & $m_B$ & $m_H$ &$E(B-V)$ &$M_B$ \\
	   & (J2000) & (J2000) & (Mpc) & &(deg) & (deg)  &(mag) & (mag)& (mag)& (mag)\\
\multicolumn{1}{c}{(1)} & (2)& (3) & (4) & (5) & (6) & (7) & (8)& (9) & (10) & (11)\\
\hline
AM0605-341&06h07m20.3s &	-34d12m04s & 7.2 & H & 85${}^{d}$ & 65${}^{d}$& 14.3${}^{f}$ &-- & 0.036 & -15.1\\
AM0704-582 &07h05m20.1s &	-58d31m28s  & 4.90 & TRGB${}^{b}$  & 45${}^{c}$ & 60${}^{c}$ &14.4${}^{i}$ & $12.72\pm0.08$& 0.119 & -14.6\\
ESO059-G001 &07h31m19.2s  &-68d11m29s & 4.57 & TRGB${}^{a}$& 160${}^{c}$ & 41${}^{c}$ &13.7${}^{g}$  & $11.28\pm0.06$ & 0.147 & -15.2\\
ESO121-G020 &06h15m53.2s & -57d43m50s  & 6.05& TRGB${}^{a}$& 45${}^{c}$ & 45${}^{c}$ &15.9${}^{g}$ & $13.87\pm0.09$& 0.040 &-13.2\\
ESO137-G018 & 16h20m59.3s  &-60d29m14s&6.40& TRGB${}^{h}$& 30${}^{e}$ & 55${}^{e}$&12.2${}^{g}$&--${}^{k}$&0.243 & -17.9\\
ESO174-G?001 &13h47m58.8s &-53d20m51s  & 6.0 &H & 165${}^{d}$& 60${}^{d}$ & 14.4${}^{f}$ &-- & 0.494 & -16.6\\
ESO215-G?009  &10h57m29.4s & -48d10m40s& 5.25 & TRGB${}^{h}$ & 40${}^{j}$& 30${}^{j}$& 16.0${}^{g}$&--& 0.221 & -13.4\\
ESO245-G005 & 01h45m04.7s &-43d35m47s & 4.43 & TRGB${}^{b}$& 127${}^{c}$ & 52${}^{c}$ &12.8${}^{g}$ &$11.10\pm0.10 $ & 0.016 &-15.6 \\
ESO308-G022 &06h39m33.1s &-40d43m13s   & 7.6 & H& 130${}^{c}$ & 37${}^{c}$ &16.2${}^{g}$ & $13.40 \pm 0.10$& 0.089 & -13.6\\
ESO325-G?011& 13h45m00.8s & -41d51m32s& 3.40& TRGB${}^{b}$& 130${}^{d}$ & 60${}^{d}$ &14.0${}^{g}$&--& 0.087 & -14.0\\
ESO381-G020  &12h46m00.4s & -33d50m17s  & 5.44 & TRGB${}^{h}$& 130${}^{d}$ & 40${}^{d}$ & 14.2${}^{g}$  &--& 0.065 & -14.7\\
IC5152  & 22h02m41.3s & -51d17m53s  & 2.07 & TRGB${}^{b}$   & 95${}^{c}$ & 50${}^{c}$ & 11.1${}^{g}$& $8.26\pm 0.03$& 0.025 & -15.6\\
\hline
\end{tabular}
\flushleft
(a) \cite{karachentsev06}, (b) \cite{karachentsev04}, (c) \cite{kirby08}, (d) measured independently using DSS image, (e) measured independently using a previously unpublished Local Sphere of Influence Survey \citep{kirby08} image. (f) \cite{doyle05}, (g) ESO-LV catalog, \cite{lauberts89}, (h) \cite{karachentsev07}, (i) \cite{parodi02}, (j) measured independently using \cite{warren04} image, (k) not measured due to foreground contamination. 
\end{table*}

\begin{table*} 
\caption{Radio Properties from HIPASS}
\label{tab:hipassprops} 
\begin{tabular}{llcccccccc}
\hline
\multicolumn{1}{c}{Galaxy}  & \multicolumn{1}{c}{HIPASS ID.}& $F_{HI}$& $v_{hel}$ & $w_{50}$ & $w_{20}$ & Ref.&$M_{HI}$& $M_{HI}/L_{B,0}$\\
	   & & (Jy\kms)   &(\kkms)& (\kkms)& (\kkms) &&$(10^8 M_{\odot})$&($M_{\odot}/L_{\odot,B}$)\\
\multicolumn{1}{c}{(1)} &\multicolumn{1}{c}{(2)} & (3) & (4) &(5) &(6) &(7)& (8)&(9) \\
\hline
AM0605-341& HIPASS J0607-34 &$ 9.0\pm1.4$ &$765\pm6$  &$123 \pm 12$ & $168\pm18$&1 &$1.1\pm0.2$ &$0.6\pm0.1$ \\
AM0704-582 &  HIPASS J0705-58&$34.8 \pm  4.4$ &$564\pm 2$ & $68\pm 4$& $ 84\pm6$ &2& $2.0\pm0.4$ & $1.9\pm0.4$\\
ESO059-G001 & HIPASS J0731-68&$17.7  \pm 2.5$ &$530\pm 3$&  $82\pm6$ & $104\pm 9$&2&$0.9\pm 0.2$& $0.5\pm0.1$\\
ESO121-G020${}^{\ast}$ & HIPASS J0615-57 &$14.1 \pm  2.9$ &$ 577\pm 5$&  $65\pm10$ & $96\pm15$&2&$ 1.2\pm0.3$& $4.1\pm1.0$\\
ESO137-G018 & HIPASS J1620-60&$37.4 \pm  4.9$ &$605 \pm 3$& $139\pm6$ &$155\pm9$ &2&$3.6\pm 0.7$&$ 0.16\pm0.03$\\
ESO174-G?001 & HIPASS J1348-53 &$55.1\pm   5.9$ &$688 \pm 3$ & $71\pm6$ &$103\pm9$&2&$4.7 \pm 0.8$& $0.7\pm0.1$\\
ESO215-G?009  & HIPASS J1057-48 & $104.4 \pm  11.5$  &$598\pm2  $ & $67\pm4$ & $83\pm6$ &2&$6.8\pm 1.2$ & $16.8\pm3.0$\\
ESO245-G005 & HIPASS J0145-43 & $81.0 \pm 9.1$ &$391\pm  2 $& $60\pm4$& $ 85\pm6$ &2& $3.9\pm0.7$ &$1.5\pm0.3$\\
ESO308-G022 & HIPASS J0639-40 &$3.8\pm1.0$ &$822 \pm5$&$52\pm10$& $74\pm15$&1&$0.5\pm0.2$ &$1.2\pm0.4$\\
ESO325-G?011& HIPASS J1345-41& $26.6 \pm  3.7$ &$ 545\pm 2$&  $59\pm4$&  $75\pm6$ &2&$0.7\pm0.1$&$1.1\pm0.2$ \\
ESO381-G020  & HIPASS J1246-33& $30.9 \pm  3.7$ &$589 \pm 2$& $ 83\pm4$ &$100\pm6$ &2&$2.2\pm0.4$ &$1.8\pm0.3$\\
IC5152  & HIPASS J2202-51& $97.2 \pm  9.5$ & $122  \pm2 $& $84\pm4$& $100\pm6$&2&$ 1.0 \pm0.2$ &$0.4\pm0.1$\\
\hline
\end{tabular}
\flushleft
($\ast$) \cite{warren06} identified a nearby companion, ATCA J061608-574552, which is not resolved from ESO121-G020 by HIPASS.
(1) \cite{meyer04}, (2) \cite{koribalski04}
\end{table*}

\section{Observations and Data Reduction}\label{s:HIobsdata}
A detailed description of the LVHIS project, including observations, data 
reduction and analysis as well as first results is the subject of an 
upcoming paper \citep{koribalski09}.  

The \HI\ line observations analysed in this paper were obtained with the 
Australia Telescope Compact Array (ATCA) as part of the LVHIS project.
The data was taken between January 2005 and January 2009 using three arrays,
EW352/EW367-m, 750-m, and 1.5-km ($\sim$12-h each), to ensure excellent 
$uv$-coverage and sensitivity to large-scale structure.
Each sample galaxy was observed for a full synthesis (12-h) in each of the 
three arrays, unless equivalent archival observations were available. In addition, all other available 
archival data were used where available, including the ATCA observations in 
the 6km array for ESO215-G?009 (no 1.5-km were taken for this galaxy).

The first frequency band was centred on 1418 MHz with a bandwidth of 8 MHz, 
divided into 512 channels. This gives a channel width of 3.3\kms\ and a 
velocity resolution of 4\kms. The ATCA primary beam is 33.6\arcmin\ at 
1418 MHz. 

\begin{figure*} 
\begin{tabular}{cccc}
  \mbox{\epsfig{file=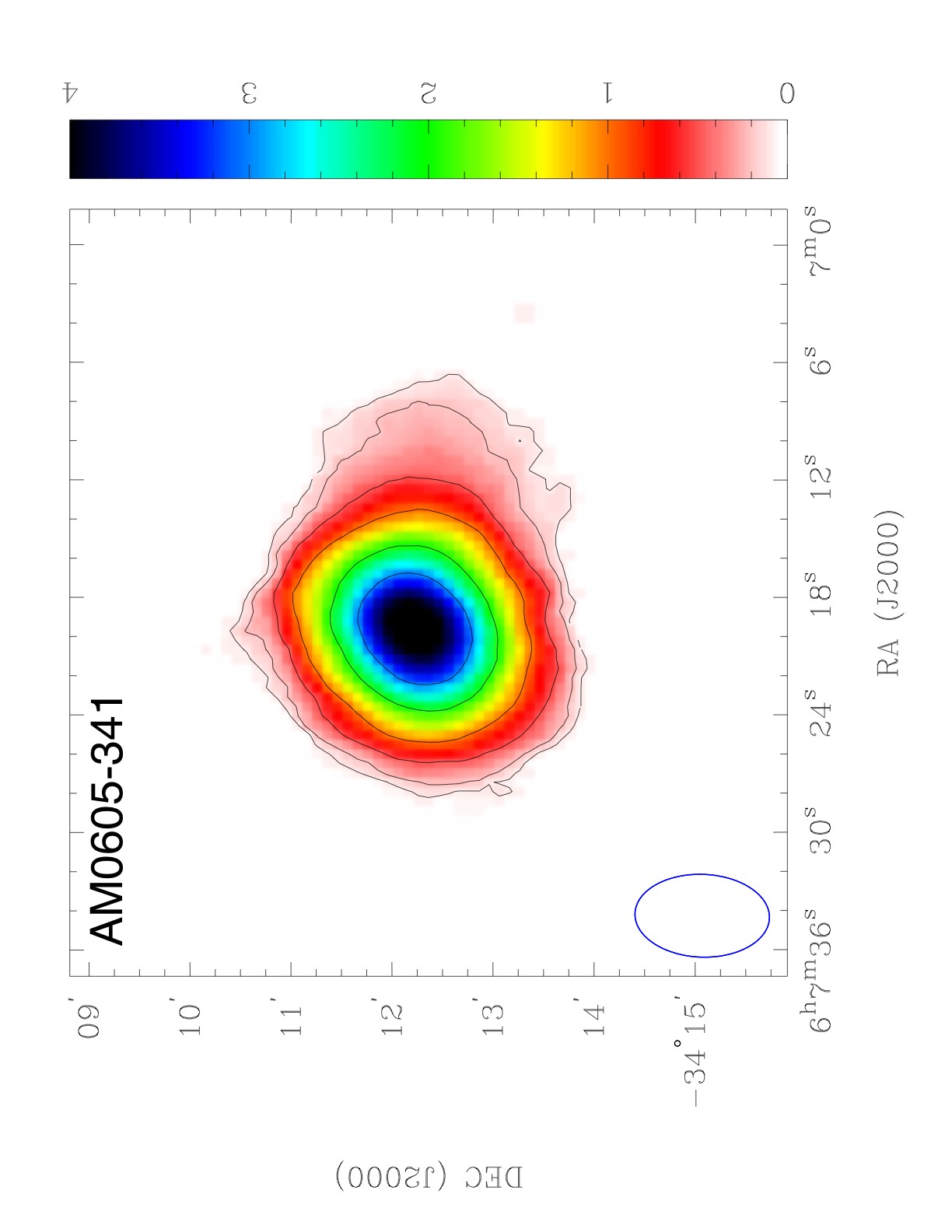,width=3.1cm,angle=-90}} &
  \mbox{\epsfig{file=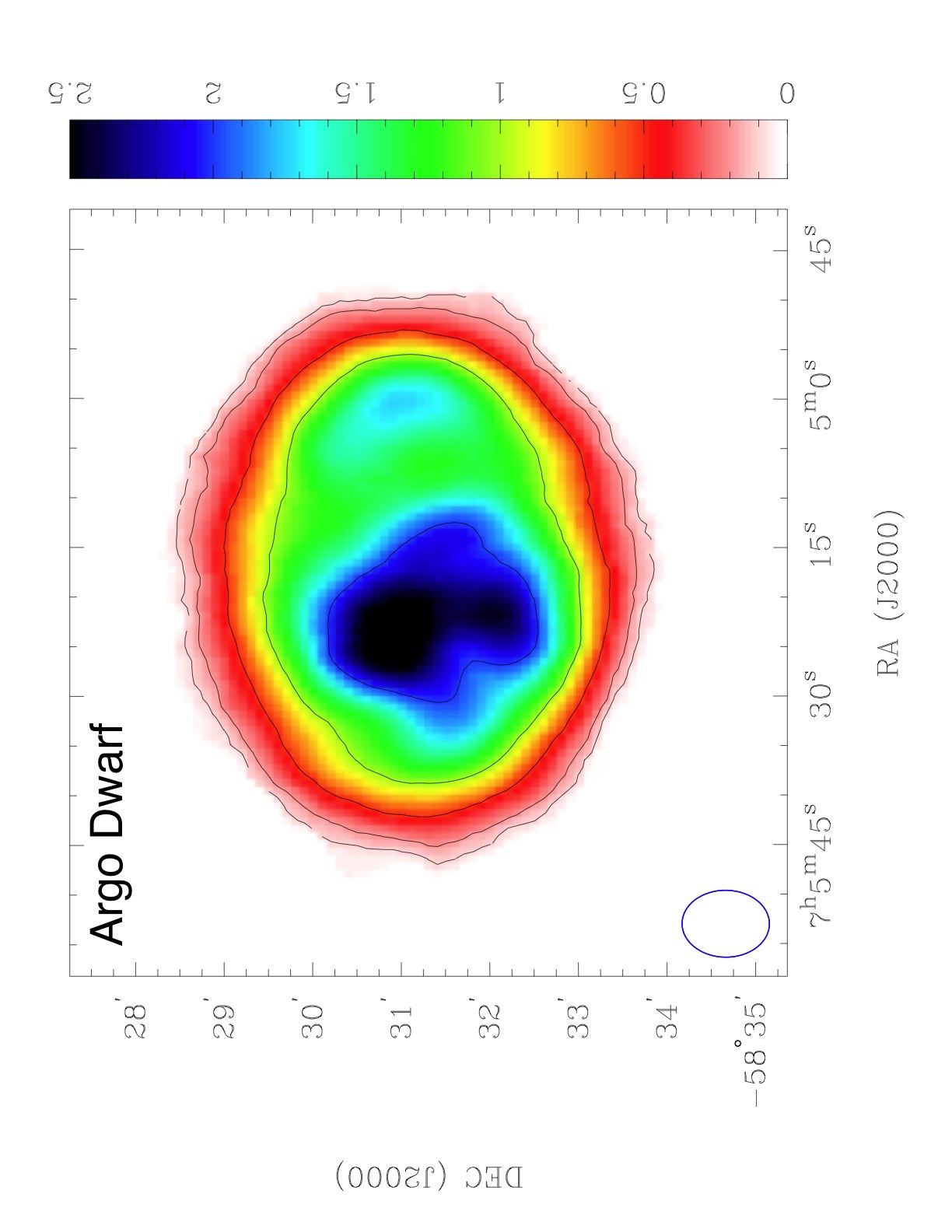,width=3.1cm,angle=-90}} &
  \mbox{\epsfig{file=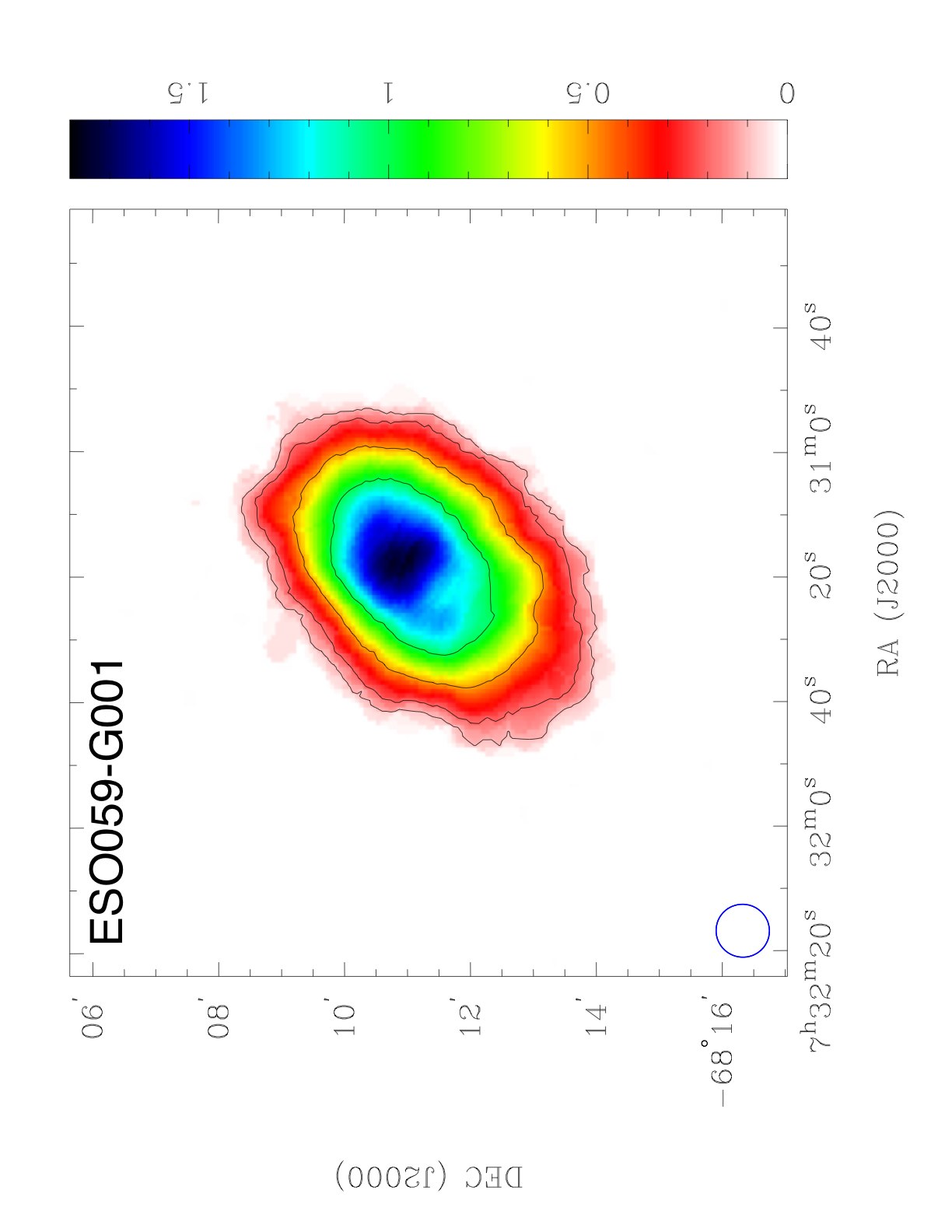,width=3.1cm,angle=-90}} &
  \mbox{\epsfig{file=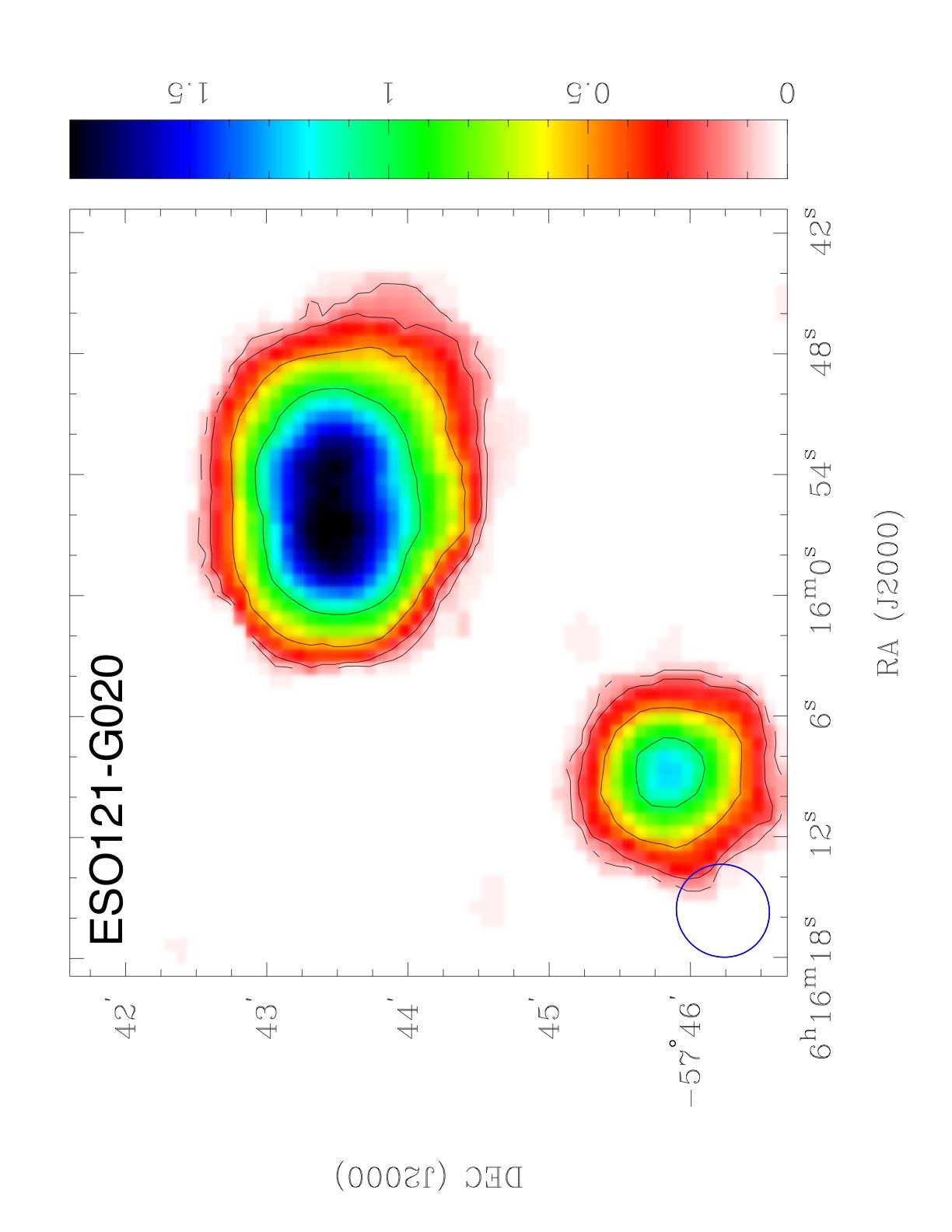,width=3.1cm,angle=-90}} \\
  \mbox{\epsfig{file=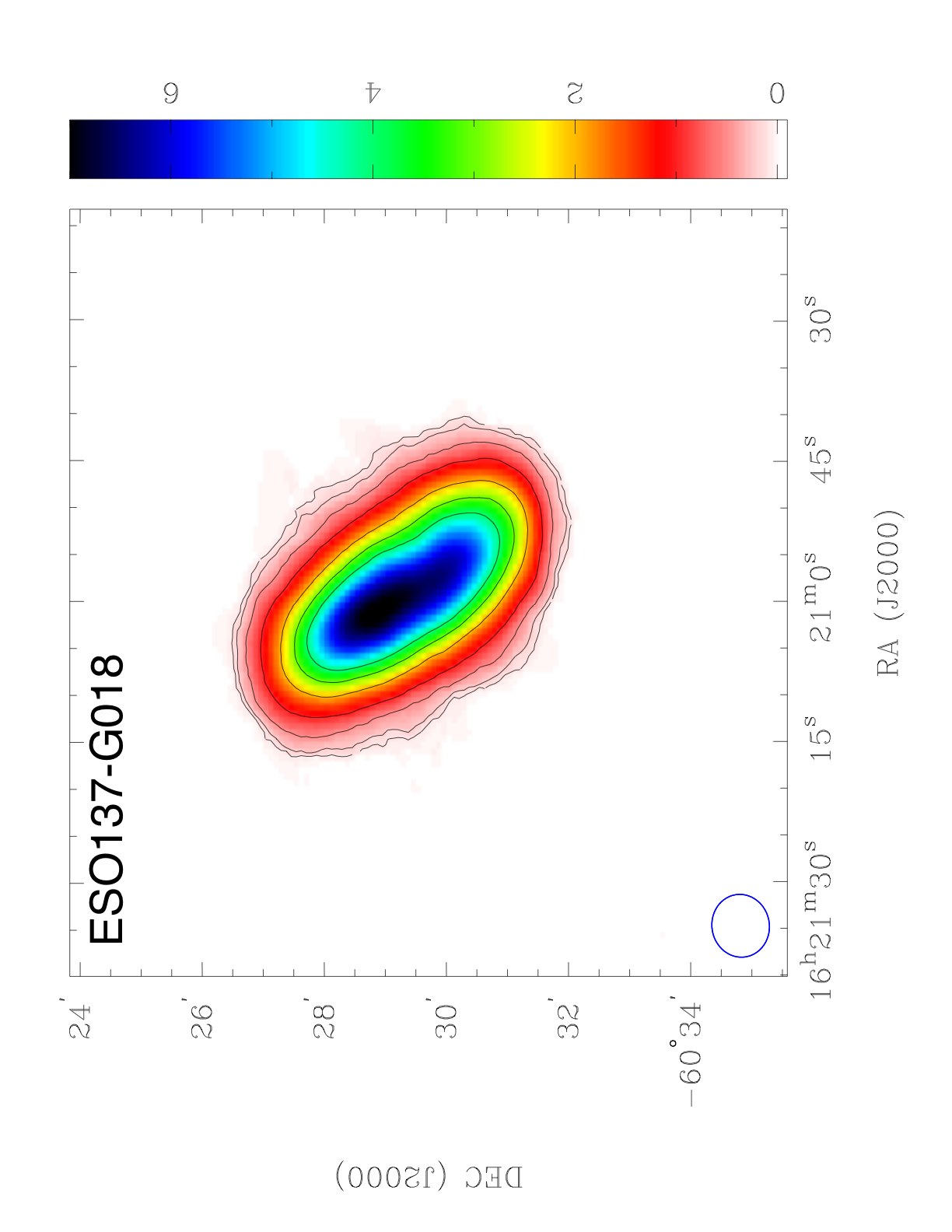,width=3.1cm,angle=-90}} &
  \mbox{\epsfig{file=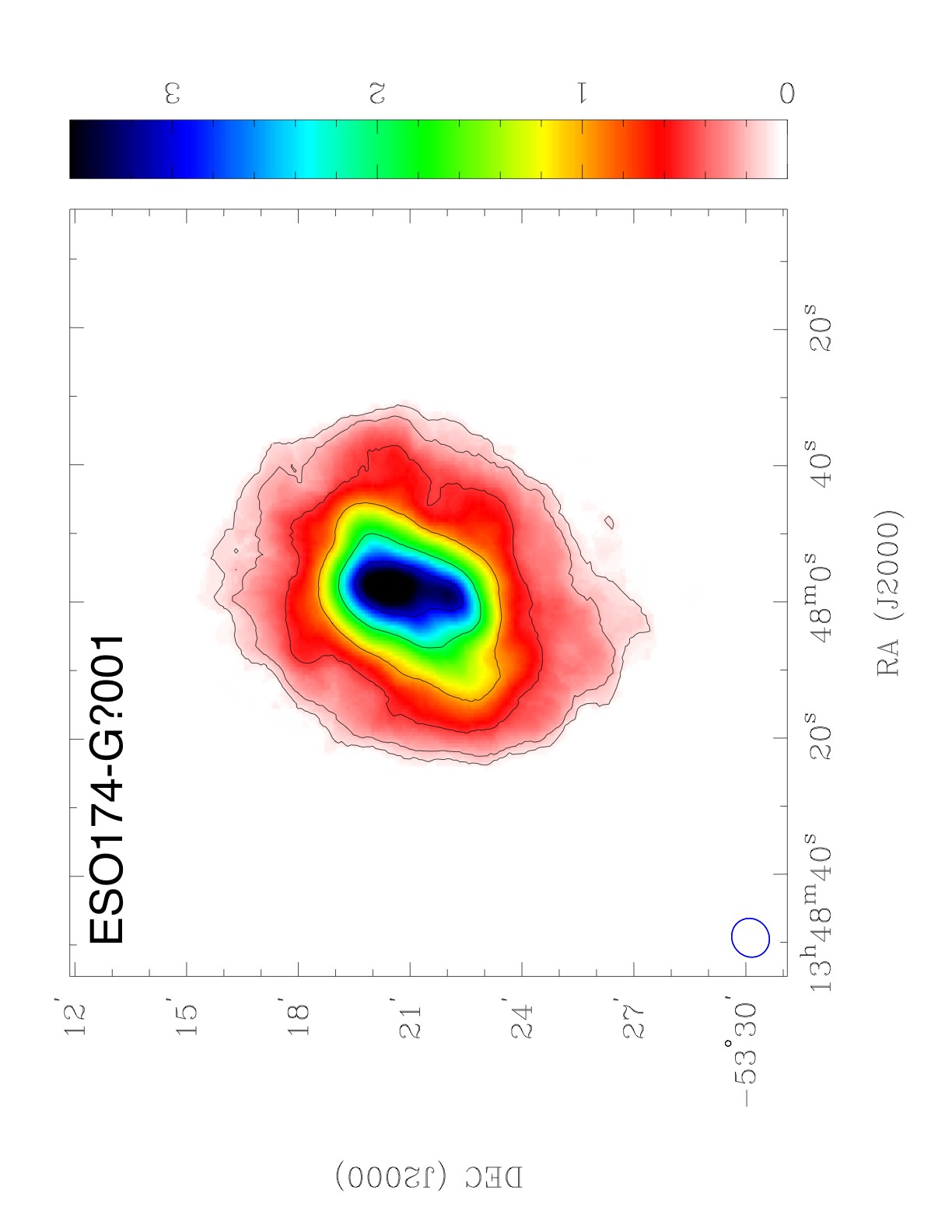,width=3.1cm,angle=-90}} &
  \mbox{\epsfig{file=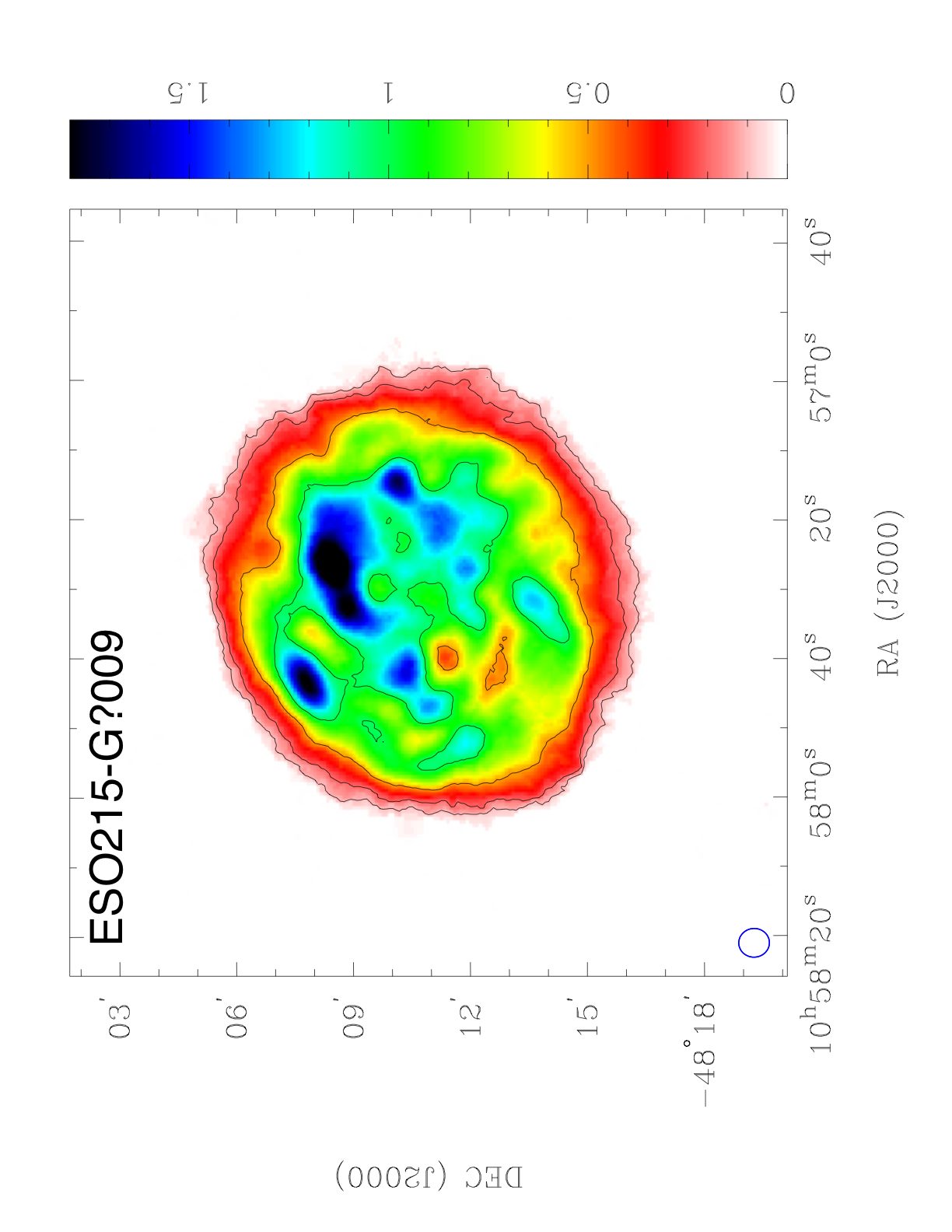,width=3.1cm,angle=-90}} &
  \mbox{\epsfig{file=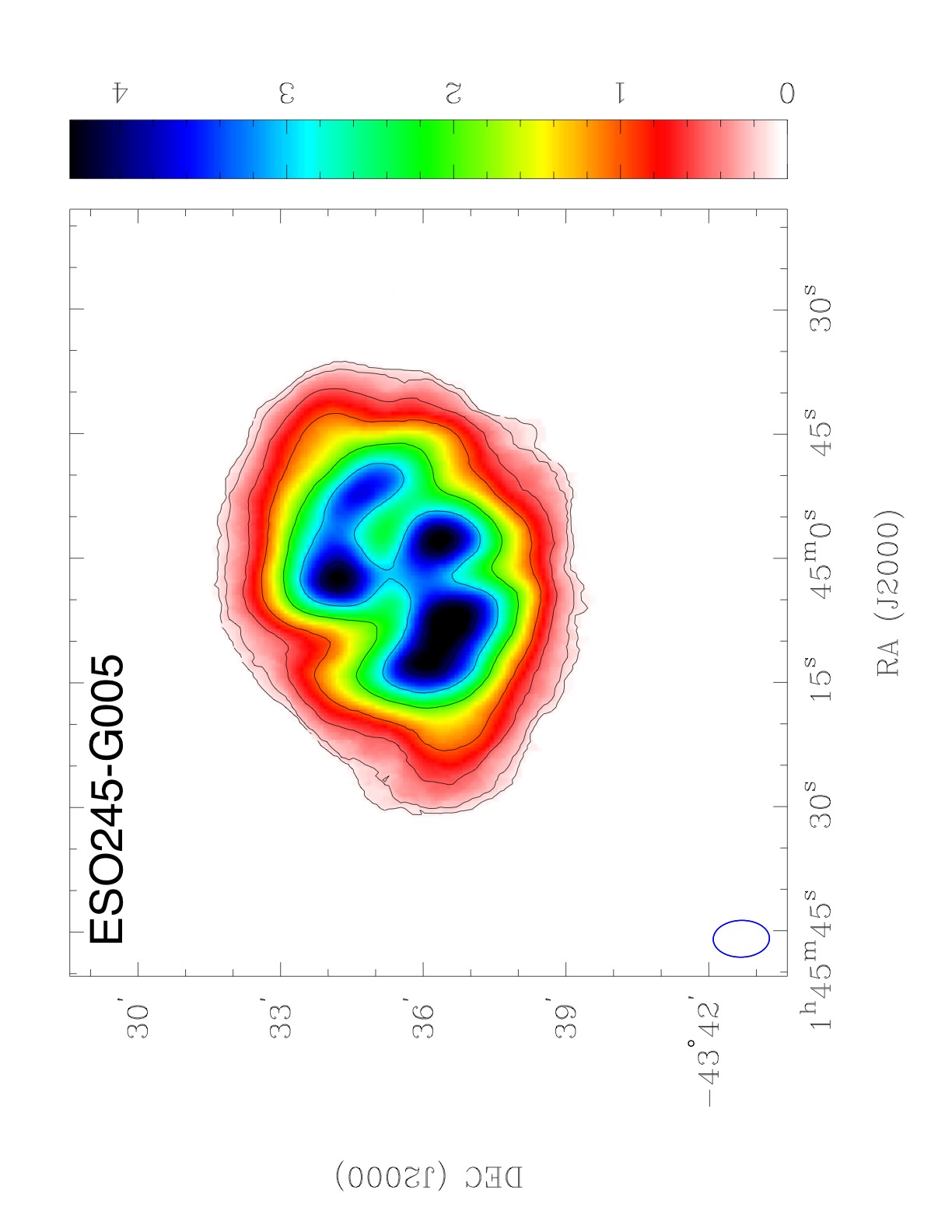,width=3.1cm,angle=-90}} \\
  \mbox{\epsfig{file=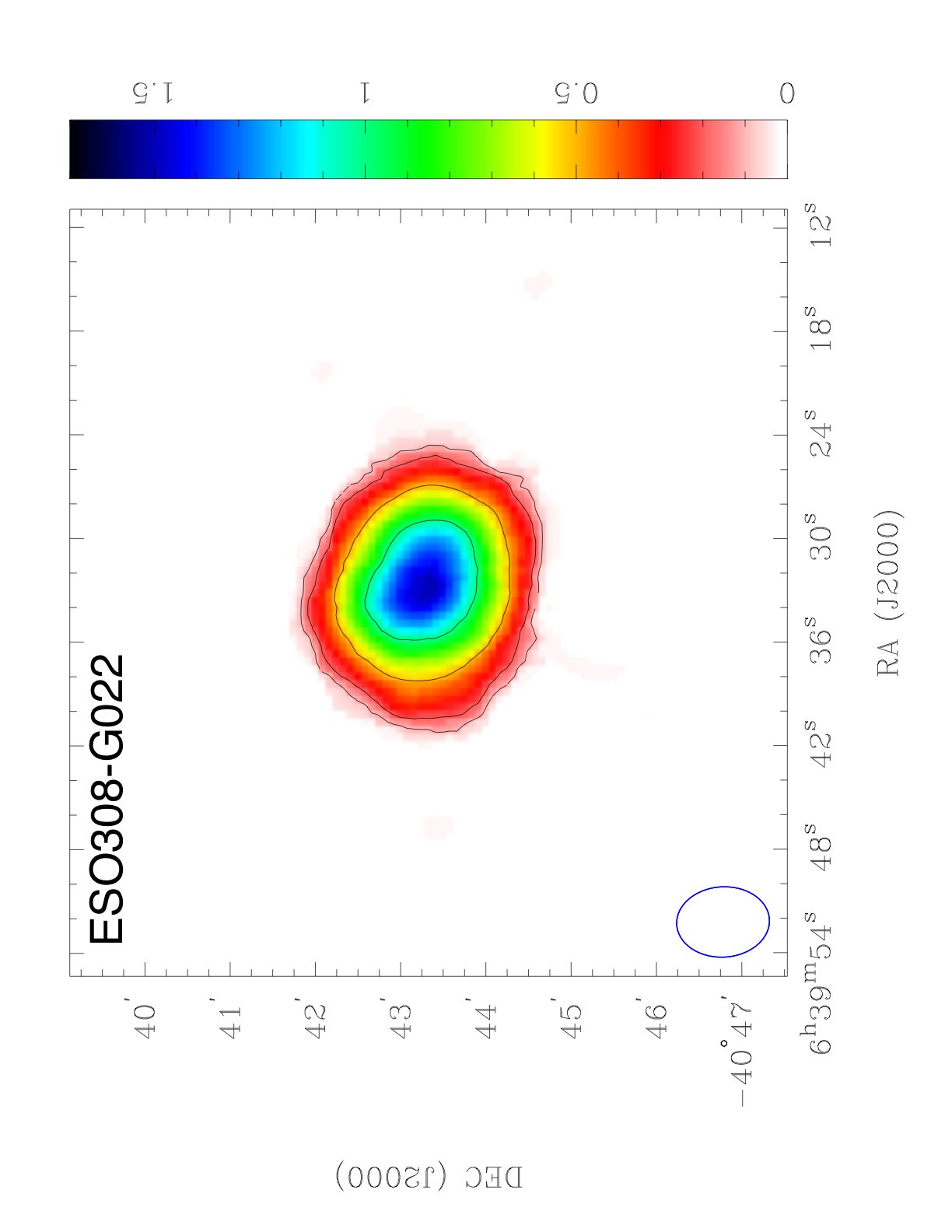,width=3.1cm,angle=-90}} &
  \mbox{\epsfig{file=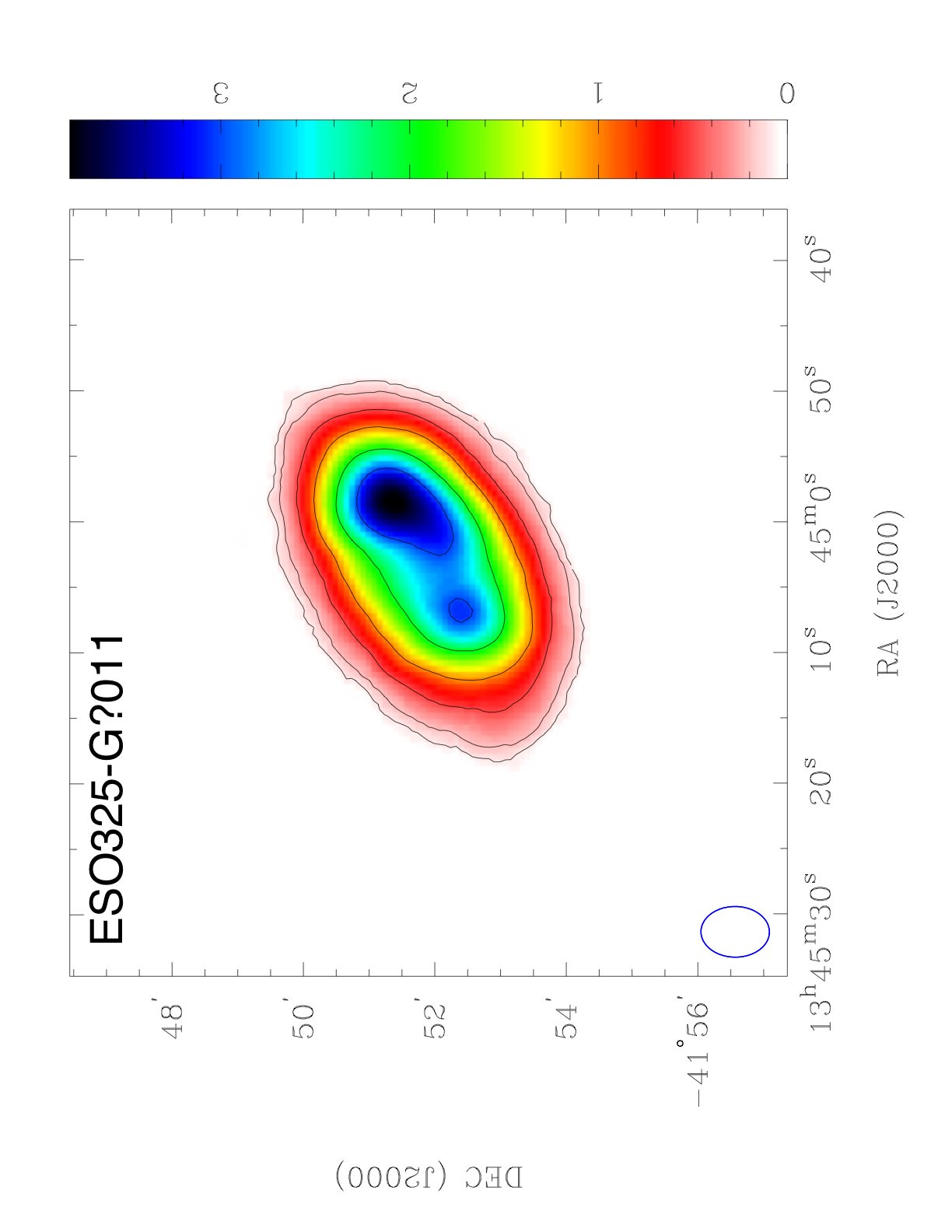,width=3.1cm,angle=-90}} &
  \mbox{\epsfig{file=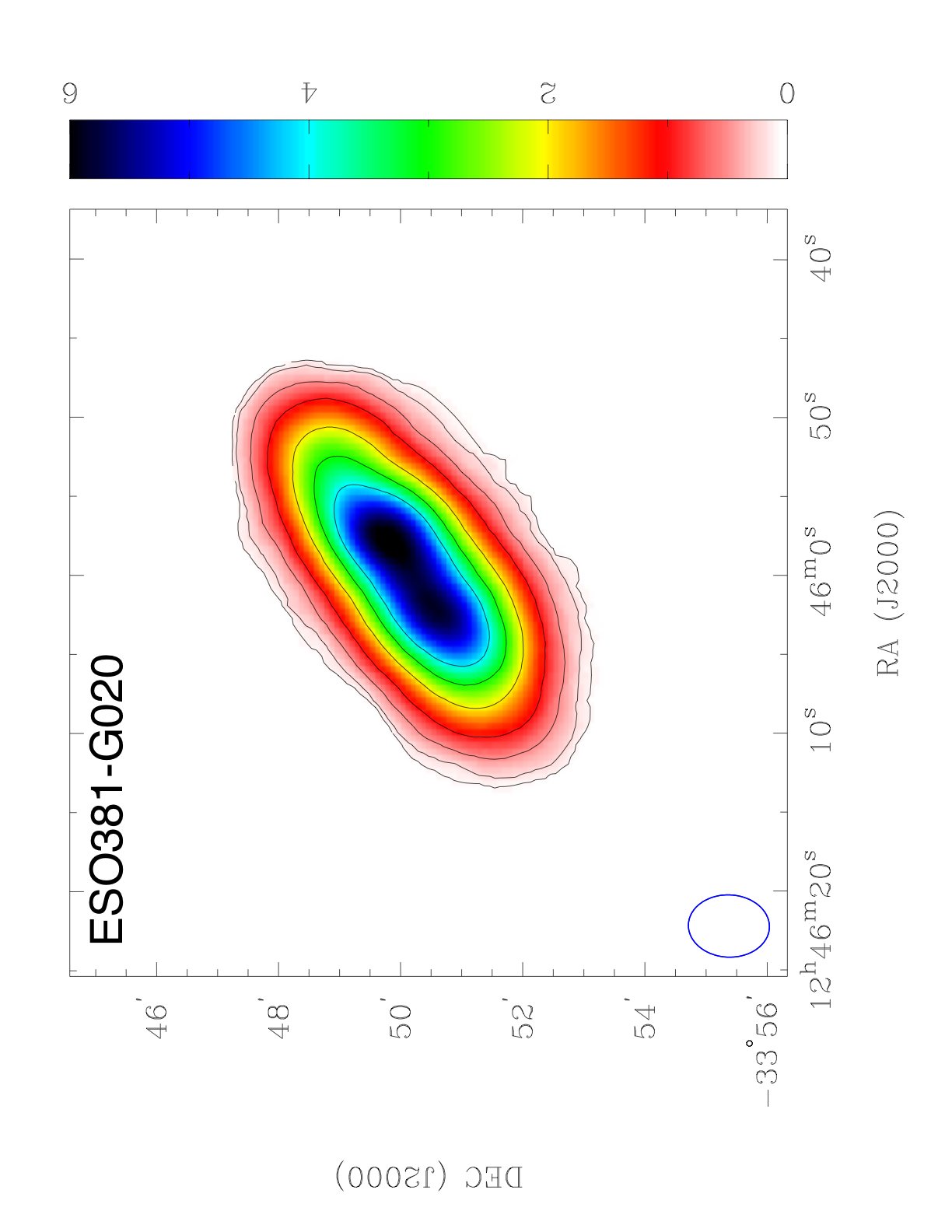,width=3.1cm,angle=-90}} &
  \mbox{\epsfig{file=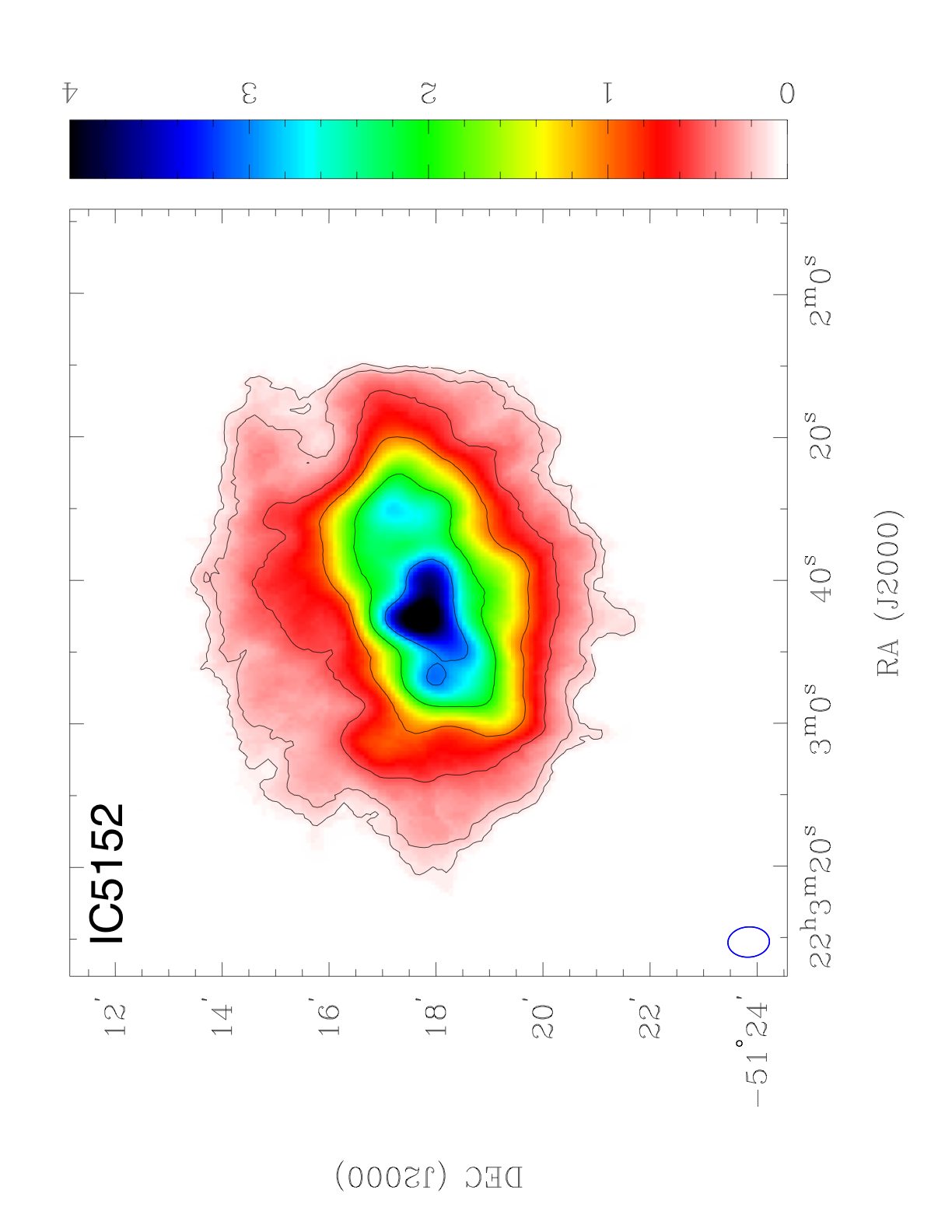,width=3.1cm,angle=-90}} \\
 \end{tabular}
\caption{The  integrated \HI\ intensity distribution with contour levels at 0.1 ($\sim 0.3\sigma$), 0.2, 0.5, 1, 2, 3 and 4 Jy beam${}^{-1}$\kms.  The synthesised beam is displayed in the bottom left corner of each panel. The image of ESO121-G020 also shows its dwarf companion, ATCA J061608-574552, which was  identified by Warren et al.\ (2006). Note that the intensity scale is adjusted for each galaxy.}
\label{fig:mom0}
\end{figure*}

Data reduction was carried out with the {\sc miriad}  (Multichannel Image Reconstruction, Image Analysis and Display; \citealt{sault95}) software  package using standard procedures.  Here we use the \HI\ moment maps made using `natural' weighting of the {\em uv}-data in the velocity range covered by the \HI\ emission using steps of 4\kms. To obtain low-resolution maps we 
excluded the longest baselines, to the distant antenna six. The average synthesized beam size is 45 arcseconds. The integrated \HI\ intensity distribution is shown in Figure~\ref{fig:mom0} for each galaxy .

\section{Galaxy Spectra}\label{s:spectra}

\begin{figure*} 
\begin{tabular}{cccc}
  \mbox{\epsfig{file=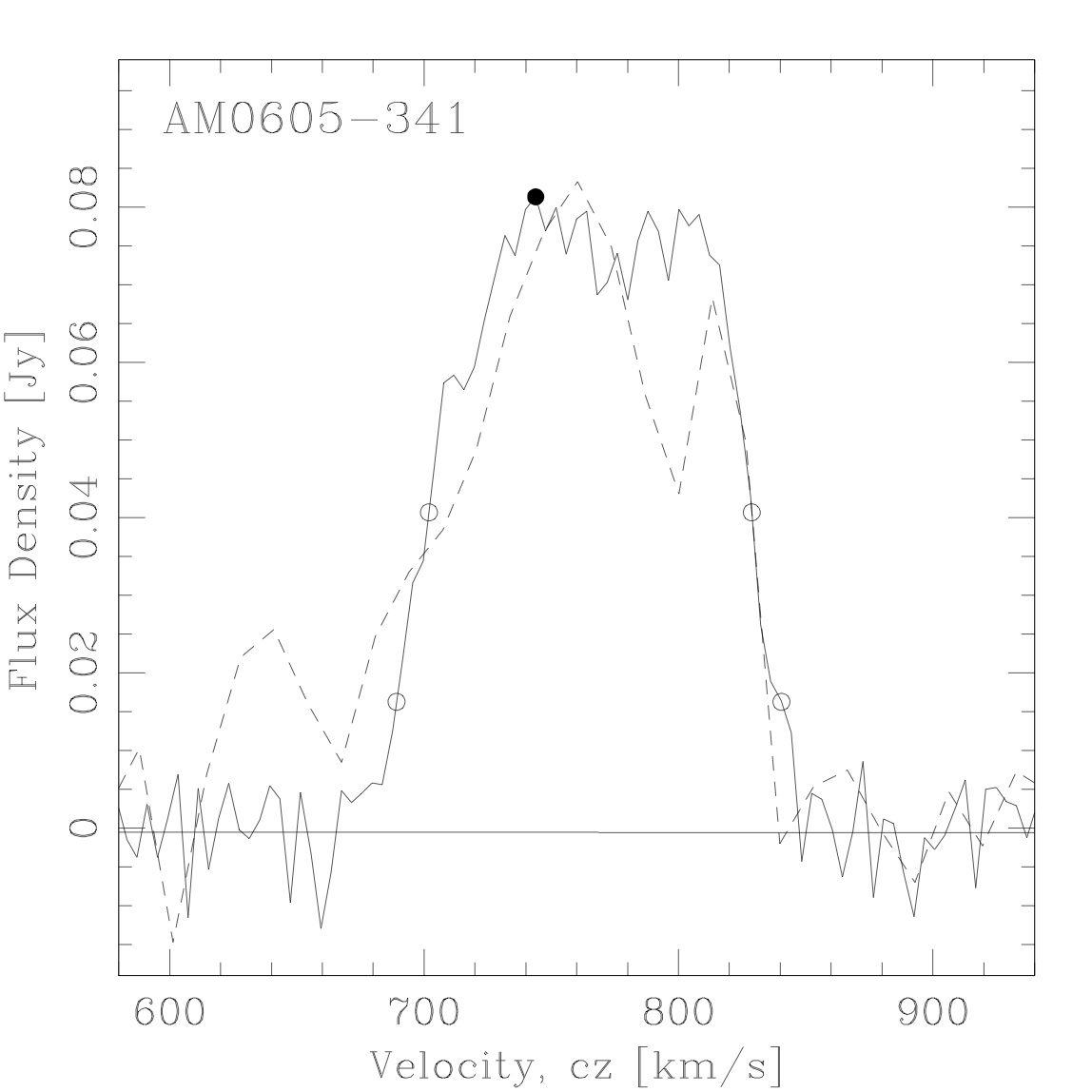,width=3.6cm}} &
  \mbox{\epsfig{file=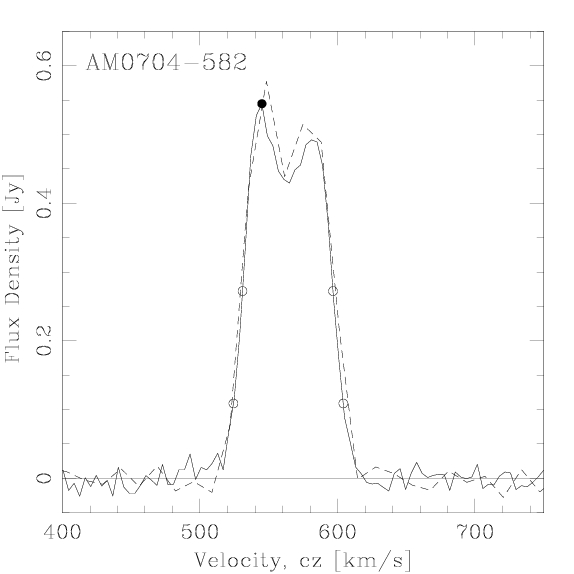,width=3.6cm}} &
  \mbox{\epsfig{file=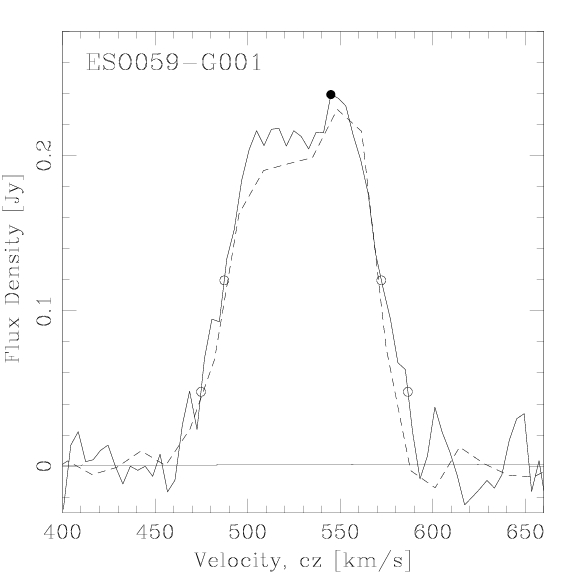,width=3.6cm}} &
  \mbox{\epsfig{file=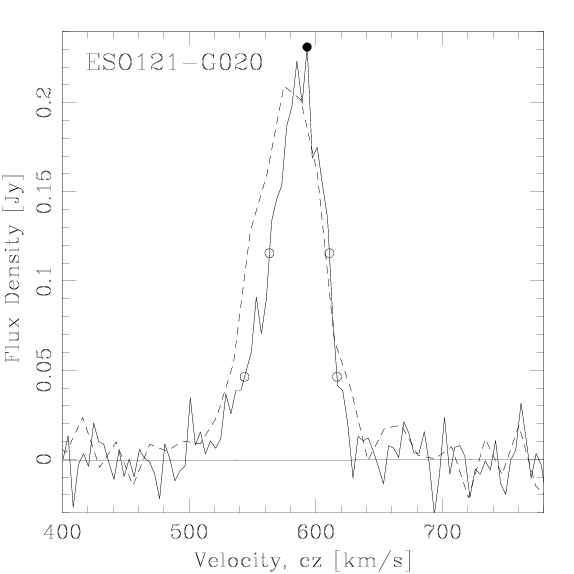,width=3.6cm}} \\
  \mbox{\epsfig{file=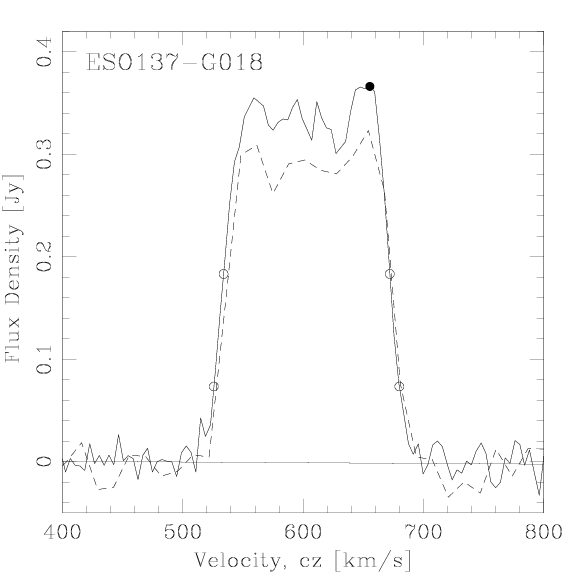,width=3.6cm}} &
  \mbox{\epsfig{file=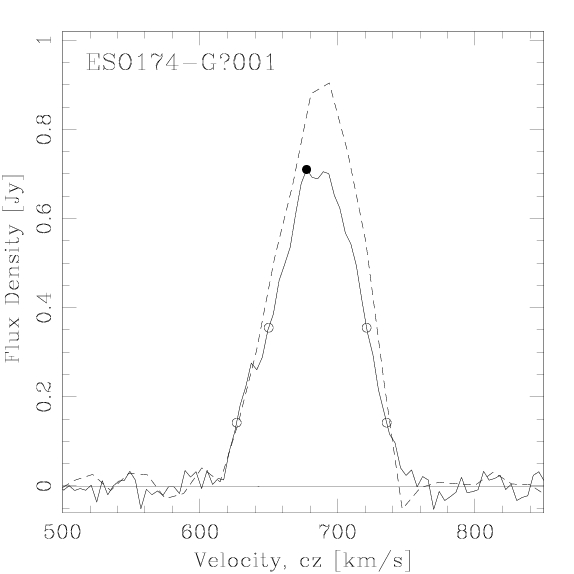,width=3.6cm}} &
  \mbox{\epsfig{file=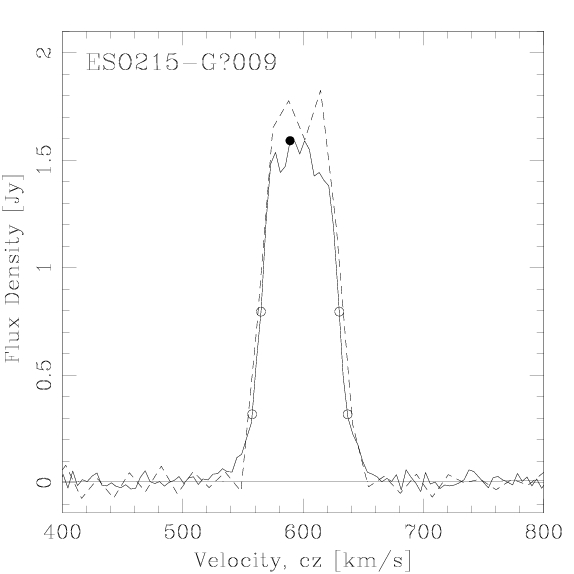,width=3.6cm}} &
    \mbox{\epsfig{file=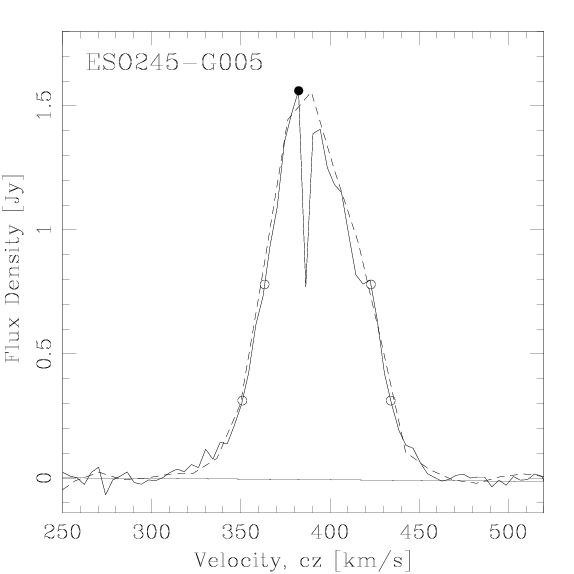,width=3.6cm}} \\
    \mbox{\epsfig{file=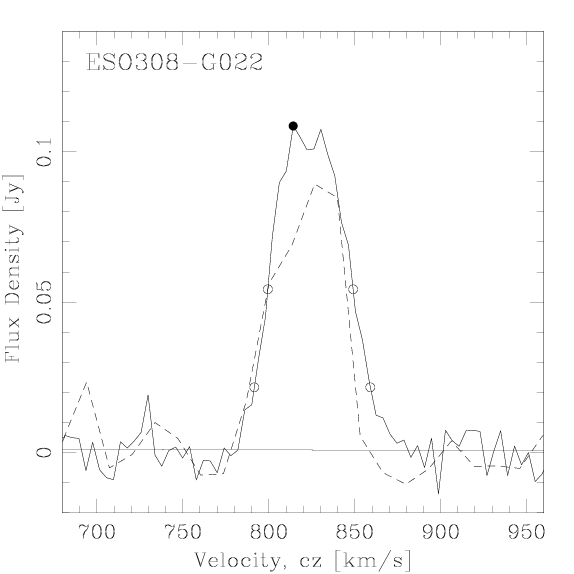,width=3.6cm}} &
    \mbox{\epsfig{file=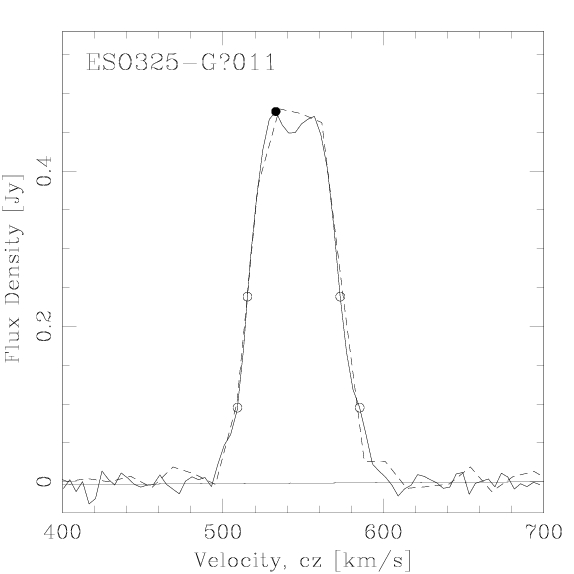,width=3.6cm}} &
    \mbox{\epsfig{file=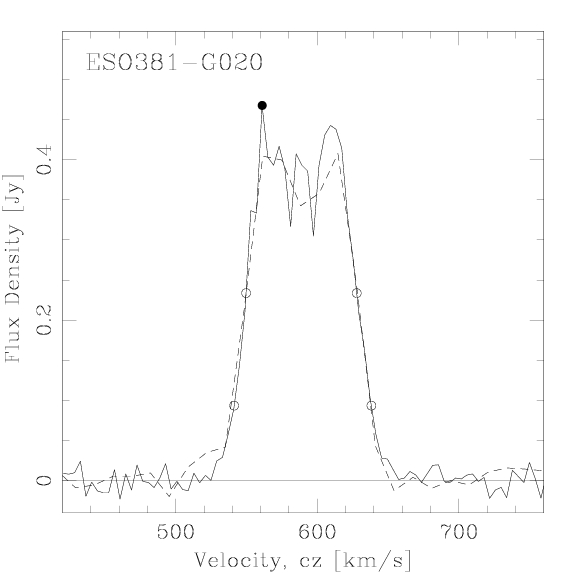,width=3.6cm}} &
  \mbox{\epsfig{file=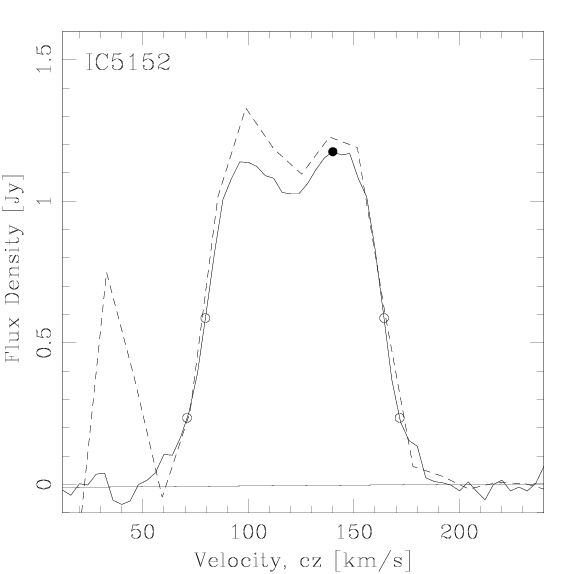,width=3.6cm}} \\
 \end{tabular}
\caption{The global HI line spectra as obtained from LVHIS (solid line) and HIPASS (dashed line). The solid dot indicates the peak flux, and the open circles show the points used to measure the \HI\ velocity line widths at the 50\% and 20\% peak flux density levels. The zero baseline is shown for the LVHIS data only. 
 }
\label{fig:spect}
\end{figure*}

The \HI\ line spectra for the selected LVHIS data were obtained using the  \emph{mbspect}  task in miriad. For each galaxy, the spectrum was integrated over the coordinates with detected \HI\ emission (defined by the region of emission in the first moment map) for the entire range of observed velocities. In Figure~\ref{fig:spect}, the LVHIS spectrum for each sample galaxy is provided (solid line) with its HIPASS spectrum for comparison (dotted line).

The HIPASS spectrum was extracted from the HIPASS datacubes. We derive the spectral parameters to ensure  that an accurate comparison can be made between the HIPASS and the LVHIS data. For each galaxy the \HI\ spectrum was integrated over the area and velocity range of the detected \HI\ signal. A first-order polynomial was fitted to the line-free channels and subtracted. Table~\ref{tab:spectralprops} lists the measured \HI\ properties as well as the ratio of single dish (HIPASS) to interferometric $F_{HI}$ (LVHIS). Figure~\ref{fig:w20} shows a comparison of HIPASS and LVHIS spectra of each of the 12 galaxies in our sample. A detailed description of each galaxy is given in Section~\ref{s:individualkinematics}

Various quantities have been obtained to describe the spectral properties of each galaxy.  The spectral properties for the sample galaxies obtained from the LVHIS data are listed in Table~\ref{tab:spectralprops}. The columns are organised as follows: 
\begin{itemize}
 \item[] {\emph{Column}} (1) --  galaxy name.
 \item[] {\emph{Column}} (2) --  the \HI\ peak flux density, $S_{\textrm{peak}}$ with its uncertainty given by $\sigma(S_{\textrm{peak}})^2 = \textrm{rms}^2 +(0.05S_{\textrm{peak}})^2$, following that described in ~\cite{koribalski04} and \cite{barnes01}.
 \item[] {\emph{Column}} (3) -- Integrated (spatially and spectrally) \HI\ flux density, $F_{HI}$, and its uncertainty in Jansky kilometres per second.  The uncertainty has been calculated as $\sigma(F_{HI})=4 \sigma(S_{\textrm{peak}})/S_{\textrm{peak}}(S_{\textrm{peak}}F_{HI}\Delta v)^{1/2}$ \citep{koribalski04, fouque90}, where  $\Delta v = 4$\kms ~is the velocity resolution of LVHIS.
 \item[] {\emph{Column}} (4) -- the \HI\ heliocentric velocity, $v_{hel}$ which has been measured as the midpoint of the 50\% level of peak flux density.  The uncertainty in $v_{hel}$ was calculated as $\sigma(v_{hel})=3 \sigma(S_{\textrm{peak}})/S_{\textrm{peak}}(P\delta v)^{1/2}$, where $P=0.5(w_{20} - w_{50})$ is a measure of the steepness of the profile edges  \citep{koribalski04, fouque90}.
  \item[] {\emph{Column}} (5) and (6) -- The velocity line widths, $w_{50}$ and $w_{20}$ which are the widths of the \HI\ line profile measured at the 50\% and 20\% level of peak flux density respectively. The uncertainties are calculated as $\sigma(w_{50})=2\sigma(v_{hel})$ and $\sigma(w_{20})=3\sigma(v_{hel})$ following that of \cite{schneider86} \citep[using the same definition as][]{koribalski04}.
   \item[] {\emph{Column}} (7) -- The \HI\ mass in solar units calculated as $M_{HI}=2.36\times10^5\,D^2\,F_{HI}\,M_{\odot}$ \citep{roberts75,roberts94}.
   \item[] {\emph{Column}} (8) -- The ratio of the flux detected by LVHIS to the flux detected by HIPASS.
  \end{itemize}

\begin{table*} 
\caption{Measured ATCA \HI\ Properties}
\label{tab:spectralprops} 
\begin{tabular}{lccccccc}
\hline
\multicolumn{1}{c}{Galaxy} &$S_{\textrm{peak}}$ & $F_{HI}$&$v_{hel}$ & $w_{50}$ & $w_{20}$&$M_{HI}$ &$\frac{F_{HI} (LVHIS)}{ F_{HI} (HIPASS)}$\\
	     &  (mJy beam${}^{-1}$)    &(Jy\kms)    &(\kkms)     &(\kkms)       & (\kkms) &$(10^8 M_{\odot})$      \\
\multicolumn{1}{c}{(1)} & (2)& (3) & (4) &(5)&(6)&(7)&(8) \\
\hline
 AM0605-341 & $81\pm4$ &$9.6  \pm0.7$& $765\pm3$ &$127 \pm 4 $ & $ 151\pm6$ & $ 1.2\pm0.2$&$1.1\pm0.3$ \\
AM0704-582   &$545\pm27$&$33.2\pm1.5$&$564\pm 1$ & $66\pm 1$& $80\pm2$ & $1.9\pm0.3 $ &$1.0\pm0.1$\\
ESO059-G001 & $239\pm12$ & $19.6 \pm 2.2$ & $ 530 \pm 3 $ & $ 85 \pm 6 $ & $112\pm 9$  & $1.0\pm0.2 $& $1.1\pm0.2  $  \\
ESO121-G020 & $231\pm12$ & $10.0 \pm 1.2$ &  $587\pm 2$& $ 47\pm 4$ & $73 \pm 6$ & $0.9\pm0.2$& $  0.7\pm0.2$ \\
ESO137-G018   &$366\pm18$& $ 47.8 \pm 1.5$ & $603 \pm 1$ & $ 138\pm2$ & $154\pm2$ & $4.6\pm0.7  $& $ 1.3\pm0.2 $  \\
ESO174-G?001 &$710\pm36$ & $ 48.8 \pm 2.6$ & $686\pm1$ & $71\pm3$ & $ 109\pm4$ & $4.1\pm0.6 $& $ 0.9\pm0.1 $   \\
ESO215-G?009 &$1591\pm80$& $100.9\pm3.1$ & $598\pm 1$ & $65 \pm 1$ & $79\pm 1$ & $6.6\pm0.9 $ & $ 1.0\pm0.1 $    \\
ESO245-G005  &$1561\pm78$& $81.6\pm2.6$ & $393\pm1 $ &$ 60\pm 1$ & $83\pm2$  & $3.8\pm0.5$ & $ 1.0\pm0.1 $ \\
ESO308-G022  &$109\pm6$ & $ 5.3\pm0.8$ &$824\pm 2$ & $ 50\pm5$ & $ 67\pm7$ & $ 0.7\pm0.1 $& $ 1.4\pm0.4 $    \\
ESO325-G?011 & $477\pm24$& $ 27.1\pm0.9 $ & $ 544\pm1 $ & $ 58 \pm 1$ & $76 \pm 2$  & $0.7\pm0.1  $ &  $ 1.0\pm0.1 $   \\
ESO381-G020  & $468\pm23$ & $ 33.4\pm1.7 $ & $589\pm1$ & $ 78\pm2$ & $ 97 \pm 3$ & $ 2.3\pm0.4$& $ 1.1\pm0.1 $  \\
IC5152 & $1175\pm59$ & $ 96.4\pm3.0 $ & $122\pm1 $ & $ 85\pm 1$ & $ 101\pm 2$ & $ 1.0\pm0.1$ & $ 1.0\pm 0.1$ \\
\hline
\end{tabular}
\flushleft
\end{table*}

\subsection{A Comparison with HIPASS}

\begin{figure}
\epsfig{file=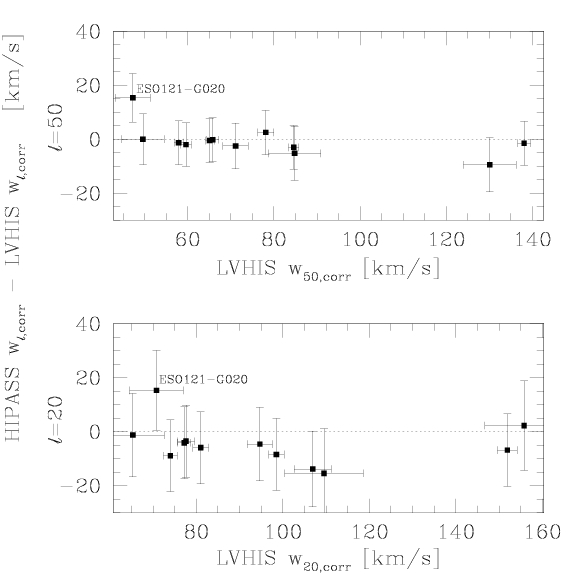, width=\linewidth}
\caption{Comparison between the derived HI velocity line widths from the LVHIS galaxies and the offset from the HI velocity line widths measured by HIPASS, corrected for instrumental broadening.  }
 \label{fig:w20} 
\end{figure}

 Interferometric measurements can underestimate the total flux because of the lack of \emph{uv-} sampling at small spatial frequencies resulting from the minimum antenna separation. Due to the missing short baselines any extended, diffuse \HI\ emission may not have been observed by LVHIS. Thus, it is imperative that a comparison to single dish observations is made. In order to make an accurate comparison between the single dish data and the interferometric data from LVHIS, the \HI\ line widths must be corrected for instumental broadening.  We use the method of \cite{bottinelli90} which is based on comparing line widths at different resolutions and  is given by the linear relations:
 \begin{eqnarray}
 w_{20,corr}&=&w_{20}-0.55R\\
 w_{50,corr}&=&w_{50}-0.13R
 \end{eqnarray}
where $w_{20}$ is the observed line width and $w_{20,corr}$ is the line width corrected for the instrumental resolution $R$  in \kms.  In Figure~\ref{fig:w20} we show the comparison between the \HI\ velocity line widths obtained by LVHIS and the HIPASS Bright Galaxy Catalogue \citep{koribalski04}.  The velocity resolution of LVHIS is 4\,\kms\ compared to the 18\,\kms\ resolution HIPASS. The average rms for the LVHIS data is 1.5\,mJy compared to 13\,mJy for HIPASS, thus the uncertainty in the LVHIS measurements is much lower.

The HIPASS measurement of the \HI\ line width of ESO121-G020  (96\kms\ at the 20\% level of peak flux density) is significantly larger than the LVHIS measurement ($73\pm6$\kms). \cite{warren06} identified a nearby companion ATCA J061608-574552 at a projected distance of 3 arcmin from ESO121-G020. HIPASS measured these two galaxies as a single point source  because the intrinsic size is much smaller than the  angular resolution of 15.5 arcmin and hence the HIPASS spectrum is broadened by the companion galaxy.

The mean difference between the HIPASS and LVHIS \HI\ velocity line widths for all sample galaxies excluding ESO121-G020  is $-2.0\pm 3.1$\,\kms and $-6. 1\pm 6.2$\,\kms (or $-0.6\pm 5.9$\,\kms and $-4.6\pm 8.0$\,\kms including ESO121-G020) at the 50\% and 20\% levels of peak flux density respectively. HIPASS measurements of the \HI\ line width have a median uncertainty of 8\,\kms\ and 12\,\kms\ for $w_{50}$ and $w_{20}$ respectively. We conclude that the two  HI surveys obtained consistent measurements of the \HI\ velocity line widths, however the agreement is clearly better for  $w_{50}$. We note that the use of the \cite{verheijen01a} method to correct for instrumental broadening does not change the overall results. 

The HIPASS spectrum at the coordinates of IC5152 also contains \HI\ emission from Galactic high velocity clouds at a velocity of 33\,\kms\ with a peak flux of 0.7 Jy. Fortunately this does not overlap with the velocity range of IC5152 and does not effect the HIPASS measurements.  This is resolved and excluded from the spectrum in the LVHIS observations. The LVHIS spectrum of ESO245-G005 shows a strong \HI\ absorption line at a redshift of 386\,\kms. Similarly, the LVHIS spectrum of ESO381-G020 shows two weak absorption lines at 581\,\kms\ and 597\,\kms.  LVHIS measured the integrated flux density of ESO174-G?001 11\% lower than HIPASS. This indicates that the interferometer filtered out the more extended, diffuse \HI\ emission due to the missing short baselines but the flux was observed by the single dish observations of HIPASS. Note that although LVHIS measured the integrated flux density of ESO137-G018 to be higher than HIPASS,  it appears to be due to uncertainty in the HIPASS baseline.

\section{Rotation Curves}\label{s:rotcurs}
\subsection{Fitting Procedure}
The \HI\ rotation curves were derived by fitting a tilted ring model to the  mean \HI\ velocity fields.  We utilised the  tilted ring fitting algorithm, ROTCUR, which is incorportated in the Groningen Image Processing System (GIPSY; \citealt{gipsy01,gipsy02}). The width of each tilted ring was set to 2/3  that of the synthesised beam (following that adopted by \citealt{verheijen01a}) with an average value of 30 arcseconds. 

First, the systemic velocity and the dynamic centre  of the galaxy were determined by fitting one large ring that encompasses all of the data. 
All data points within the ring were given equal consideration, i.e., were uniformly weighted.  The uncertainty in the systemic velocity and the dynamic centre was estimated by fitting multiple  tilted rings to the data and taken to be the range in  the values obtained. Once the systemic velocity and dynamic centre were determined, they were kept at that value for the rest of the analysis.

Next, the kinematical position angle and inclination of the galaxy's \HI\ disk  was obtained. The data was weighted by $|\cos\theta|$, where $\theta$ is the angle from  the semi-major axis of the galaxy. Hence data near the semi major axis is given the most consideration whereas data near the semi-minor axis is given the least consideration to avoid large deprojection errors. The position angle was fixed first, either to the average value for all tilted rings if appropriate, or to a general trend, if it was found to vary systematically as a function of radius. Once the position angle was fixed, the inclination was investigated. The inclination could sometimes be highly variable. Hence at this step, the inclination would either be obtained via the titled ring model, or the inclination was set such that the model has the same semi-major/semi-minor axis ratio as the observed distribution of the \HI\ gas. We indicate in the results when the titled ring model failed and the inclination is obtained to reproduce the galaxy's axis ratio and correct this value for the thickness of the disk \citep{maller09} using $q_z=0.1$ as the galaxy compression \citep{nedyalkov93}.  The uncertainty in the position angle and inclination was estimated from the range in values obtained for the multiple rings fitted. 

Finally, the rotation curve was obtained by fitting a final tilted ring model to the $|\cos\theta|$ weighted data with systemic velocity, dynamic centre, position angle and inclination fixed to the values obtained in the previous steps.  The rotation curve for the receding and approaching sides of a galaxy were obtained separately by fixing the dynamic centre and the systemic velocity to that found when analysising the whole galaxy, and obtaining the position angle and inclination separately for each side. The difference between the approaching and receding sides of the rotation curve is representative of the uncertainties due to asymmetries in the  galaxy rotation \citep{swaters99,deblok08}.

To accurately obtain a true measure of the rotational velocity, the effect of pressure on the rotation curve needs to be considered (ie.,  an asymmetric drift correction). However, this is only important if the rotational velocity is comparable to the velocity dispersion \citep{begum03}, which is not the case for our sample galaxies.  \cite{swaters09} calculated the correction required for a sample of similar galaxies and showed that the correction is typically smaller than 3~\kms; less than the uncertainties in our derived velocities. Therefore we have not corrected the rotation curves for asymmetric drift.

Any deviations from the method stated above will be noted in the detailed discussion for each galaxy.

\subsection{Rotation Curve Results}

The \HI\ rotation curve parameters obtained by the titled ring analysis for each galaxy are listed in Table~\ref{tab:rotcurresults}. There we list:
\begin{itemize}
\item[]\emph{Column} (1). --  galaxy name.
\item[]\emph{Column} (2). -- beam size, in arcsec.
\item[]\emph{Columns} (3), (4) and (5). --  dynamic centre position and uncertainty.
\item[]\emph{Column} (6). -- the systemic velocity, $v_{sys}$.
\item[]\emph{Column} (7). -- the position angle, taken in anti-clockwise direction, between the north direction on the sky and the major axis of the receding half of the galaxy. Here if two values are given, then the position angle was found to vary as a function of radius, with the inner position angle given by the first value and the outer position angle given by the second.
\item[]\emph{Column} (8). -- the inclination of the tilted ring model. Here, for galaxies where the tilted ring model failed to obtain a solution for the inclination, the value was chosen such that the physical dimensions of the tilted ring model matched the observed distribution of the \HI\ gas. This is indicated by listing the value obtained in parentheses.
\item[]\emph{Column} (9). -- the maximum rotational velocity obtained via the tilted ring model. The uncertainties, if  given, represent the difference between the maximum rotational velocity obtained by fitting the tilted ring model to the entire galaxy and the maximum rotational velocity obtained by fitting the tilted ring model to the approaching and receding sides.
\item[]\emph{Column} (10). -- the rotational velocity at which the flat part of the rotation curve is obtained. Note that this value may be lower than the maximum rotational velocity as it is taken as the average of the flat part of the rotation curve (if sufficient data points are available). 
\end{itemize}

\begin{table*}
\caption{Rotation Curve Results}
\begin{tabular}{llllccccccc}
\hline
\multicolumn{1}{c}{Galaxy} &\multicolumn{1}{c}{Beam} & \multicolumn{1}{c}{centre R.A.} & \multicolumn{1}{c}{centre Decl.} & $\Delta$centre & $v_{sys}$${}^{a}$ & P.A.${}^{a}$ & $i$${}^{a}$          & $v_{max}{}^{b}$ &$v_{flat}$ \\
	   &\multicolumn{1}{c}{(arcsec${}^{2}$)}& \multicolumn{1}{c}{(J2000)}          &\multicolumn{1}{c}{(J2000)}   &  (arcsec)        & (\kkms)      & (degrees)  & (degrees) & (\kkms) & (\kkms) \\
\multicolumn{1}{c}{(1)} &\multicolumn{1}{c}{(2)} &\multicolumn{1}{c}{(3)}  & \multicolumn{1}{c}{(4)} & (5) & (6) & (7) & (8)& (9)& (10) \\
\hline
AM0605-341 & $80\times52$    & 06h07m20.2s    & -34d12m22s& 7.5&$757 \pm 2$ &$274 \pm 5$ & (50) & $85$&A\\
AM0704-582  &$59\times 53$   & 07h05m18.0s   & -58d31m10s& 10.5  &$564\pm 3$& $276\pm 2$ & (35) & $57$&A\\
ESO059-G001 &$51\times 47$    &07h31m18.6s   &-68d11m17s& 8 &$527 \pm 2$ & 329 -- 319 & $45 \pm 10$ &$61.8$&B\\
ESO121-G020   & $40\times37$ &06h15m54s${}^{\ast}$            & -57d43m32s${}^{\ast}$& -- &$584 \pm 1$&$265 \pm 5$&$(40)$&$48.7$&A\\
ESO137-G018&$57 \times 49$& 16h20m58.5s        & -60d29m23s& 19.5  &$601\pm 2$ & 33 -- 28& $50 \pm 6$ & $80.1^{+1.2}_{-1.3}$ & $ 80\pm 2$\\
ESO174-G?001&$61\times 50$   & 13h47m57.4s & -53d21m08s   &10& $682\pm 5$& 233 -- 202 &$ 40\pm5$& $66^{+7.4}_{-2.1}$&$65\pm9$\\
ESO215-G?009&$47\times41$   &10h57m31.0s       &-48d10m45s& 30&$598 \pm 3$ & 123 -- 116 & $35\pm3$& $53.8^{+6.0}_{-5.2} $& $54\pm5$\\
ESO245-G005  &$71\times 48$  &01h45m03.7s     & -43d36m38s &20& $392 \pm 3$ & 70 -- 98 &  $36\pm6$ &51&51\\
ESO308-G022 &$65\times46$    &06h39m33.0s        & -40d43m12s &7 &$823 \pm 1$ &$82\pm2$& $(40)$ & $40$&A\\
ESO325-G?011&$62\times 43$   &13h45m01.8s&-41d51m50s & 15 &$546\pm5$& $302\pm4$& $42\pm10$& $46.0^{+3.9}_{-2.9}$&  A -- B\\
ESO381-G020  &$80\times49$  & 12h46m00.0s&-33d50m11s& 10 &$587\pm 2$ & 295 -- 314 & $55\pm10$&$46.7$ & --\\ 
IC5152      &$47\times40$  & 22h02m42.3s & -51d17m 50s &16&$121 \pm 2$ & 271 -- 198 & $49\pm6$&$59.5^{+6.3}_{-5.7}$& A -- B\\
\hline
\end{tabular}
\flushleft
(a) value for tilted ring model fitted to both approaching and receding sides, inclinations given in parenthesis indicate that the tilted ring analysis failed and the \HI\ dimensions were used to constrain this parameter; 
(b) the uncertainty in the measurement represents the tilted ring model for the approaching and receding sides;
(A) the flat part of the rotation curve is not reached; 
(B) the rotation curve appears to be flattening at the outermost radii with measured rotational velocity;
($\ast$) could not be fixed using the tilted ring analysis and was set to the optical centre.\\
\label{tab:rotcurresults} 
\end{table*}

\begin{center}
\begin{figure*} 
\begin{tabular}{cccc}
{ \mbox{\epsfig{file=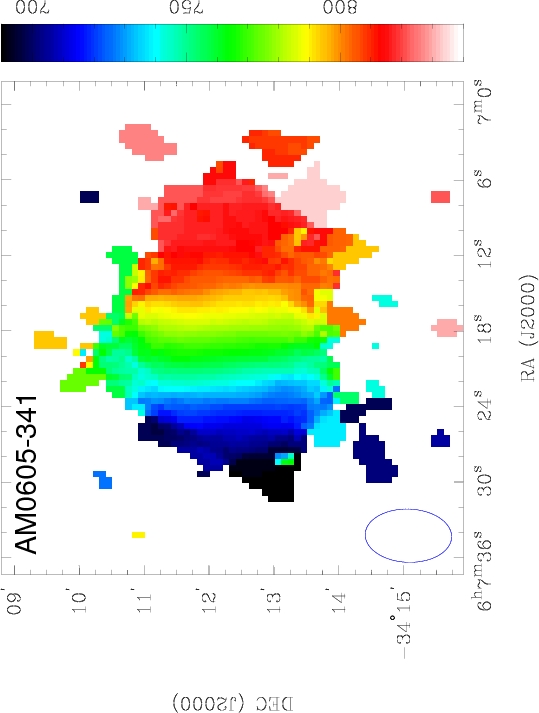, height =5.5cm,angle=-90}}} &
  \mbox{\epsfig{file=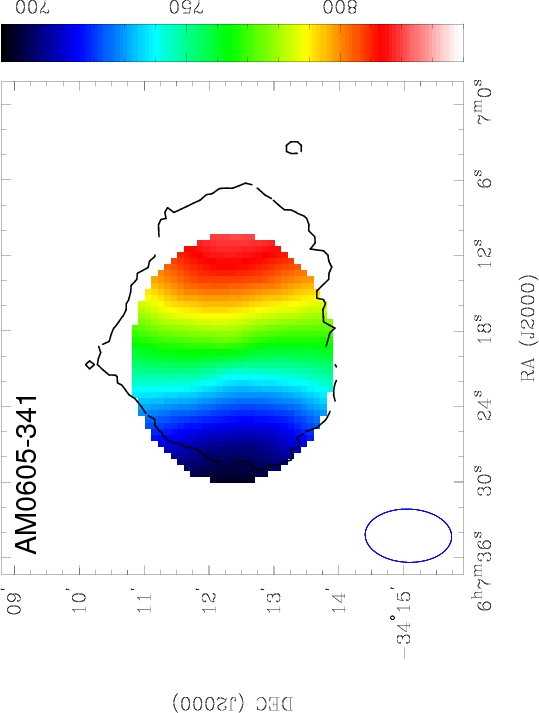, height =5.5cm,angle=-90}} &
  \mbox{\epsfig{file=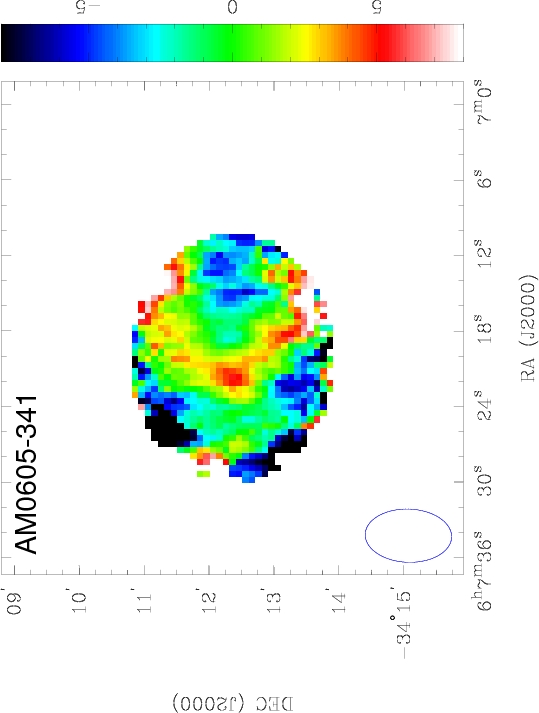, height =5.5cm,angle=-90}} \\
  \mbox{\epsfig{file=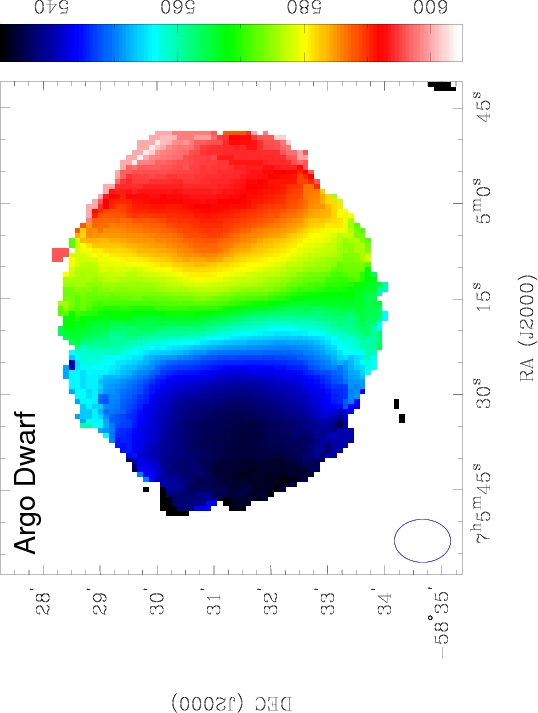, height =5.5cm,angle=-90}} &
  \mbox{\epsfig{file=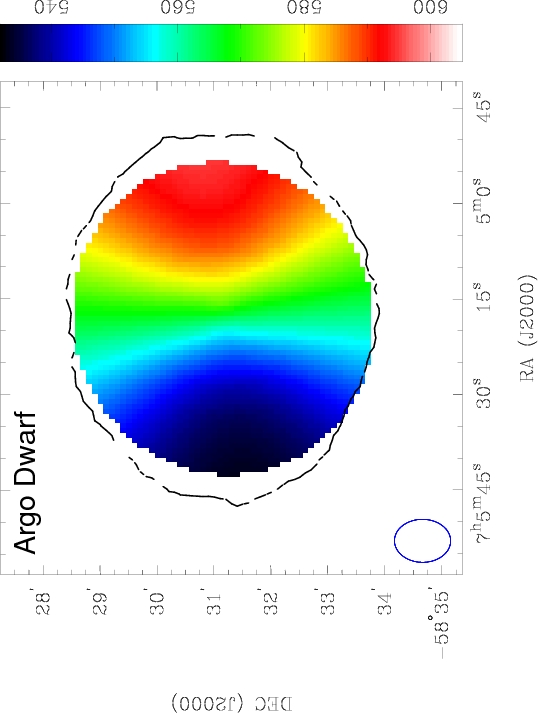, height =5.5cm,angle=-90}} &
  \mbox{\epsfig{file=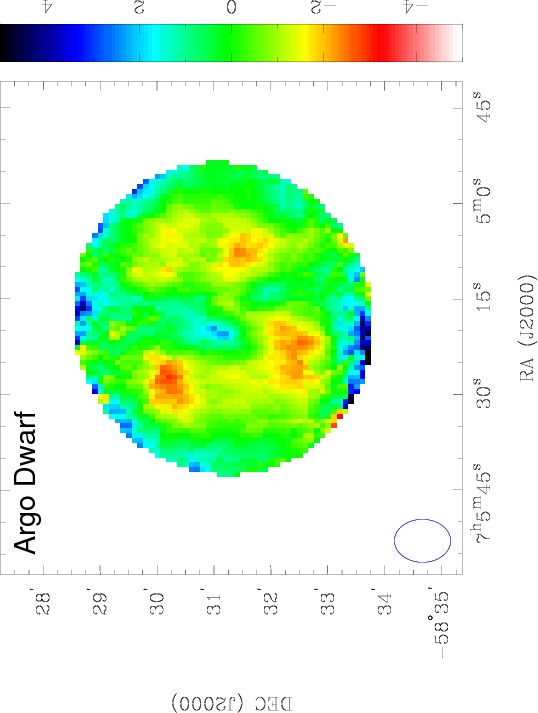, height =5.5cm,angle=-90}} \\
  \mbox{\epsfig{file=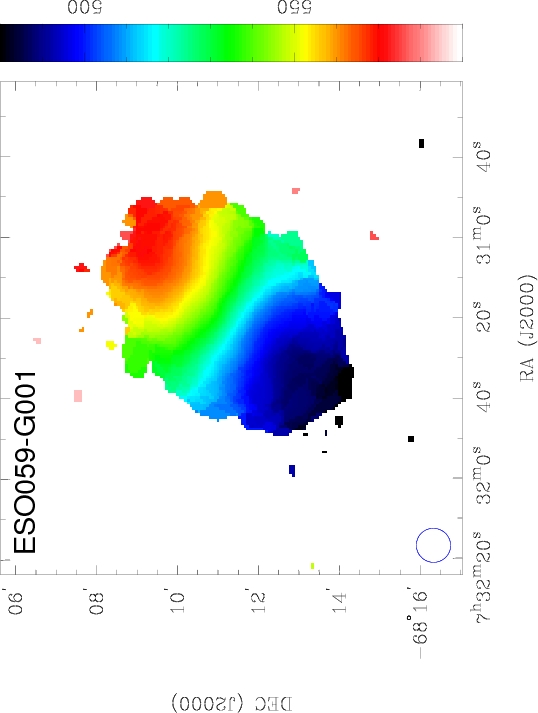, height =5.5cm,angle=-90}} &
  \mbox{\epsfig{file=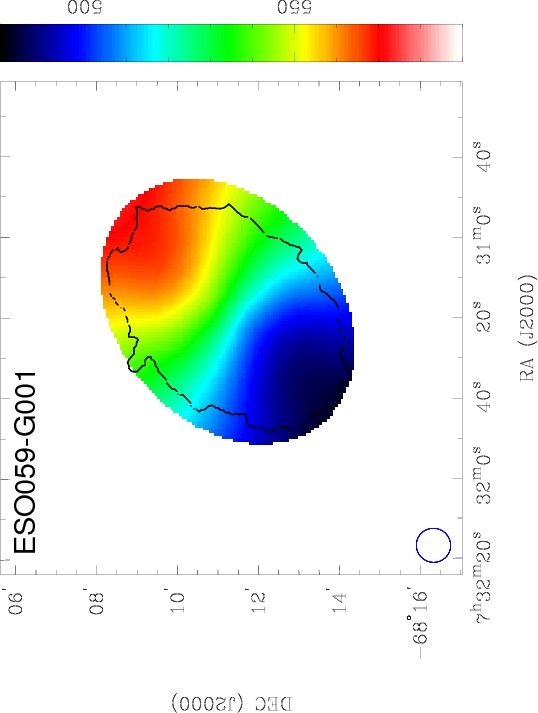, height =5.5cm,angle=-90}} &
  \mbox{\epsfig{file=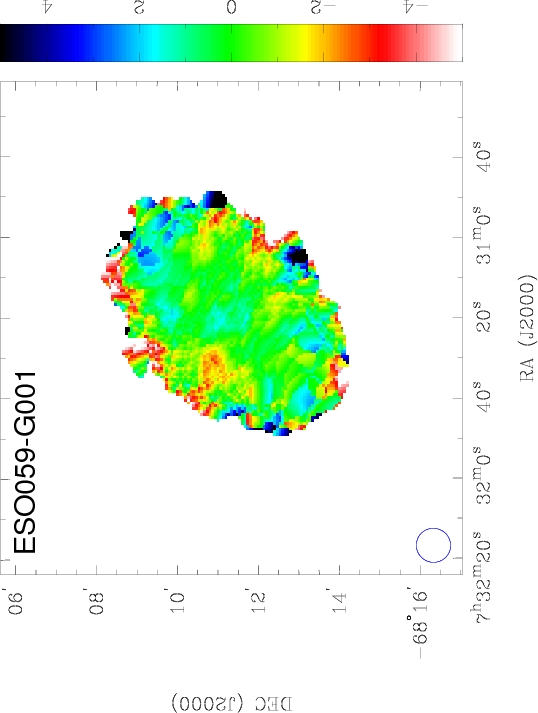, height =5.5cm,angle=-90}} \\
  \mbox{\epsfig{file=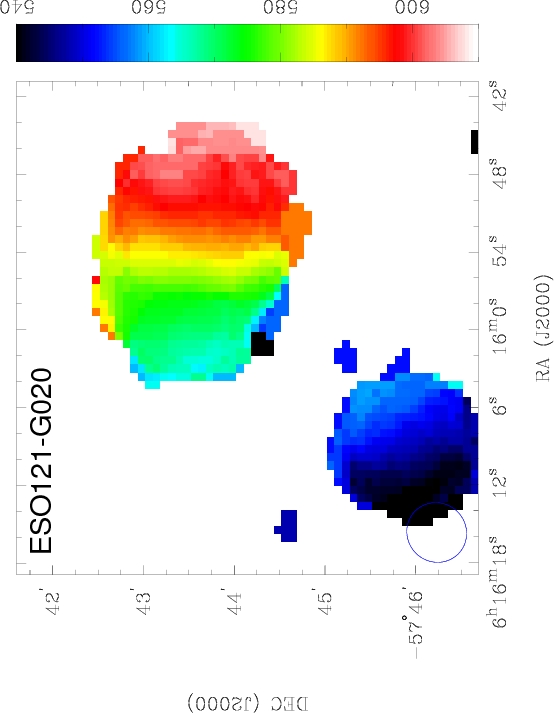, height =5.5cm,angle=-90}} &
  \mbox{\epsfig{file=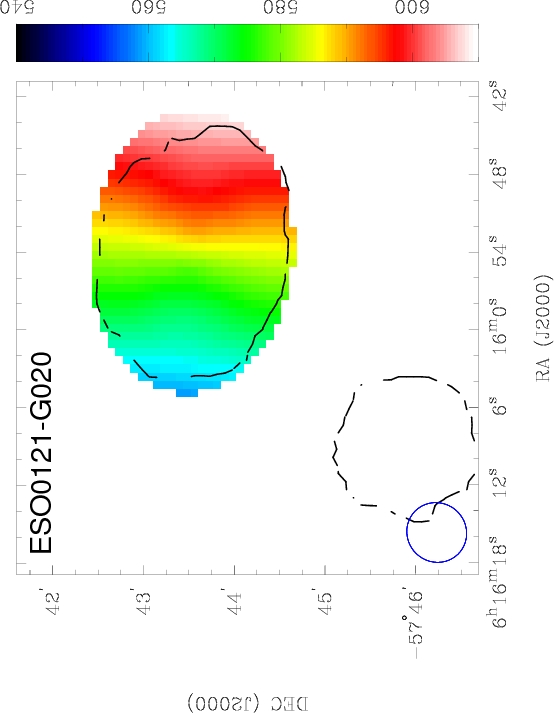, height =5.5cm,angle=-90}} &
  \mbox{\epsfig{file=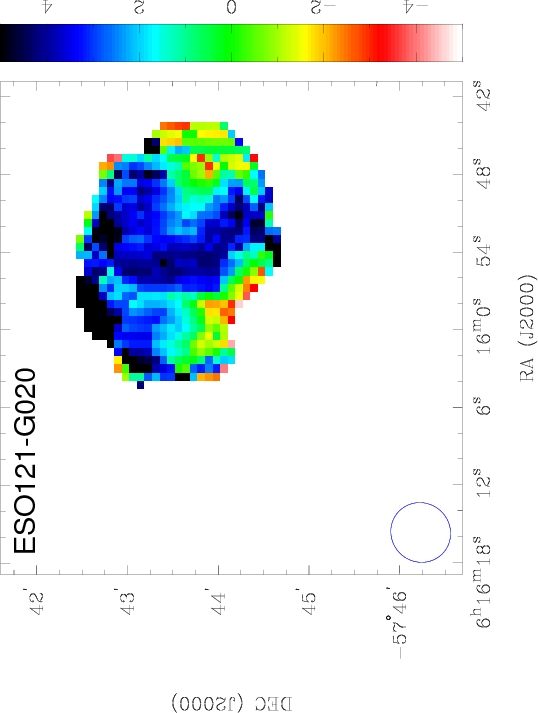, height =5.5cm,angle=-90}} \\
  \mbox{\epsfig{file=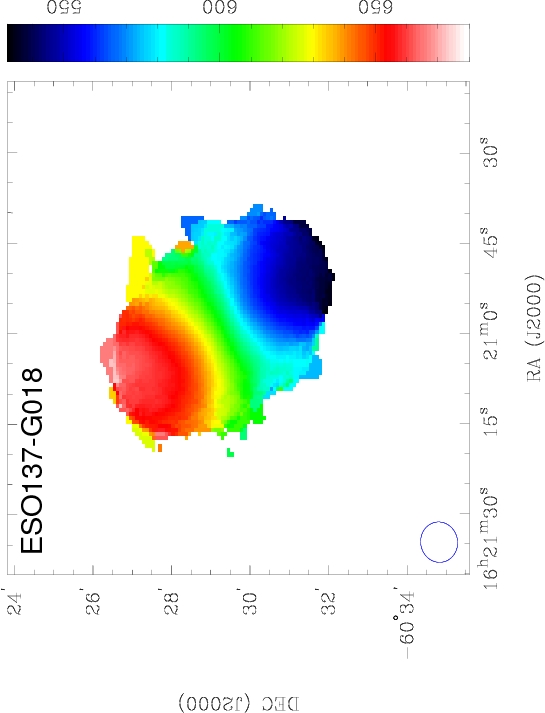, height =5.5cm,angle=-90}} &
  \mbox{\epsfig{file=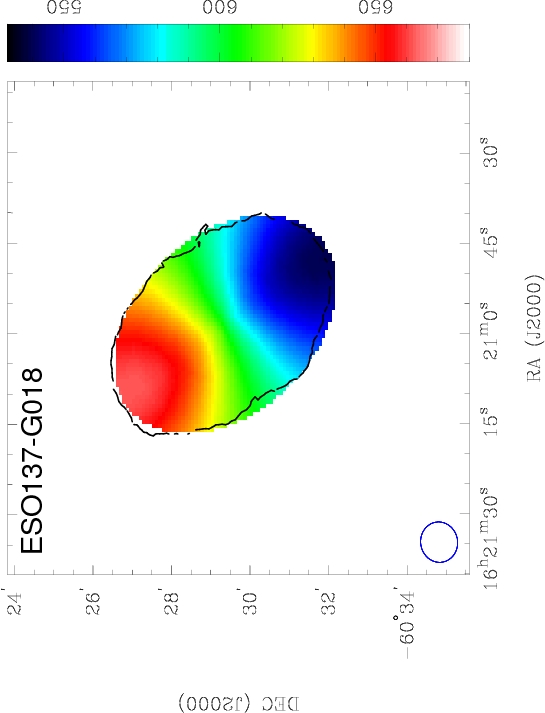, height =5.5cm,angle=-90}} &
  \mbox{\epsfig{file=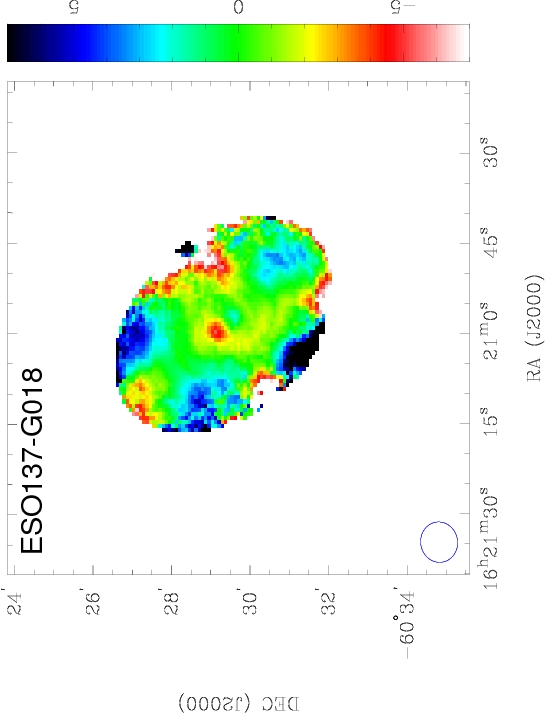, height =5.5cm,angle=-90}} \\
\end{tabular}
\caption{From left to right: the HI velocity field, the derived tilted ring model and the difference between the mean \HI\ velocity field and the model velocity field. The black contour overlaid on the derived tilted ring model shows the distribution of \HI\ flux at the level of 0.5 Jy beam${}^{-1}$. Note that the velocity range chosen for the residual \HI\ velocity field is adjusted for each galaxy.}
\label{fig:rotcur1}
\end{figure*}
\end{center}

\begin{center}
\begin{figure*} 
\begin{tabular}{ccc}
    \mbox{\epsfig{file=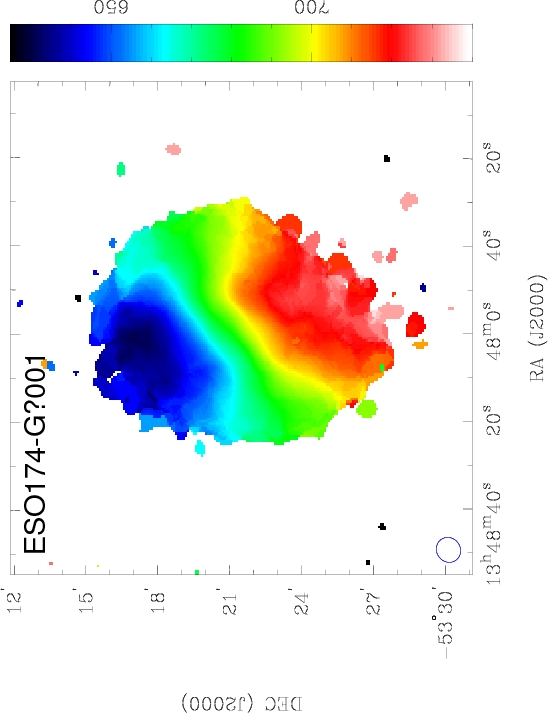, height =5.5cm,angle=-90}} &
  \mbox{\epsfig{file=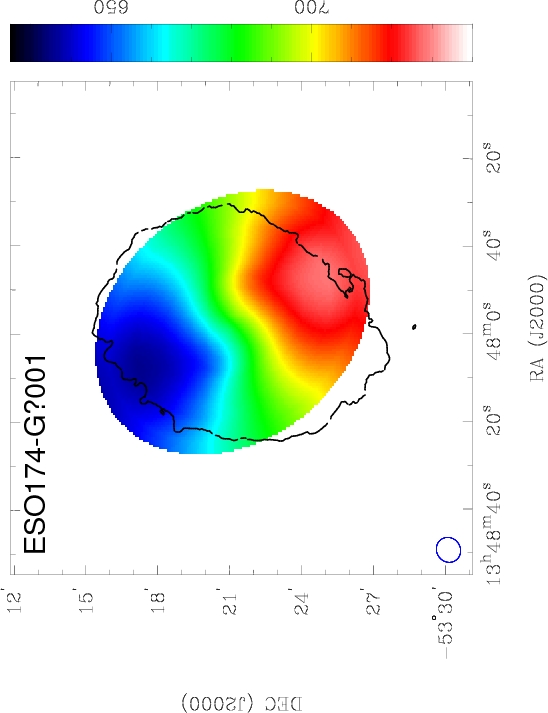, height =5.5cm,angle=-90}} &
  \mbox{\epsfig{file=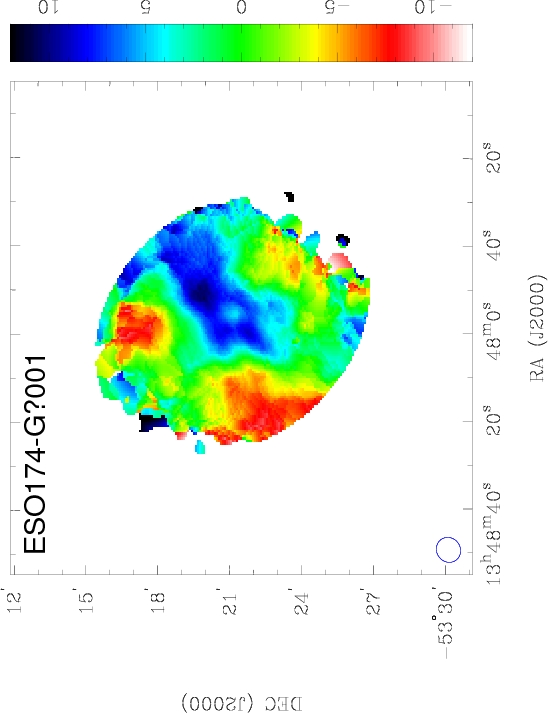, height =5.5cm,angle=-90}} \\
    \mbox{\epsfig{file=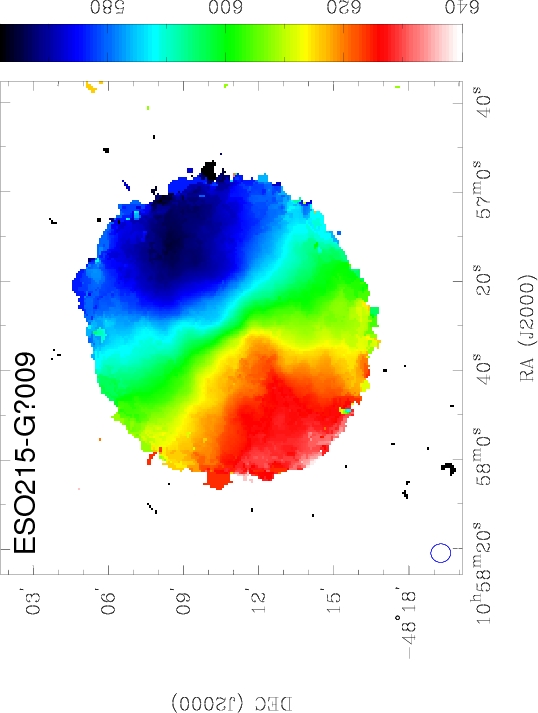, height =5.5cm,angle=-90}} &
  \mbox{\epsfig{file=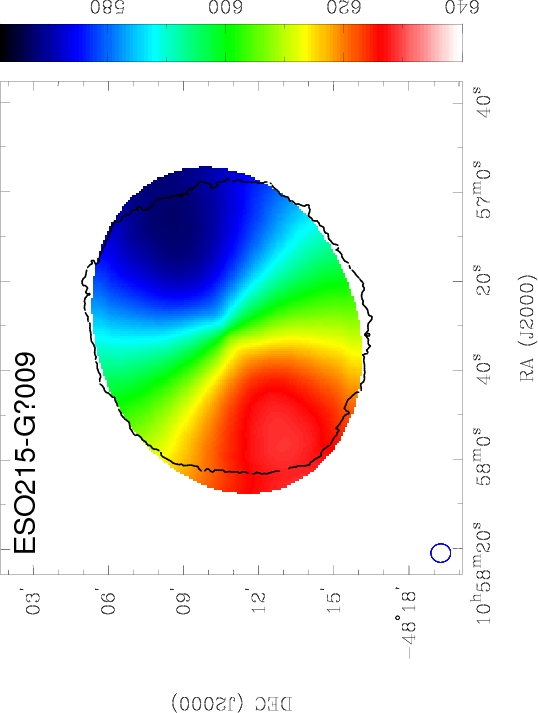, height =5.5cm,angle=-90}} &
  \mbox{\epsfig{file=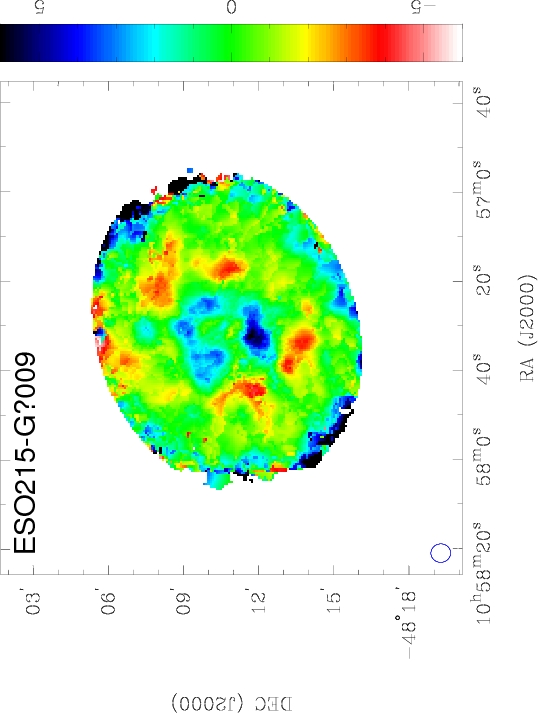, height =5.5cm,angle=-90}} \\
    \mbox{\epsfig{file=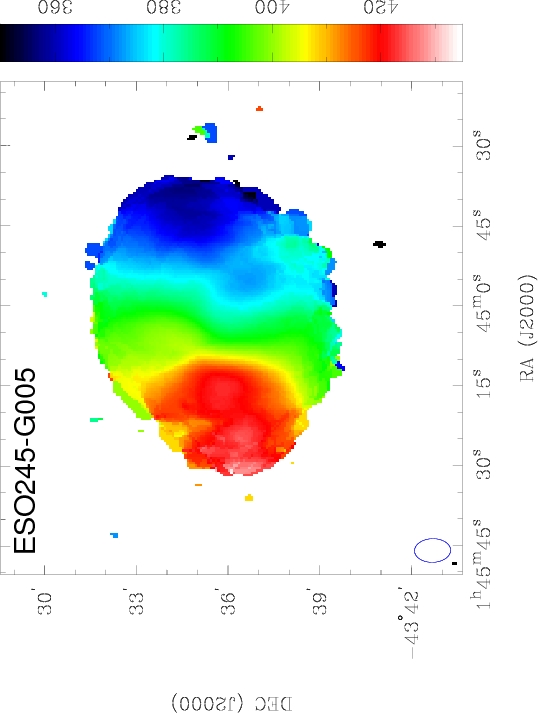, height =5.5cm,angle=-90}} &
  \mbox{\epsfig{file=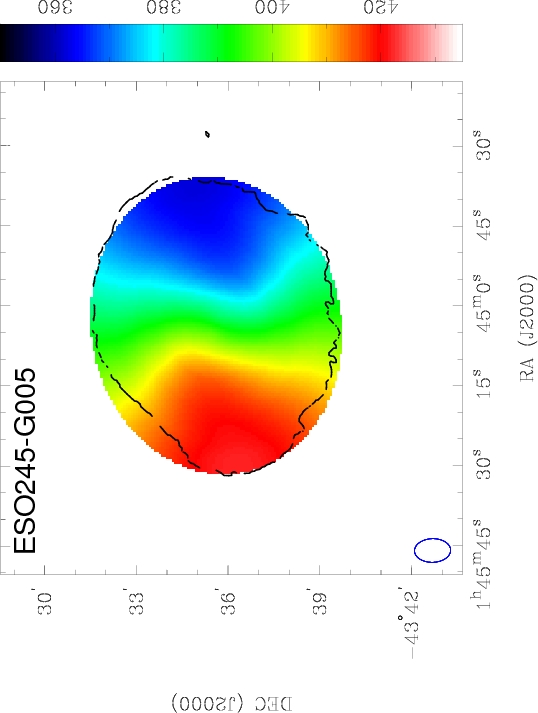, height =5.5cm,angle=-90}} &
  \mbox{\epsfig{file=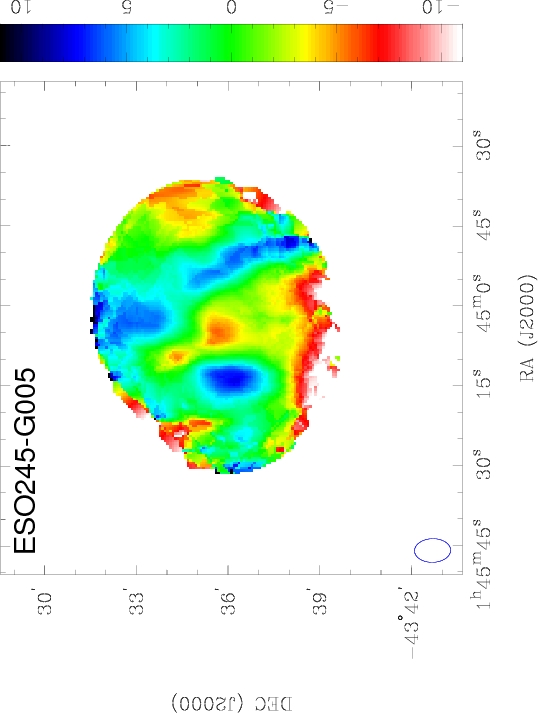, height =5.5cm,angle=-90}} \\
    \mbox{\epsfig{file=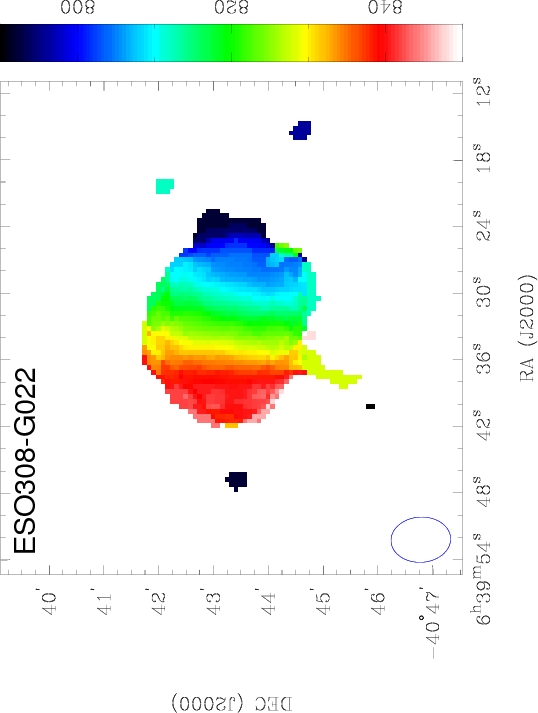, height =5.5cm,angle=-90}} &
  \mbox{\epsfig{file=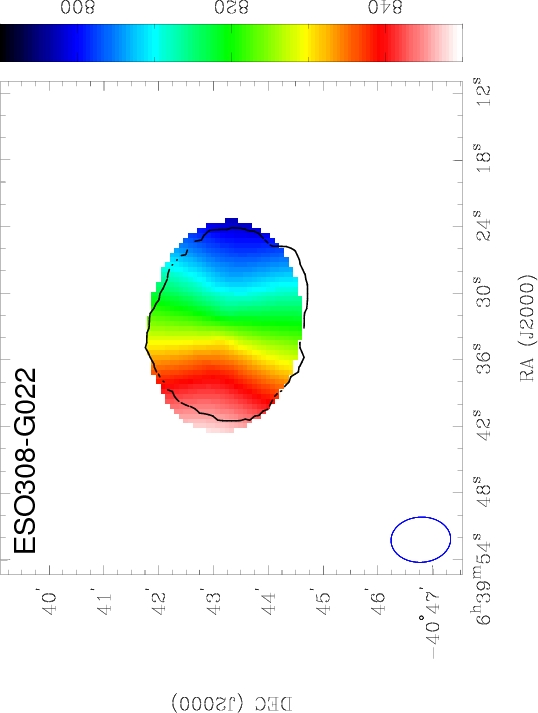, height =5.5cm,angle=-90}} &
  \mbox{\epsfig{file=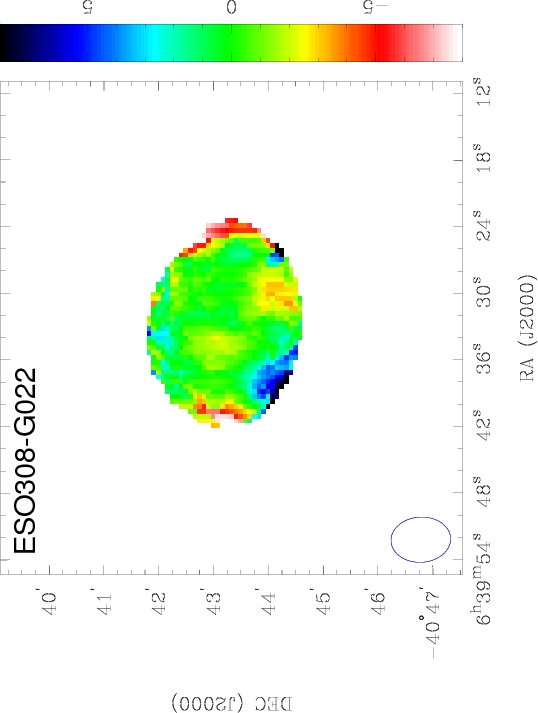, height =5.5cm,angle=-90}} \\
    \mbox{\epsfig{file=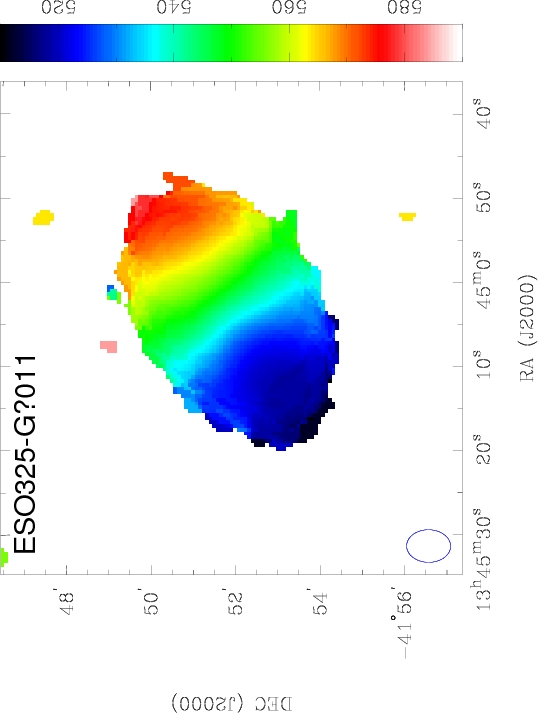, height =5.5cm,angle=-90}} &
  \mbox{\epsfig{file=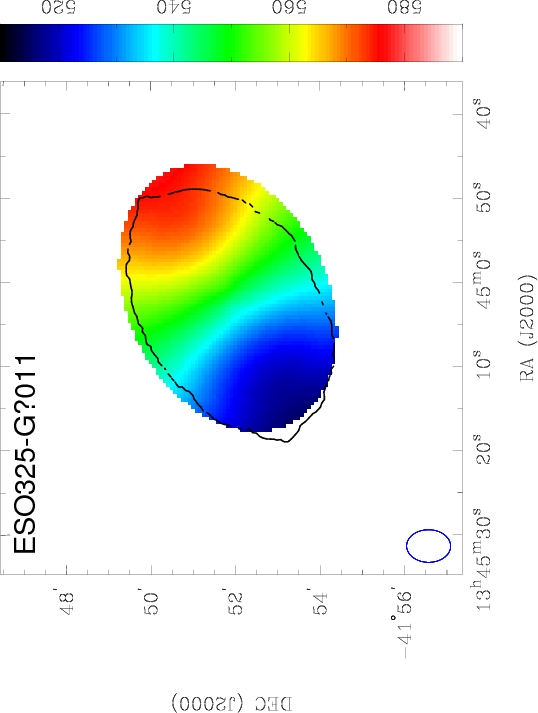, height =5.5cm,angle=-90}} &
  \mbox{\epsfig{file=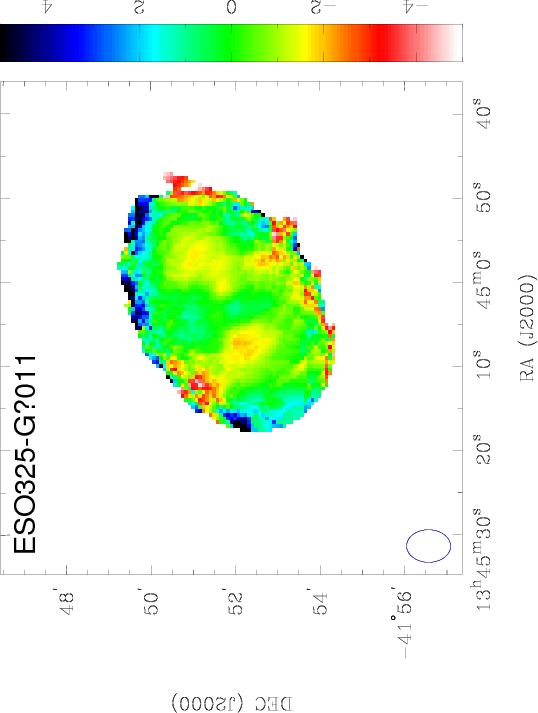, height =5.5cm,angle=-90}} \\
  \end{tabular}
\caption{From left to right: the HI velocity field, the derived tilted ring model and the difference between the mean \HI\ velocity field and the model velocity field. The black contour overlaid on the derived tilted ring model shows the distribution of \HI\ flux at the level of 0.5 Jy beam${}^{-1}$. Note that the velocity range chosen for the residual \HI\ velocity field is adjusted for each galaxy.}
\label{fig:rotcur2}
\end{figure*}
\end{center}

\begin{center}
\begin{figure*} 
\begin{tabular}{ccc}
    \mbox{\epsfig{file=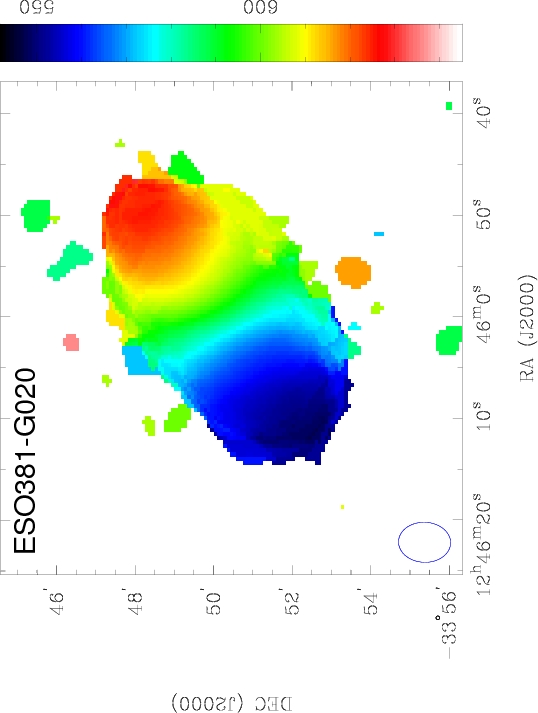, height =5.5cm,angle=-90}} &
  \mbox{\epsfig{file=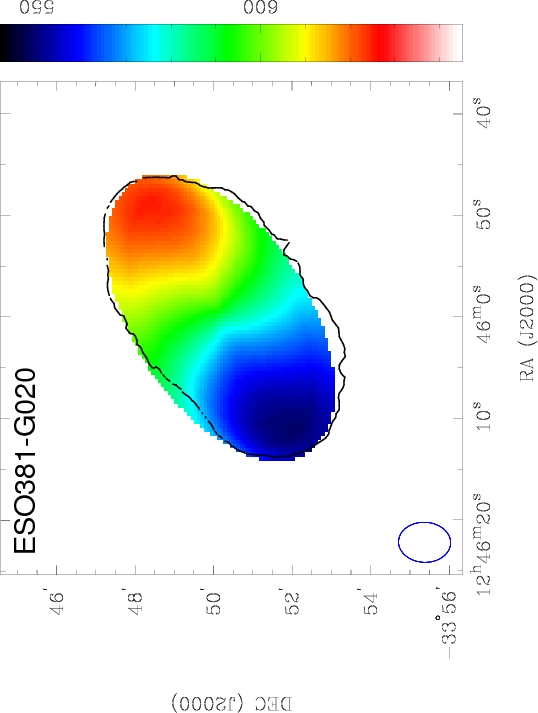, height =5.5cm,angle=-90}} &
  \mbox{\epsfig{file=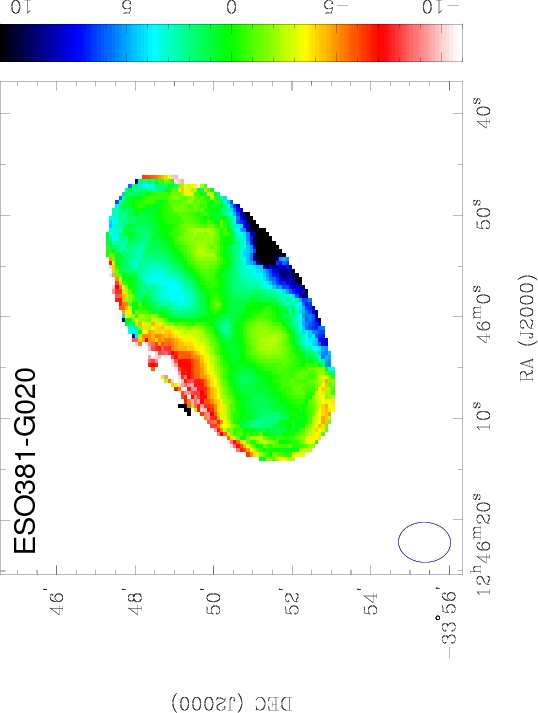, height =5.5cm,angle=-90}} \\
    \mbox{\epsfig{file=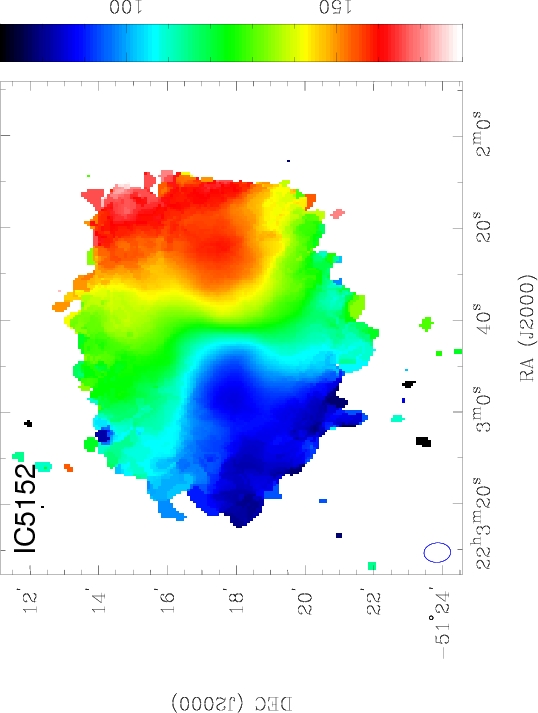, height =5.5cm,angle=-90}} &
  \mbox{\epsfig{file=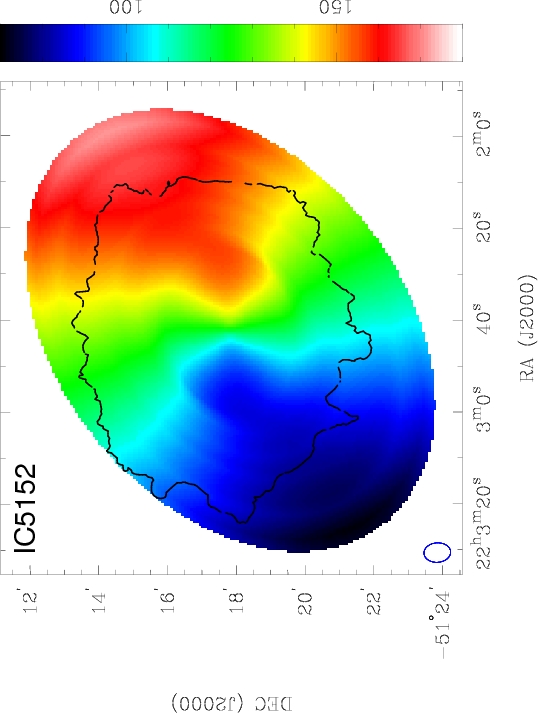, height =5.5cm,angle=-90}} &
  \mbox{\epsfig{file=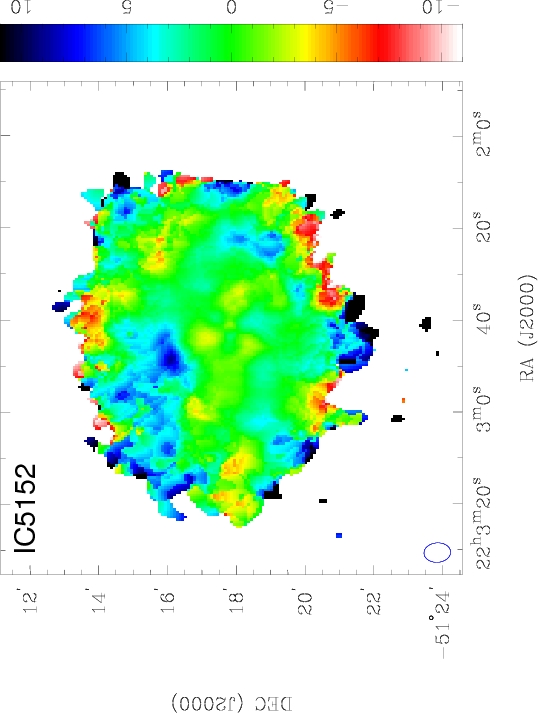, height =5.5cm,angle=-90}} \\
   \end{tabular}
\caption{From left to right: the HI velocity field, the derived tilted ring model and the difference between the mean \HI\ velocity field and the model velocity field. The black contour overlaid on the derived tilted ring model shows the distribution of \HI\ flux at the level of 0.5 Jy beam${}^{-1}$. Note that the velocity range chosen for the residual \HI\ velocity field is adjusted for each galaxy.}
\label{fig:rotcur3}
\end{figure*}
\end{center}

\begin{center}
\begin{figure*} 
\begin{tabular}{cccc}
{ \mbox{\epsfig{file=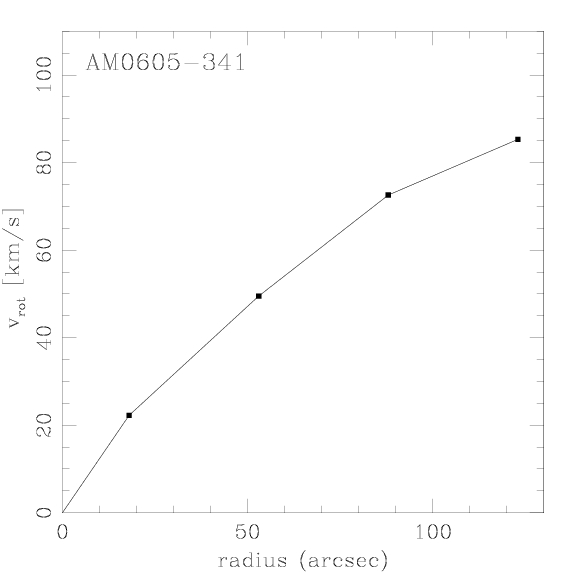,height=4cm}}} &
  \mbox{\epsfig{file=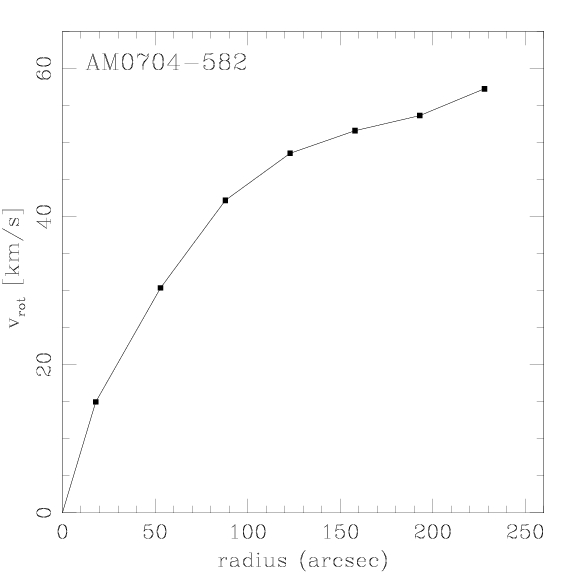, height=4cm}}  &
  \mbox{\epsfig{file=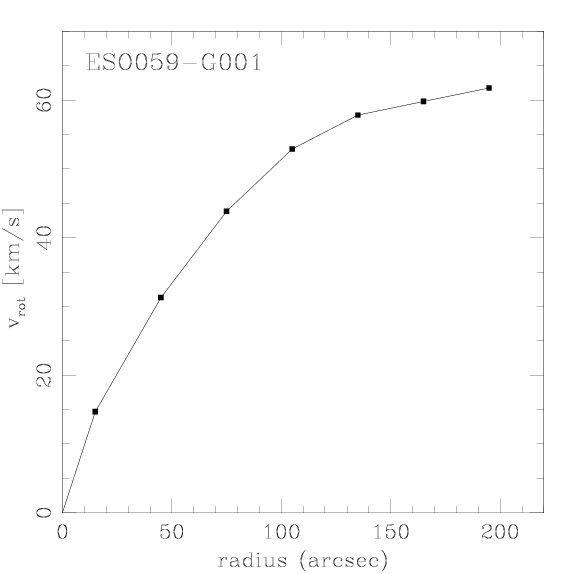,width=4cm}}  &
  \mbox{\epsfig{file=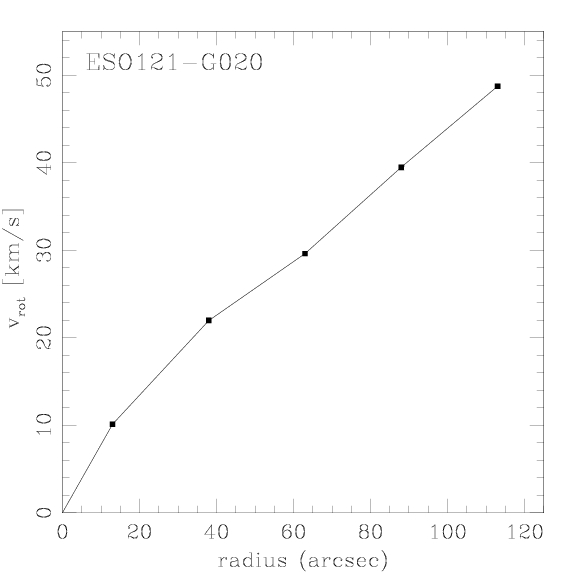,width=4cm}} \\
  \mbox{\epsfig{file=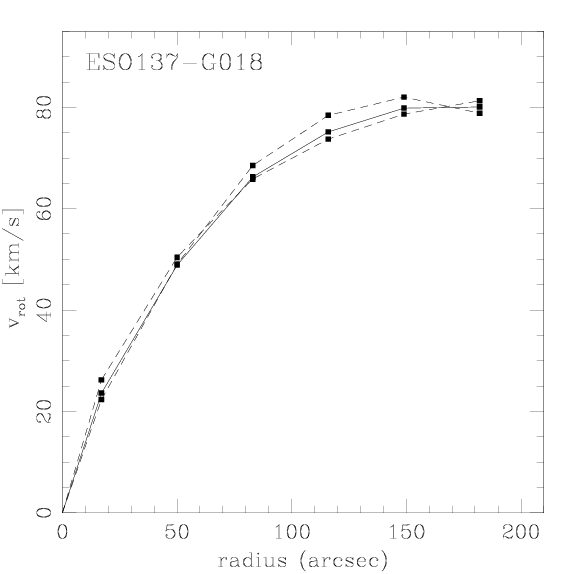,width=4cm}}  &
  \mbox{\epsfig{file=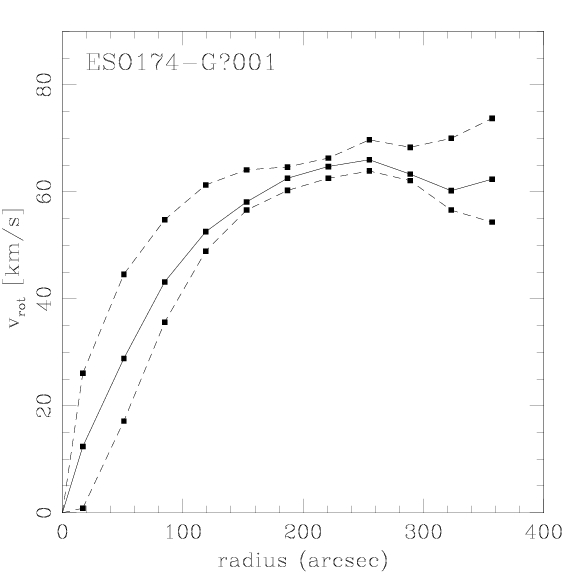,width=4cm}}  &
  \mbox{\epsfig{file=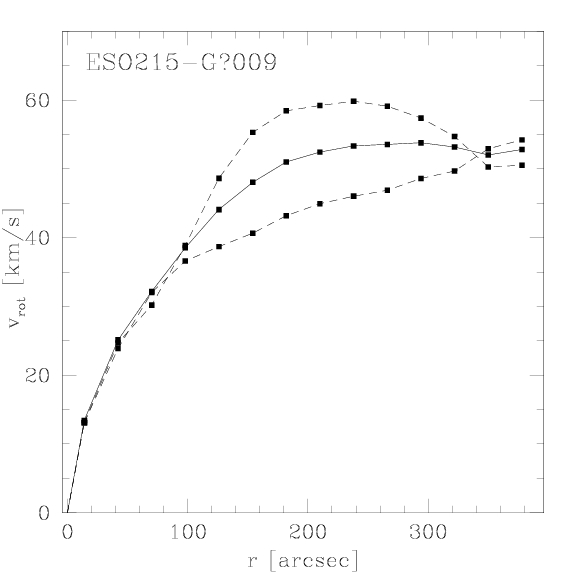,width=4cm}}  &
    \mbox{\epsfig{file=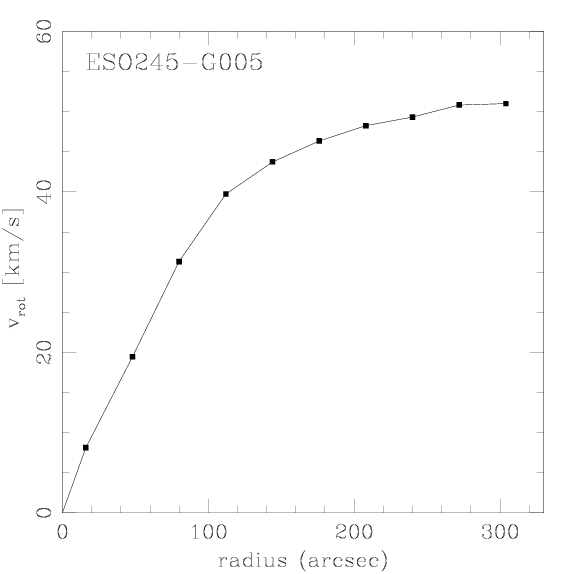,width=4cm}}\\
    \mbox{\epsfig{file=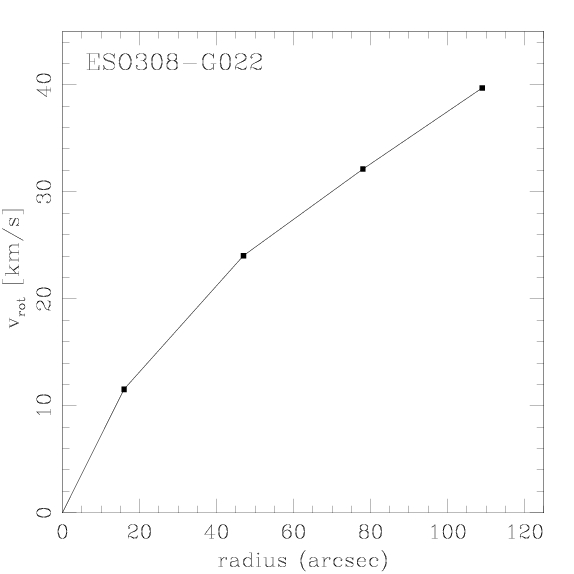,width=4cm}}  &
  \mbox{\epsfig{file=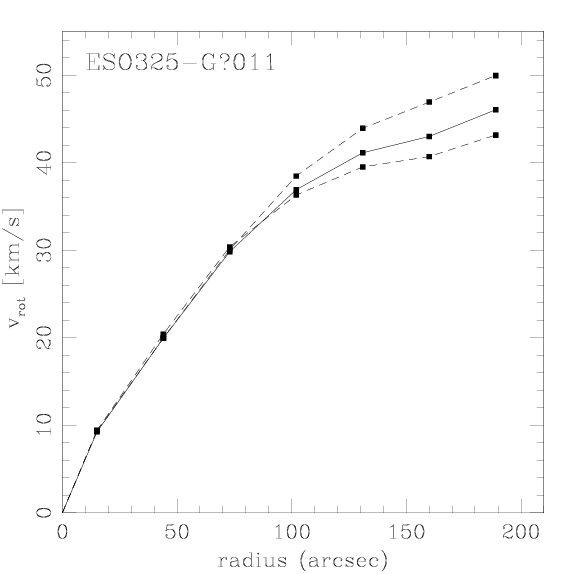,width=4cm}}  &
  \mbox{\epsfig{file=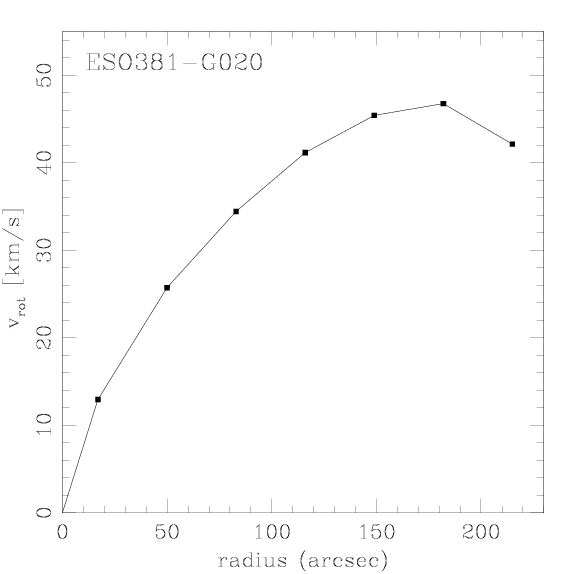,width=4cm}}  &
  \mbox{\epsfig{file=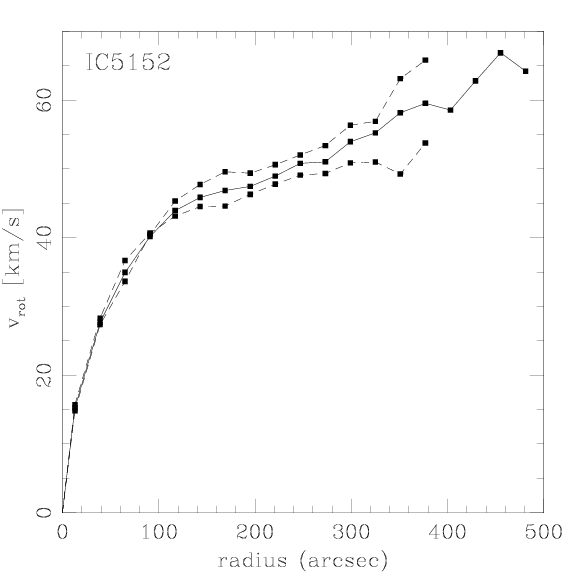,width=4cm}} \\
\end{tabular}
\caption{The derived rotation curve for sample galaxies. The solid line represents the titled ring model fitted to the entire galaxy and the dashed lines represent the approaching and receding sides. }
\label{fig:rotcur}
\end{figure*}
\end{center}

\begin{center}
\begin{figure} 
{ \mbox{\epsfig{file=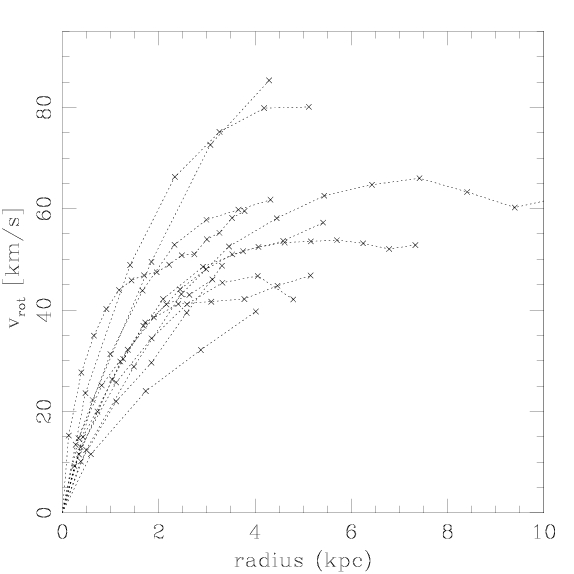, width=\linewidth}}} 
\caption{The derived rotation curve for sample galaxies overlaid for comparison. Here, only the average rotation curve for each galaxy is presented. The rotation curves for the individual approaching and receding sides are omitted.}
\label{fig:combrot}
\end{figure}
\end{center}

The derived tilted ring models are best compared visually to the mean \HI\ velocity field of the galaxy. In Figures~\ref{fig:rotcur1} to \ref{fig:rotcur3}, we present the mean \HI\ velocity field (the first moment map), the model velocity field (the tilted ring model) and the residual field (the difference between the observed field and the model). In Figure~\ref{fig:rotcur} the derived rotation curve for each of the sample galaxies is presented. The derived rotation curve is presented with the solid line representing the tilted ring model fitted to the entire galaxy and the dashed lines represent the approaching and receding sides. In Figure~\ref{fig:combrot} we show the rotation curve of all sample galaxies overlaid for comparison.

\section{Analysis} \label{s:rotcurspect}

\subsection{The kinematic systemic velocity of LVHIS galaxies.}

\begin{figure}
\epsfig{file=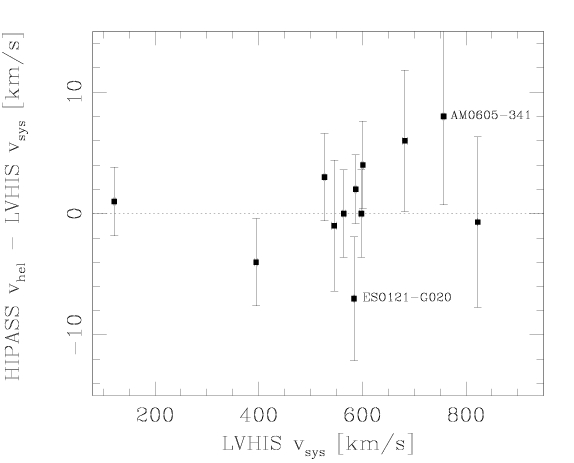, width=\linewidth}
\caption{Comparison between the derived systemic velocity from the LVHIS galaxies (the centre of the tilted ring model) and the offset from the heliocentric velocity measured by HIPASS (the midpoint of the \HI\ line profile). This shows that HIPASS data can be used to obtain the true systemic velocity of a galaxy to an accuracy of $\sim10$\kms.} \label{fig:vel} 
\end{figure}

The heliocentric velocity (the central velocity of the \HI\ spectrum), is fundamentally different from the systemic velocity derived from detailed studies of the kinematics of these galaxies. For example, a spectrum with a classic, symmetrical, double horn profile will likely have the centre of the spectrum coincide with the true systemic velocity. However in galaxies with an asymmetric spectral profile, this may not be the case if there is inflowing gas or if their dynamic centre is not aligned with the centre of the gas distribution (axisymmetric or peculiar HI distributions). In Figure~\ref{fig:vel} we show the comparison between our derived systemic velocity (given in Table~\ref{tab:rotcurresults}) and the centre of the \HI\ spectrum where the latter is defined as the midpoint of the 50\% level of the peak flux ~\citep{koribalski04}. The vertical errorbars are calculated as the sum in quadrature of the uncertainties in the derived systemic velocity and the HIPASS heliocentric velocity.

 It can be seen in Figure~\ref{fig:vel} that the heliocentric velocities are similar to the true systemic velocities. The two furthest outliers are ESO121-G020 and AM0605-341. The  heliocentric velocity of ESO121-G020 can not be correctly measured using HIPASS data due to the inclusion of an unresolved  nearby companion \citep{warren06} as mentioned before.  The offset of the galaxy AM0605-341 is due to the dynamical centre being offset from the centre of the \HI\ distribution. This galaxy has an extension of  redshifted atomic hydrogen, thus the heliocentric velocity is larger than the true systemic velocity. The distribution of the \HI\ gas in AM0605-341 is discussed in further detail in section \ref{ss:am0605}. Figure~\ref{fig:vel} shows that there is no significant difference between the centre of the \HI\ line profile and the true systemic velocity, with HIPASS data suitable to obtain the systemic velocity to an accuracy of approximately 10\kms.

\subsection{Measuring rotation velocities from \HI\ line profiles}
Information on the rotational velocities of disk galaxies are of great importance for the study of
their evolution and the reconstruction of the underlying dark matter potential 
(see for example \citealt{bosma78,rubin78} and \citealt{bosma81}). Key areas of 
research in this context are the classical and baryonic Tully-Fisher relations, two empirical 
relations between the luminous or baryonic mass of a spiral galaxy and its peak rotation velocity 
(for recent studies see \citealt{pfenniger05, begum08, trachternach09} and \citealt{stark09}). 
These relations can be used to measure distances, constrain properties of dark matter and 
study galaxy evolution as a function of redshift \citep{combes09}. In this section we compare the rotational velocity obtained from \HI\ line profiles and those obtained from kinematic modelling.

\begin{figure*}
\begin{tabular}{cc}
\epsfig{file=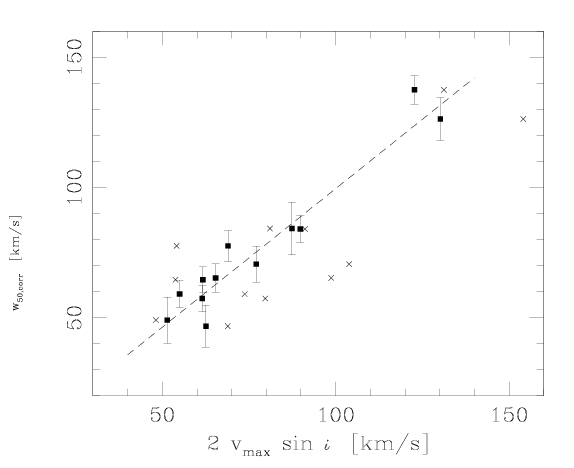, width=0.5\linewidth}&
\epsfig{file=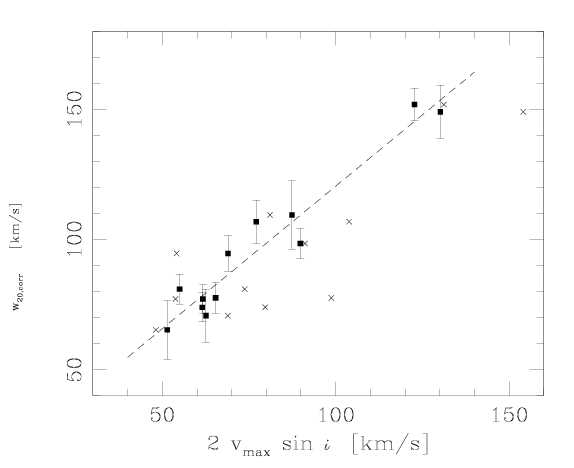, width=0.5\linewidth}\\
\end{tabular}
\caption{Comparison between the derived rotational velocity using the tilted ring analysis and the \HI\ line width 
measured at the 50\% and 20\% level of peak flux denisty. The \HI\ line width is corrected for instrumental broadening. The black squares show the rotational velocity corrected using the inclination of the tilted ring model. The crosses show the considerable scatter introduced when the optical inclination is used. The dashed lines lines show the weighted least squares fit to the data (equations~\ref{eq:w50} and~\ref{eq:w20}).} \label{fig:vrotw2050} 
\end{figure*}

In Fig.~\ref{fig:vrotw2050}  we show the relationship between the \HI\ line width  (from the LVHIS spectrum; Table~\ref{tab:spectralprops}) measured at the 50 percent  
($w_{50}$, left panel) and 20 percent ($w_{20}$, right panel) level of peak flux density and the maximum 
rotational velocity obtained by tilted ring modelling (see Table~\ref{tab:rotcurresults}). The black squares represent the rotational velocity corrected using the inclination of the tilted ring model whereas the crosses show the rotational velocity corrected using the optical inclination.
The \HI\ line width has been corrected for instrumental broadening using the \cite{bottinelli90} method. 
We also show the weighted least squares fits to the data as 
dashed lines which are described by:
\begin{eqnarray}
w_{50} &=& (1.1 \pm 0.2)\cdot 2 v_{max}\sin i - (4.7 \pm 14.6)\label{eq:w50}\\
w_{20} &= &(1.1 \pm 0.2) \cdot 2 v_{max}\sin i+ (9.3 \pm 15.3)\label{eq:w20}
\end{eqnarray}
The slope and y-intercept of the two relations are fully consistent within the measured uncertainties, 
with a one-to-one relationship between the line width and the rotational velocity (ie., a slope of unity 
and a y-intercept = 0). The line widths measured at the 20\% and the 50\% level of peak flux density 
are highly correlated to the rotational velocity (correlation coefficient = 0.97 and 0.98, respectively). 

From the tight correlations in eqns. \ref{eq:w50} and \ref{eq:w20} we conclude that $w_{20}$ and $w_{50}$ both are 
highly suitable quantities to infer the rotational velocity in the velocity range $50<v<150\,$\kms. 
Previous studies \citep{corbelli97,meyer08} were suggesting that $w_{50}$  is the optimal 
quantity. This preference is not surprising as the HIPASS data on which the \citealt{meyer08} study was based
on had a much lower signal-to-noise than our data (average rms noise of 13mJy compared to 1.5mJy 
for our sample). Using $w_{50}$ naturally is less affected by the noise in the radio continuum. 
The \cite{corbelli97} study also recommended adopting the $w_{50}$ to minimize the effects of 
warped outer disks, a feature that was not observed in our sample. 

The rotational velocity can be retrieved from the \HI\ line widths by using equations~\ref{eq:w50} or 
\ref{eq:w20} but this simple approach does not take into account possible turbulent motions 
within the disk, which are particularly important when studying dwarf galaxies where the maximum rotation
velocities have comparable amplitudes \citep{patterson96}. A physical model that addresses this issue
was put forward by \cite{tully85}. Gaussian random motions are subtracted linearly for fast rotators 
and in quadrature for dwarf galaxies using the equation:
\begin{eqnarray}
v^2_{max} &=&w_l^2 + w_{t,l}^2\left[1-2e^{-(w_l/w_{c,l})^2}\right] \nonumber\\
&~&- 2w_l w_{t,l}\left[1-e^{-(w_l/w_{c,l})^2}\right]\label{eq:turb}
\end{eqnarray}
where $l$ is the 20\% or 50\% level of peak flux density, $w_{t,l}$ characterises the amount of profile 
broadening caused by random motions and $w_{c,l}$ is chosen to give a 
smooth transition between the boxy shaped HI profiles of large spiral galaxies and the Gaussian profiles of 
dwarf galaxies. We follow \cite{verheijen01a} and adopt the values $w_{c,50}=100$\,\kms\ and 
$w_{c,20}=120$\,\kms. Using our sample we find that the optimal values which allow an accurate 
retrieval of the maximum rotational velocity from the line widths to be:
\begin{eqnarray}
w_{t,50}&=&2\pm 14~\textrm{km s}^{-1}\nonumber\\
w_{t,20} &=&34\pm 10~\textrm{km s}^{-1}\nonumber
\end{eqnarray}

Our value of $w_{t,20}$ is slightly larger than $w_{t,20} =22$\,\kms ~obtained by \cite{verheijen01a}. 
However, given that no uncertainties were quoted and the large scatter in the data (approximately 50\kms) 
in their plot of global profile width versus rotational velocity (see their Figure 2), it is reasonable 
to say that the two studies are in agreement.  Our value for $w_{t,50}=2\pm 14$\,\kms\ is in good agreement 
with their value of $w_{t,50}=5$\,\kms ~within the measured uncertainties. The two improved equations of the 
\cite{tully85} model with the newly derived parameters are:
\begin{eqnarray}
v^2_{max} &=&w_{50,corr}^2 + (2\pm 14)^2\left[1-2e^{-(w_{50,corr}/100)^2}\right] \nonumber\\
&~&- w_{50,corr} (4\pm 28)\left[1-e^{-(w_{50,corr}/100)^2}\right]\label{eq:newtub1}\\
v^2_{max} &=&w_{20,corr}^2 + (34\pm 10)^2\left[1-2e^{-(w_{20,corr}/120)^2}\right] \nonumber\\
&~&- w_{20,corr} (68\pm 20)\left[1-e^{-(w_{20,corr}/120)^2}\right]\label{eq:newtub2}
\end{eqnarray}

This analysis has demonstrated that it is indeed possible to estimate the disk rotation 
velocity from the \HI\ line width measurements either using  the simple model given by 
equations~\ref{eq:w50} and \ref{eq:w20} or equations~\ref{eq:newtub1} and \ref{eq:newtub2}
when broadening by turbulent motions is a concern. However, there is considerable scatter in Figure~\ref{fig:vrotw2050} making resolved observations essential if an accuracy of greater than $\sim10$\,\kms\ in the rotational velocity measurement is required. We also note that using an optical inclination rather than the \HI\ inclination can introduce significant error as these two values are not always identical (optical emission only traces the inner region of a galaxy).
It is also important to note that the rotational velocity derived is not necessarily the peak velocity of the galaxy disk
because HI observations for low surface brightness dwarf galaxies often lack the sensitivity to trace 
the \HI\ emission in the low column density regime, typically a few galactic-disk scale radii where 
the rotation curve becomes flat. Finally, the above discussion regarding the calculation of the rotational 
velocity from the \HI\ line width is  valid for galaxies in the nearby Universe. Measurements for galaxies at higher redshifts require 
an extra line width correction to take into account relativistic broadening of the \HI\ profile using the standard 
factor of $(1+z)$.

\section{Individual Galaxies}\label{s:individualkinematics}
\subsection{AM0605-341} \label{ss:am0605}

\begin{figure} 
\epsfig{file=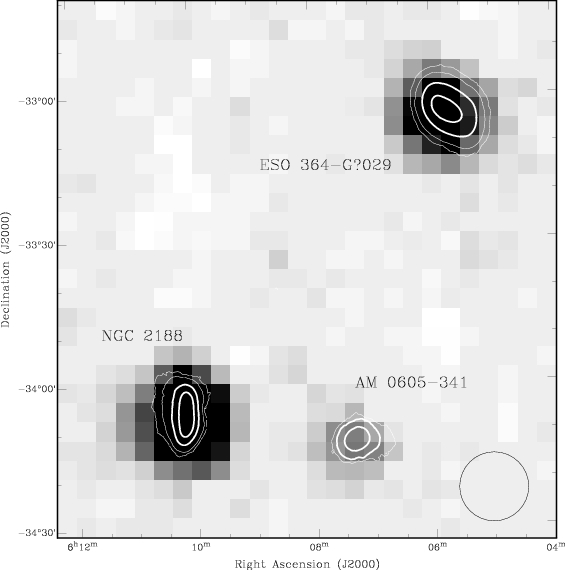, width=\linewidth}
\caption{The HIPASS image of the galaxy group containing AM0605-341,  NGC2188 and ESO364-G?029.  Due to the large beam size (shown at bottom right), these galaxies were unresolved in the HIPASS image. Overlaid are contours of the LVHIS integrated \HI\ intensity distribution which have been enlarged (x3) so that they are visible on the widefield HIPASS image. Both NGC2188 and AM0605-341 have extensions of \HI\ gas which we propose is of tidal origin. }
\label{fig:group}
\end{figure}

AM0605-341 is a nearby ($D=7.2$\,Mpc) magellanic type spiral galaxy  (see Figure~\ref{fig:opt}).
It has a structureless stellar disk and a very bright bar \citep{matthews97}, possibly with a starburst
nucleus \citep{matthews02}. This galaxy forms a small ensemble with the late-type spiral NGC2188
(separation of 35.6 arcmin and 18\kms) and the irregular galaxy ESO364-G?029 (separation of 70.2 arcmin and 22\kms). The HIPASS image of this group is shown in Figure~\ref{fig:group} along with contours showing the \HI\ distribution measured by LVHIS (enlarged by a factor of 3 so that it is visible on the widefield HIPASS image).

The small angular diameter of AM0605-341 ($\sim4$\,arcmin) makes resolved \HI\ imaging extremely difficult to obtain.
Previous \HI\ observations \citep{gallagher95,matthews98,meyer04} have been from single dish telescopes and
were able to measure only a point source spectrum. Our \HI\ synthesis imaging of AM0605-341 thus reveals the atomic
hydrogen distribution for the first time.

We find an asymmetric   \HI\ distribution  with respect to
the optical and  dynamic centres (see Figure~\ref{fig:rotcur1}). There is an extension of redshifted \HI\ located
to the west of the galaxy. This is opposite to the direction of NGC2188 (located to the east of AM0605-341). \cite{domgeorgen96} have
already noticed that NGC2188  shows a similar extension directly opposite to AM0605-341.  These authors
suggested that the extension observed in NGC2188 is unlikely to be due to some  interaction. However, they identify 
only ESO364-G?029 as a potential disturber although they do point out that an interaction with an unknown neighbour was
possible. We propose that the extensions of NGC2188 and AM0605-341 are of tidal origin due
to their interaction. The angular distance of the two galaxies is 77\,kpc.

There is no distortion to the AM0605-341 velocity field indicating a bar feature. However the bright bar observed
by \cite{matthews97} is smaller than the synthesized beam, so the non-detection in the \HI\ is to be expected.

We derive a \HI\ rotation model in Figure \ref{fig:rotcur1}. Because of the galaxy's small angular size,
the inclination measurement did not converge using a tilted ring model. Thus a value was chosen such that the model
had the same semi-major and minor dimensions as the observed \HI\ distribution. This corresponds to an inclination
of 50 degrees which was adopted to derive the AM0605-341 rotation curve. Setting the inclination
to slightly higher or lower values ($\pm 10$ degrees) does not significantly increase the residual difference  between the observed \HI\ velocity field
and the model due to the degeneracy between the rotational velocity and the inclination \citep{begeman89}. Also, the chosen inclination is higher than the optical inclination of 26 degrees from \cite{matthews02}.
A rotation curve could not be obtained separately for the approaching and receding sides of the galaxy.

A H$_{\alpha}$ position velocity curve was measured for AM0605-341 out to $r=20\arcsec$ by \cite{matthews02}
who concluded that the galaxy must have extended \HI\ and reach its maximum rotational velocity well outside its
stellar disk. Our result, with the rotation curve still increasing at the observed \HI\ column density limit ($r=120\arcsec$),
is in good agreement with the Matthews \& Gallagher result.

\subsection{AM0704-582/ Argo}

The near-IR image of the Argo dwarf (Figure~\ref{fig:opt}, top panel, second to left) shows a very low surface brightness
galaxy with no visible structures such as a bar or spiral features. The optical emission appears very extended and has
almost no compact region \citep{parodi02}. The galaxy is extremely isolated in space with a tidal index of -2 using
the \cite{karachentsev99} measure of interaction. The atomic hydrogen emission
of this galaxy has been observed previously by HIPASS \citep{koribalski04} and a point source spectrum was obtained.
Here we present the first resolved \HI\ synthesis imaging of the Argo Dwarf.

Our \HI\ map reveals a regular velocity field. We also see that the peak integrated density is offset from the centre
of the \HI\ emission. The stellar component is located at the centre of the \HI\ emission which is also the location
of the dynamic centre as derived by our rotation curve analysis.

The orientation  parameters of the \HI\ disk are inconsistent with orientation parameters for the stellar disk in the literature,
but this is due to the large uncertainty in the optical measurements rather than a detection of a warp. The position
angle of the \HI\ disk derived by rotation curve fitting was effectively constant as a function of radius and determined
at $276\pm 2$ degrees. This is different to the values of 225 degrees \citep{kirby08} and 354 degrees \citep{parodi02}
however due to the extreme low surface brightness of the optical emission, these values have considerable uncertainty.
Hence deeper optical imaging is required to determine whether the optical position angle and \HI\ position angle are aligned.
We found that the inclination could not be determined using the tilted ring analysis. Thus the inclination was chosen so that the model had
the same semimajor and semiminor dimensions as the observed HI distribution. This yielded an inclination of 35 degrees
which was used to derive the rotation curve. Once again, it should be noted that the inclination and the rotational velocity are degenerate parameters  \citep{begeman89}, therefore changing the inclination does not significantly change the residual difference  between the observed \HI\ velocity field
and the model.  \cite{kirby08} measured the inclination
of the outermost isophote to be 65 degrees by fitting ellipses as a function of radii, however once again, it should be
noted that due to the extreme low surface brightness, this value has considerable uncertainty.

The kinematic model obtained by rotation curve analysis is in good agreement with the observed velocity field. The typical differences between the observations and the model are less than 4\kms. A rotation curve could not be obtained separately for the approaching and receding sides of the galaxy.

\subsection{ESO059-G001}

ESO059-G001 is an isolated low-surface brightness dwarf irregular galaxy 4.57\,Mpc away from the Milky Way \citep{karachentsev06}.
The galaxy has a tidal index of $-1.5$ using the \cite{karachentsev99} measure of interaction. \cite{parodi02}
found rudiments of spiral arms in their $B$-band imaging which were subsequently confirmed by ACS-HST imaging
\citep{karachentsev06}. \cite{helmboldt04,helmboldt05} reported two HII regions and a total H$\alpha$ flux
of $\log(F_{H\alpha})=-12.19\,$ergs\,s$^{-1}$\,cm$^{-2}$.

An \HI\ point source spectrum
was initially obtained by HIPASS \citep{koribalski04} while our resolved \HI\ synthesis image of ESO059-G001 reveals
details of the atomic hydrogen distribution for the first time.
The galaxy shows a regularly rotating \HI\ velocity field. The kinematic model obtained by the rotation curve analysis is in
excellent agreement with the observed velocity field with a typical residual less than 4\kms.  We found that the inclination
varied between 35 and 55 degrees with no clear trend. Thus the inclination was set to 45 degrees. We also note that this average
value optimised the dimensions of the model compared  to the observed \HI\ distribution. A rotation curve could not be obtained separately for the approaching and receding sides of the galaxy.

\subsection{ESO121-G020}
ESO121-G020 is a dwarf irregular galaxy located 6.05\,Mpc from the Milky Way. The galaxy is about 2.6\,mag fainter in the near-IR
than ESO059-G001. It has a small companion, ATCA J061608-574552, located approximately 3\arcmin\ to the south east.
\cite{warren06} detected neither an \HI\ nor a stellar bridge between the two galaxies.

The galaxy ESO121-G020 was analysed using new LVHIS data and ATCA archival data that
was originally obtained by \cite{warren06}. The combined data set easily resolves the two galaxies in angular and velocity space.
A slight distortion is observed in the southeastern region of the velocity field most likely caused by the companion,
however the strength of this feature is low compared to the overall bulk rotation. The companion galaxy
was masked from the data during the kinematic analysis of ESO121-G020.

The dynamic centre could not be obtained with the help of a tilted ring model and was thus assumed  to be the optical centre.
This assumption was justified by the observation that the dynamic centres for all galaxies in our sample
agree with the optical centres as defined by the outermost isophotes (see \citealt{trachternach08} for similar result).
In contrast to the \cite{warren06} kinematic model, the inclination value was statistically unstable.
The value of $78\pm 5$ used by Warren et al.\ implies that the galaxy would be almost edge-on
which is clearly not the case (see Figure \ref{fig:rotcur1}). This discrepancy was investigated more closely and we
concluded that the high inclination obtained by Warren et al.\ was in fact the result of fitting a small number of data
points rather than a genuine better fit. The \HI\ image had very low resolution and the fitting routine
tended to higher inclination values to reduce the number of data points used in the fitting process.
While the residuals were lower, this was a direct consequence of comparing fewer data points, not that the
obtained model was  intrinsically more accurate.  We choose to model ESO121-G020 by keeping the
inclination fixed at 40 degrees which results in a model with the same semimajor and semiminor
dimensions as the observed HI distribution.

The overestimation of the inclination in the Warren et al.\ result has a direct effect on some of the
values quoted in their paper. We measure a rotation velocity of 36\kms ~at the maximum
radius of 80\,arcsec compared to their value of 21\kms. At our outermost radius (113 arcsec)
we measure a rotation velocity of 49\kms\ with the rotation curve still rising. These new parameters
and the improved TRGB distance measurement \citep{karachentsev06} imply that the lower limit for the
dynamical mass is $1.7\times 10^9$\Msun\ ($M=v^2rG^{-1}$). Using the \cite{warren06} total
$B$-band luminosity of $(2.39\pm0.13)\times 10^7$\Lsun, we get a lower limit for the dynamical
mass-to-light ratio  of $\sim70$\Msun/\LLsun\ suggesting that this galaxy is a dark matter
dominated object. Other low luminosity dwarf galaxies are known to have mass-to-light ratios similar to this  and higher (see for example ESO215-G?009 which has a dynamical mass-to-light ratio of $200\pm110$\Msun/\LLsun, \citealt{warren04})

The incorrect inclination obtained by \cite{warren06} highlights the need for visual inspection
of the model rotation field compared to the observed rotation field. This is particularly important
for upcoming surveys in the Square Kilometre Array (SKA) era, where due to the sheer volume
of data automated pipelines will be employed to do rotation curve fitting.

\subsection{ESO137-G018}
ESO137-G018 is an isolated late-type galaxy of type Sm or Im  located behind the
Galactic plane ($b=-7.4^{\circ}$). It has several bright superimposing foreground stars (Figure~\ref{fig:opt})
and hence the optical properties are poorly constrained. The current distance estimate ($D=6.4$ Mpc) is based on the
TRGB magnitude \citep{karachentsev07}.

The only previously available \HI\ observations of ESO137-G018 were obtained by HIPASS \citep{koribalski04}
which measured point source spectrum. Our \HI\ synthesis observations show that the integrated density field is
very symmetric about the major and minor axes of the galaxy. The velocity field resembles an undisturbed,
regularly rotating system. The kinematic model obtained by rotation curve analysis is in good agreement
with the observed velocity field. The typical velocity residuals are less than 5\kms. Our rotation curve shows
that the \HI\ is tracing the galaxy out to the radius at which the peak rotation velocity at $80\pm 2$\kms\ is reached.

We find that the position angle of ESO137-G018 decreases from 33 to 28 degrees, but the inclination is
steady at $50\pm 6$ degrees. The position angle on the approaching and receding sides are well constrained
at $30 \pm 2$  and $33 \pm 2$ degrees, respectively. The inclination on the approaching and receding sides
are $50 \pm 7$ and $48 \pm 4$ degrees, respectively. Thus the approaching and receding sides of ESO137-G018
exhibit similar kinematic behaviour.

\subsection{ESO174-G?001}
ESO174-G?001 is a nearby ($D=6$\,Mpc) low-surface brightness (LSB) galaxy  located near the Galactic plane ($b=8.6^{\circ}$).
The galaxy is a member of the Centaurus A Group \citep{banks99} and has poorly constrained optical properties.
The optical image of ESO174-G?001 shows a uniform and elongated distribution of stars\footnote{Note that the optical image currently available in NED is incorrect. The optical image available is that of ESO174-G001 (not ESO174-G?001) which has the coordinates 13h33m19.7s, $-53$d21m17s.}.

Previous \HI\ studies of this galaxy \citep{banks99,huchtmeier01,koribalski04} have been from single dish telescopes and were
only able to measure a point source spectrum. Here we present the first resolved \HI\ imaging of ESO174-G?001.
We observe a disturbed \HI\ velocity field (Figure~\ref{fig:rotcur2}). The kinematic major axis  of ESO174-G?001 is not
perpendicular to the kinematic minor axis indicating that the gas is moving in elliptical orbits in the plane of the
galaxy \citep{bosma78, simon03}.
Thus it was to be expected that the tilted ring model will not  accurately reproduce the observed
velocity field. Due to this, ESO174-G?001 has not been included in the Tully-Fisher analysis presented in section~\ref{s:tfr}. It should be noted that  a tilted ring model was fitted to a galaxy showing non-circular motions, which means 
that the derived rotational velocity is likely to have been underestimated~\citep{rhee04,oh08}.
While a rotation curve could be derived, the residual between the observed velocity field and the model velocity
field in some areas was up 11\kms, almost 3 times higher than the velocity resolution. The flat part of the rotation curve was reached. 

The rotation curve was derived for the approaching and receding sides of the galaxy. It was found that the position angle
  decreased from 233 to 202 degrees as a function of radius both for the individual sides
and for the galaxy as a whole. The position angle of the \HI\ is not aligned with the position angle of the
stellar component (165 degrees). The inclination was
found to be $40\pm5$ degrees overall or $46 \pm7$ and $40\pm5$ degrees on the approaching and
receding sides, respectively, which is lower than the inclination of the stellar component (60 degrees).

\subsection{ESO215-G?009}
ESO215-G?009 is a LSB galaxy located near the Galactic plane.  \cite{warren04} have studied
the optical and \HI\ properties of this galaxy extensively and we refer the reader to their study. They found that ESO215-G?009 has a \HI\ mass-to-light ratio of $22\pm4$\Msun/L$_{\odot,B}$, the highest known for any galaxy.\footnote{In Table~\ref{tab:hipassprops} we list a lower value for the the \HI\ mass-to light ratio as $16.8\pm3.0$\Msun/L$_{\odot,B}$. The difference is  introduced by using the \cite{lauberts89} $B$-band magnitude (16.03 mag) whereas \cite{warren04} used a value taken from LEDA (16.43 mag). Also, their value is using ATCA data whereas Table~\ref{tab:hipassprops} is quoting HIPASS properties. Our value using ATCA measurement is fully consistent with the \cite{warren04} value.}  We modeled the kinematics using a tilted ring analysis and our results are generally consistent with those
of \cite{warren04}.

It is interesting to note that although the velocity field of ESO215-G?009 was reasonably well
modeled, the ellipticity of the model does not agree well with the observed \HI\ distribution.
The highly circular gas distribution suggests that the galaxy must have an inclination of approximately
20 degrees but this is inconsistent with the high line-of-sight rotation ($w_{20}=79\pm1$\kms)
and the derived inclination of the model ($35\pm3$ degrees). This may well be evidence that
the assumption of the gas being located in an infinitely thin disk is incorrect. Such extraplanar \HI\ was
first observed in NGC891 \citep{swaters97} and has since been observed in other nearby
galaxies \citep{boomsma05,hess09}.  Any extension of gas along the vertical component of the disk will
naturally result in an increase of the observed semi-minor axis and could explain the highly
circular observed distribution of \HI\ gas in ESO215-G?009.

\subsection{ESO245-G005}

ESO245-G005 is a barred magellanic galaxy located 4.4 Mpc away from the Milky Way. It has a star formation rate of $0.02\,M_{\odot}$\,yr${}^{-1}$ \citep{oey07}. \cite{cote00}
suggest that due to a strong abundance oxygen gradient along the bar (oxygen abundance changes by a factor of three; \citealt{miller96}),
either an \HI\ cloud of a different metallicity has been accreted or that the whole object is the product
of a recent merger. \cite{cote00} have obtained  \HI\ synthesis imaging and observed a very peculiar velocity field. Despite the indications that ESO245-G005 has had past interactions, no shock waves are detected through the study of its diffuse ionised gas \citep{hidalgo06}.

Our \HI\ imaging shows that the integrated density field has several regions of peak flux density.
The velocity field is  disturbed. A kinematic model was derived by tilted ring analysis but
the residual between the observed velocity field and the model velocity field in some areas was up 10\kms.

We find evidence that ESO245-G005 has a warp in its outer regions. At a radius of $r\approx200$ arcsec there is a rise in inclination and an associated change in the
position angle. The detection of a warp gives further support to the \cite{cote00} hypothesis that ESO245-G005 has undergone recent  accretion \citep{ostriker89,jiang99}.

The \HI\ position angle was found to rise from 70 degrees in the inner portion of the galaxy ($r<200$ arcsec)
 to 98 degrees in the outer regions. This is not aligned with the position angle of  the stellar
disk (127 degrees; \citealt{kirby08}). The \HI\ inclination was modeled $36\pm6$ degrees which is also
not aligned with the inclination of the stellar disk (52 degrees; \citealt{kirby08}). The \HI\ inclination
increases and becomes highly variable  in the outer parts  ($56\pm12$ degrees  for $r>200$ arcsec).
Due to this variability, the entire galaxy was modelled using the inclination derived
for the inner region.   Kinematic modelling could not be
achieved for the approaching and receding sides separately.

The \HI\ kinematics of ESO245-G005 has been studied previously by  \cite{cote00} whose results
differ somewhat from those presented here. \cite{cote00} observed a similar trend in
position angle however chose to model the position angle as a constant value of $88\pm8$ degrees.
They find a higher inclination of $54\pm10$ degrees which is similar to that which we obtain for the
inclination at larger radii.

\subsection{ESO308-G022}
ESO308-G022 is a dwarf irregular galaxy located $\sim 7.6$ Mpc away from the Milky Way. The deep near-IR image (see Figure~\ref{fig:opt}) shows an extended stellar distribution with no visible structures such as a bar or spiral features.  The optical image of \cite{parodi02} shows an extended and diffuse galaxy with a few regions of brighter emission.

The new LVHIS \HI\ synthesis imaging of ESO308-G002 reveals the atomic hydrogen distribution for the first time. Previous \HI\ observations \citep{matthews95,huchtmeier00,meyer04} of this galaxy have been obtained by single dish telescopes and were only able to measure a point-source spectrum. Our \HI\ imaging is able to resolve ESO308-G002  despite its small angular diameter ($\sim 4$ arcmin).

 The kinematic model obtained by rotation curve analysis is in good agreement with the observed velocity field. The typical differences between the observations and the model are less than 6\kms. A rotation curve could not be obtained separately for the approaching and receding sides of the galaxy. The rotation curve does not flatten at the last points measured suggesting that the \HI\ is not tracing the galaxy out to the radius at which the maximum rotation velocity is reached.

Due to the small angular diameter of  ESO308-G022 our observations have low resolution and the inclination could not be determined using a tilted ring model. Thus it was chosen such that the model had the same semi-major and semi-minor dimensions as the observed HI distribution (40 degrees). The position angle was found to be constant at $82\pm2$ degrees. This is inconsistent with measurements  of the position angle of the stellar disk (130 and 160 degrees by \citealt{kirby08} and \citealt{parodi02} respectively). However, the alignment parameters for the stellar component have considerable uncertainty as the galaxy has very low surface brightness. Therefore, deeper near-IR or optical imaging is required before the alignment between the stellar and \HI\ disks can be confirmed.

\subsection{ESO325-G?011}
ESO325-G?011 is a dwarf irregular galaxy  and a member of the Centaurus A Group. It has a TRGB distance of 3.4 Mpc which is consistent with its group membership \citep{karachentsev02}. The optical image shows that the stellar distribution is asymmetric; the central area of highest surface brightness is offset from the centre defined by the outer isophotes.

The atomic hydrogen of this galaxy has been observed on numerous occasions with both single dish telescopes (\citealt{longmore82,huchtmeier86,cote97,banks99,koribalski04}) and with \HI\ synthesis  imaging \citep{cote00}. While the \cite{cote00} imaging was able to resolve the atomic hydrogen distribution of ESO325-G?011, their data had significantly lower sensitivity with an rms noise of 4.6 mJy beam${}^{-1}$ than the LVHIS value of 1.3 mJy beam${}^{-1}$.

Our \HI\ imaging shows that the integrated density is asymmetric with the peak flux density offset from the centre of the \HI\ emission. The position of the peak integrated density is located at the position of the optical core. ESO325-G?011 has  an extremely regular velocity field. The kinematic model obtained by the rotation curve analysis is in excellent agreement with the observed velocity field. The typical differences between the observations and the model are less than 2\kms, the lowest for all galaxies in our sample.

We find that the optical position angle is aligned with the \HI\ position angle. The \HI\ inclination ($42\pm10$ degrees) is lower than that of the stellar component (60 degrees), but the optical inclination was obtained from a DSS image and thus deeper imaging is required. Kinematic modelling has been done previously by \cite{cote00} who used a slightly higher inclination ($52\pm5$ degrees) for their models. This value is consistent within the uncertainties with the value used here. We also note that  \cite{cote00} fitted 4 tilted rings to their \HI\ data and only the inner two agree with the value adopted in their model. The outer two tilted rings have inclinations similar to the value we measure.

\subsection{ESO381-G020}

ESO381-G020 is a dwarf irregular galaxy located at 5.4 Mpc \citep{karachentsev07} and a member of the Centaurus A Group. The optical image shows a very asymmetric stellar distribution, similar to that of ESO325-G?011.

Previous \HI\ studies of this galaxy have been carried out using single dish telescopes \citep{longmore82,huchtmeier86,cote97,banks99,koribalski04} as well as with synthesis imaging \citep{cote00}. The kinematics of this galaxy have previously been studied by \cite{cote00} using their synthesis imaging. Their observations have higher angular resolution ($13\times 13$ arcsec compared to $80\times 49$ arcsec) but slightly lower sensitivity (rms noise of 1.7 mJy beam${}^{-1}$ compared to 1.3 mJy beam${}^{-1}$). The  \cite{cote00} study allows a useful comparison to ensure that our analysis provides consistent results with those of others.

The observed \HI\  velocity field is somewhat disturbed with the contours  of constant velocity appearing asymmetric about the major axis.  While a kinematic model could be derived, the residual between the observed velocity field and the model velocity field in some areas was up 10\kms.

The \HI\ position angle was found to increase steadily from 295  to 314 degrees as  a function of radius. The latter value is consistent with the 
position angle of 310 degrees derived from the outer isophotes of the stellar light. The inclination obtained by tilted ring analysis ($55\pm10$ degrees) is higher than the inclination of the stellar component (40 degrees). The kinematic study  of this galaxy  by \cite{cote00} obtained similar results. They modelled ESO381-G020 with a position angle of $311 \pm 1$ degrees and an inclination of $57\pm 6$ degrees. Their position angle is an average of the values obtained by their tilted ring analysis (with actual values ranging from 305 to 313 degrees). 
 
\subsection{IC5152}\label{ss:ic5152}
IC5152 is a nearby  irregular galaxy that was originally considered  a Local Group (LG) candidate  \citep{yahil77}. However current distance estimates place it at 2.1 Mpc \citep{karachentsev04}, beyond the edge of the LG. Due to its close proximity to the Milky Way, it is often included in studies of the LG dynamics and evolution (see for example \citealt{chernin04,sawa05, pasetto07}), hence requires  accurate mass modelling. The near-IR image (see Figure~\ref{fig:opt}) shows a well resolved galaxy with a fairly bright centre. The central region  of IC5152 is an active site of star formation \citep{zijlstra99} and several \HII\ regions have been studied \citep{hidalgo02,lee03}. IC5152 has many blue stars and dusty patches \citep{karachentsev02b}. The north west corner has a very bright superimposed star. Deep CO emission maps have been obtained but no emission was detected ($4\sigma$ upper limit of 0.03 K \kms) despite the presence of large amounts of neutral gas \citep{buyle06}. 

The \HI\ line profile of IC5152 has been studied previously using single dish telescopes \citep{huchtmeier86,becker88,koribalski04}. \HI\ synthesis imaging has also been obtained but only the integrated intensity map has been published   \citep{buyle06}.

\begin{center}
\begin{figure} 
\begin{tabular}{cc}
  \mbox{\epsfig{file=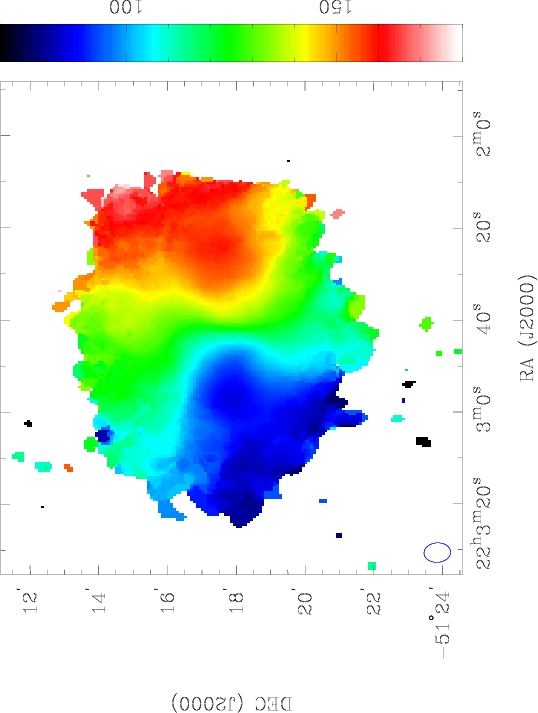,width=3cm,angle=-90}} &
  \mbox{\epsfig{file=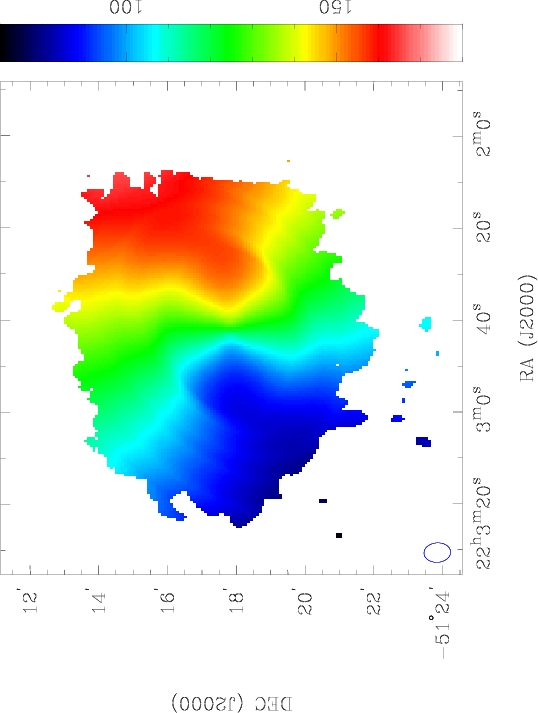,width=3cm,angle=-90}} 
\end{tabular}
\caption{[Left] The observed \HI\ velocity field of the Local Group irregular galaxy IC5152. [Right] The model velocity field masked using the region defined by the observed velocity field.}
\label{fig:ic5152}
\end{figure}
\end{center}

The \HI\ velocity field shows a twisting of the contours of constant velocity. The kinematic model obtained by rotation curve analysis is presented in Figure~\ref{fig:rotcur3}. At first glance, the model velocity field and the observed velocity field appear inconsistent as the physical dimensions are different. 
This occurs because the inclination and position angle obtained by tilted ring analysis implies a model that extends past beyond the observed \HI\ in the north west and south east regions of the galaxy. However, if the model velocity field is masked by region defined in the observed velocity field, the agreement appears to be excellent, particularly given the distortion in the observed field (see Figure~\ref{fig:ic5152}). The typical differences between the observations and the model are less than 5\kms ~with maximum deviations less than 10\kms. The physical motivation which allows for masking of the model is that the masked \HI\ gas is thought to be present but below the observed column density limit. 

The inclination of IC5152 is found to be constant as a function of radius at $49\pm6$ degrees which is consistent with the inclination of the stellar disk (50 degrees; \citealt{kirby08}). The \HI\ position angle (increasing from 271 to 298 degrees) is also consistent with the inclination of the stellar disk (275 degrees; \citealt{kirby08}).

\section{The Tully-Fisher Relation}\label{s:tfr}

\begin{figure*}
\begin{tabular}{cc}
\epsfig{file=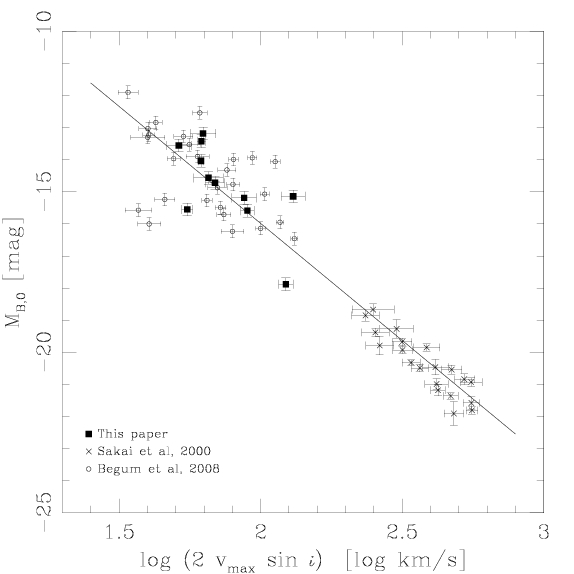, width=0.5\linewidth}&
\epsfig{file=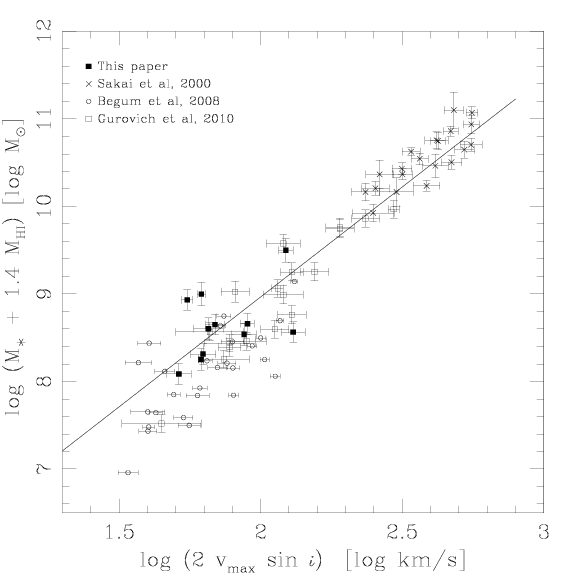, width=0.5\linewidth}\\
\end{tabular}
\caption{The classical [left] and baryonic [right] Tully-Fisher relations. Here we compare our sample to the bright galaxy sample of Sakai et al.\ (2000),  the faint galaxy sample of Begum et al.\ (2008) and the isolated disk galaxy sample of Gurovich et al.\ (2010). The weighted least squares fit to the data is given as the solid lines (see equations~\ref{eq:1} and \ref{eq:2}.)}
 \label{fig:tfr} 
\end{figure*}

The classical and baryonic Tully-Fisher relations are empirical relations between the luminous or baryonic mass of a spiral galaxy and its peak rotation velocity (for recent studies see \citealt{pfenniger05, begum08, trachternach09} and \citealt{stark09}). These relations can be used to measure distances, constrain properties of dark matter and study galaxy evolution as a function of redshift \citep{combes09} indicating the importance of accurately determining the empirical relationship, particularly in the dwarf regime. In Figure~\ref{fig:tfr} we show the classical [left panel] and baryonic [right panel] Tully-Fisher plots for our sample galaxies in the black squares. 

The classical Tully-Fisher relation shows the absolute $B$-band magnitude (calculated using the apparent magnitudes, $m_B$,  and the distances listed in Table~\ref{tab:basicprops}) plotted against the maximum rotational velocity of a galaxy obtained by rotation curve analysis (Table~\ref{tab:rotcurresults}). The $B$-band magnitudes have been corrected for Galactic extinction using \cite{schlegel98}.  Also plotted are the samples of  \cite{sakai00} which contains many bright spiral galaxies and of \cite{begum08} which contains many faint dwarf galaxies. These are indicated by open circles and crosses respectively. \cite{sakai00} provide \HI\ line widths which are used to obtain the rotational velocity by correcting for inclination and broadening due to turbulent motions (using equation~\ref{eq:newtub2}).  We determine the classical Tully-Fisher relation by a weighted least squares fit to be:
\begin{equation}\label{eq:1}
 M_{B,0}=(-7.29\pm 0.33) \log(2 v_{max}) - (1.40\pm0.71)
\end{equation}

The baryonic Tully-Fisher relation  (lower panel of Figure~\ref{fig:tfr}) shows the total baryonic mass plotted against the maximum rotational velocity of a galaxy obtained by rotation curve analysis for our sample (black squares), the Begum et al.\ sample (open circles) and the Sakai et al.\ sample (crosses). Once again, the rotational velocity  for the \cite{sakai00} sample was obtained by correcting the \HI\ linewidth for inclination and turbulent broadening (equation~\ref{eq:newtub2}).  We also plot the relatively isolated disk galaxy sample of \cite{gurovich10} in the open squares.  We determine the baryonic Tully-Fisher relation by a weighted least squares fit to Figure~\ref{fig:tfr} (right panel) as:
\begin{equation}\label{eq:2}
\log(M_{bary})=(2.51\pm 0.11) \log(2 v_{max}) - (3.94\pm0.23)
\end{equation}

 The total baryonic mass was computed by adding the total stellar mass, $\log M_{\ast}$, and the total gas mass, $1.4\, M_{HI}$. Since we have no direct measurement of the mass-to-light ratio for our sample galaxies, we estimate the stellar mass from the $B$-band luminosity using the moderate ratio $M_{\ast}/L_B=1.2\, M_{\odot}/L_{\odot,B}$ (\citealt{bell01}). The baryonic mass includes the total gas mass which we obtain using the conversion factor $M_{gas}/M_{HI}=1.4$ (following \citealt{mcgaugh00}, \citealt{geha06} and \citealt{warren07}). This takes into account primordial helium and metals but does not include molecular hydrogen. Dwarf galaxies are thought not to contain large amounts of molecular hydrogen \citep{taylor98,leroy05}. While the bright galaxies of the \cite{sakai00} sample will have significant amounts, the baryonic mass of these systems is dominated by the stellar component and the uncertainty in the mass-to-light ratio is likely to contribute a larger error than ignoring the molecular hydrogen \citep{begum08}. We note that \cite{pfenniger05} suggested that the conversion factor could be as high as 2.98. This result, however, is based on a slight reduction to the scatter of their baryonic Tully-Fisher relation rather than any observational or theoretical evidence. \cite{begum08} confirmed that the conversion factor is poorly constrained by the baryonic Tully-Fisher relation and showed that any value between $\sim2$ and 29 leads to significant  tightening of the relation. 

The tight correlation between the \HI\ line width and the true rotational velocity obtained by tilted ring analysis (see Figure~\ref{fig:vrotw2050}) provides a degree of confidence in studies which utilise the \HI\ line widths as a measure of rotation for Tully-Fisher investigations (see \citealt{meyer08} for a recent example). However, we note that obtaining accurate distance estimates rather than using the midpoint of the \HI\ line profile is imperative as both the classical and baryonic Tully-Fisher relation depend on this measurement (through the derivation of the absolute magnitude and the stellar and \HI\ mass). Also of note is the fact that 5 galaxies (AM0605-341, ESO174-G?001, ESO245-G?009, ESO325-G?011 and  ESO381-G020) in our sample were found to have an inclination of their \HI\ disk different  to the measured optical inclination (by more than 5 degrees). When deprojecting the observed rotational velocity obtained from an \HI\ line width, it is the inclination of the \HI\ disk which should be used. 
 
\section{Conclusion}\label{s:HIsummary}

We have presented a kinematic study of 12 galaxies in the Local Volume ($D<10$ Mpc). The data was obtained using the ATCA as part of the LVHIS Survey. For six galaxies in our sample (AM0605-341, Argo Dwarf, ESO059-G001, ESO137-G018, ESO174-G?001, ESO308-G022) we have presented the only resolved \HI\ imaging available to date, revealing the atomic hydrogen distribution for the first time.

The global \HI\ line spectra are presented and compared to those obtained by HIPASS. The spectrum of ESO245-G005 shows a strong \HI\ absorption line at a redshift of 386\kms. The spectrum of ESO381-G020 shows two weak absorption lines at 581\kms\ and 597\kms. The comparision to the single-dish spectra of HIPASS shows that the new \HI\ synthesis observations of IC5152 are missing flux due to missing short baselines in the interferometer. The \HI\ line widths obtained by HIPASS and LVHIS are consistent within the experimental uncertainties.

Rotation curve analysis has been carried out for all sample galaxies by applying a tilted ring model to the observed velocity field. The best fitting rotation curve parameters are provided.

We show that the \HI\ line width can be used to derive the rotational velocity of galaxies in the velocity range $50<v<150$\kms\ using the newly updated \cite{tully85} model, however resolved observations are essential if an accuracy greater than approximately 10\,\kms\ is required. 

We derive classical and baryonic Tully-Fisher relations. These relationships will be explored further in the near future using LVHIS data and new deep $H$-band imaging currently being obtained at the 3.9m Anglo-Australian Telescope.

AM0605-341 was found to have an extension of redshifted \HI\ located to its west. We propose that this is due to a tidal interaction with its nearby neighbour NGC2188, which has previously been found to have a similar tidal extension.

ESO121-G020 was found to have a much lower inclination (40 degrees) than the current value available in the literature (78 degrees; \citealt{warren06}). We derive the new lower limit for its dynamical mass as $1.7\times10^9$\Msun.

It was observed that the observed ellipticity of ESO215-G?009 is inconsistent with its kinematic inclination ($35\pm3$ degrees). The observed \HI\ distribution is highly circular implying that the galaxy is nearly face-on. This may be evidence that the assumption of the gas being located in an infinitely thin disk is incorrect. 

We find evidence that ESO245-G005 has a warp in its outer disk, located at an angular radius of 200 arcsec. This lends support to the \cite{cote00} hypothesis that ESO245-G005 has undergone recent accretion.

\section{Acknowledgements}
The authors thank Nic Bonne, Janine van Eymeren, Erwin de Blok and Juergen Ott for their help and advice during the preparation of this manuscript. We also thank the referee for their useful comments and suggestions for improvements. We acknowledge financial support from the Australian Research Council Discovery Project Grant DP0451426. This research has made use of the NASA/IPAC Extragalactic Database (NED) which is operated by the Jet Propulsion Laboratory, California Institute of Technology, under contract with the National Aeronautics and Space Administration. This research has made use of NASA's Astrophysics Data System.

\bibliographystyle{mn2e}
\bibliography{mybibfile}

\end{document}